\documentclass[aps,prb,twocolumn,amsmath,amssymb,superscriptaddress,floatfix,eqsecnum]{revtex4-2}
\usepackage{amsmath}
\usepackage{amsthm}
\usepackage{mathtools}
\usepackage{graphicx}
\usepackage{amssymb}
\usepackage{bm} 
\usepackage{ulem}   
\usepackage{slashed} 
\usepackage{color} 
\usepackage{bbm}
\usepackage{amsfonts}
\usepackage[breaklinks=true,colorlinks,citecolor=blue,linkcolor=blue,urlcolor=blue]{hyperref}
\usepackage{cases}
\usepackage{float}
\usepackage{url}
\usepackage{framed}
\usepackage{bbold}
\usepackage{cancel,soul,ulem}
\usepackage{appendix}
\usepackage{hhline}

\definecolor{bleu}{rgb}{0.0, 0.5, 0.69}

\definecolor{amar}{rgb}{0.9, 0.17, 0.31}

\definecolor{brred}{rgb}{0.8, 0.25, 0.33}
\definecolor{frblue}{rgb}{0.0, 0.45, 0.73}

\def\=={\raisebox{0.35pt}{$\mathrm{:}$}\!\!=}
\def\df{\,\raisebox{1.20pt}{.}\hspace{-3.1pt}\raisebox{3.35pt}{.}\!\!=}

\newcommand{\bea}{\begin{eqnarray}}
\newcommand{\eea}{\end{eqnarray}}

\newcommand{\ket}[1]{\lvert #1 \rangle}
\newcommand{\bra}[1]{\langle #1 \rvert}

\theoremstyle{definition}

\theoremstyle{remark}

\usepackage{pdfpages}
\makeatletter
\AtBeginDocument{\let\LS@rot\@undefined}
\makeatother

\begin{document}

\title{
Elementary derivation of the stacking rules
of invertible fermionic topological phases in one dimension
      }
\author{\"{O}mer M. Aksoy}
\affiliation{Condensed Matter Theory Group, Paul Scherrer Institute, CH-5232 Villigen PSI, Switzerland}

\author{Christopher Mudry}
\affiliation{Condensed Matter Theory Group, Paul Scherrer Institute, CH-5232 Villigen PSI, Switzerland}
\affiliation{Institut de Physique, EPF Lausanne, CH-1015 Lausanne, Switzerland}

\begin{abstract}
Invertible fermionic topological (IFT) phases are gapped phases of
matter with nondegenerate ground states on any closed spatial manifold.
When open boundary conditions are imposed,
nontrivial IFT phases support
gapless boundary degrees of freedom. Distinct IFT phases in
one-dimensional space with an internal symmetry group $G^{\,}_{f}$ have been
characterized by a triplet of indices $([(\nu,\rho)],[\mu])$. Our main result is an elementary
derivation  of the fermionic stacking rules of one-dimensional IFT
phases for any given internal symmetry group $G^{\,}_{f}$
from the perspective of the boundary, i.e.,
we give an explicit operational definition for
the boundary representation
$([(\nu^{\,}_{\wedge},\rho^{\,}_{\wedge})],[\mu^{\,}_{\wedge}])$
obtained from stacking two IFT phases
characterized by the triplets of boundary indices
$([(\nu^{\,}_{1},\rho^{\,}_{1})],[\mu^{\,}_{1}])$
and
$([(\nu^{\,}_{2},\rho^{\,}_{2})],[\mu^{\,}_{2}])$,
respectively.
\end{abstract}

\date{\today}

\maketitle

\section{Introduction}
\label{sec:Intro}

Invertible topological phases of matter are described 
by Hamiltonians that are spatially local and support
a nondegenerate and gapped ground state 
on \textit{any closed} spatial manifold, 
after the thermodynamic limit has been taken.
By convention, such a Hamiltonian is 
said to realize the trivial invertible topological phase if its
gapped and nondegenerate ground state 
is a direct product of localized states, one localized state for each 
local degree of freedom. When space is not a closed manifold,
any Hamiltonian realizing a 
\textit{nontrivial} invertible topological phase must support
gapless degrees of freedom that are localized at the boundaries.
The problem of classifying the
invertible topological phases in $d$-dimensional space has attracted
a lot of interest in the last decade%
~\cite{Chen2013,Kapustin2015,Freed2021,Gaiotto2019}.

A continuous deformation of a local Hamiltonian is defined to include both
the continuous change of short-range couplings
between all existing local degrees
of freedom or the addition (removal) of decoupled local degrees of freedom
that realize a trivial invertible topological phase of their own.
Any pair of Hamiltonians
with nondegenerate and gapped ground states on any closed manifold
are said to be equivalent if they can be continuously
deformed into one another without closing the spectral gap.
Invertible topological phases 
are then defined as the equivalence classes of such Hamiltonians 
under gap-preserving continuous deformations.
Invertible topological phases display a group structure
under a composition rule called the \textit{stacking rule}.
The stacking of any pair of invertible phases consists in
creating a new invertible phase by defining the new local degrees
of freedom to be the union of the local degrees of freedom
from a representative of each invertible phase
and by defining the new Hamiltonian acting on the new
local degrees of freedom by taking the direct sum of
the pair of representative
Hamiltonians for each invertible phase.
If the invertible topological phase
resulting from stacking is the trivial one, then the pair of
stacked invertible topological phases are inverse pairs.

The classification of invertible topological phases can be enriched by
imposing an internal (independent of space) symmetry group $G$
such that two invertible phases are
equivalent only if they can be continuously deformed to one another
without gap closing and without (neither explicitly nor spontaneously)
breaking the $G$ symmetry. Those invertible topological phases that are
equivalent to the trivial phase under the continuous deformation that
spontaneously or explicitly break the $G$ symmetry are called the
\textit{symmetry protected topological} (SPT) phases%
~\cite{Gu2009,Chen2013}. 
When open boundary conditions are imposed, SPT phases support gapless
degrees of freedom at the boundaries that are protected by the
$G$ symmetry, i.e., the boundary degrees of freedom cannot be gapped
unless $G$ symmetry is either explicitly or spontaneously broken.

Invertible topological phases are best understood in one-dimensional
space. Their classification has been conjectured based on the study
of translation-invariant and injective matrix product states (MPS)
\cite{Fidkowski2011,Chen2011a,Schuch2011,Bultinck2017,Turzillo2019}.
By taking advantage of the split property
of nondegenerate gapped ground states
in one-dimensional space \cite{Matsui13},
Bourne and Ogata in Ref.\ \onlinecite{Bourne2021}
have derived rigorously for any internal symmetry group $G$
an exhaustive classification of
invertible fermionic topological (IFT) phases
that includes their stacking rules.

In this paper, we build on the seminal work 
by Fidkowski and Kitaev in Ref.\ \onlinecite{Fidkowski2011} and provide
an operational construction of the boundary representations of any
internal $G^{\,}_{f}$ symmetry imposed on IFT phases in one-dimensional space.
We find the counterparts that characterize the boundary representations
to the topological indices used to classify one-dimensional IFT phases
from a bulk perspective
in Refs.\ \onlinecite{Turzillo2019,Bourne2021}. Moreover, we
explicitly derive their stacking rules by elementary methods.
Our stacking rules differ from the ones derived in Ref.\
\onlinecite{Bultinck2017},
but agree with the ones derived in Refs.\
\onlinecite{Turzillo2019,Bourne2021}.

This paper is organized as follows.
In Sec.\ \ref{sec:IFTP in 1D} we lay out our strategy 
and summarize the results. In Secs.\ \ref{sec:boundary rep gen}
and \ref{sec: Boundary projective representation of Gf}, we define 
the boundary degrees of freedom and the representation of the 
internal symmetry group $G^{\,}_{f}$ acting on them, respectively,
for any one-dimensional IFT phase.
In Sec.\ \ref{sec: Definition of indices}, we define a triplet 
of indices $([(\nu,\rho)],[\mu])$ that characterizes the 
boundary properties of a one-dimensional IFT phase. 
In Sec.\ \ref{sec: Fermionic Stacking Rules}, we derive the
fermionic stacking rules of one-dimensional IFT phases by only using 
elementary means. In Sec.\ \ref{sec: Ground-state degeneracies},
we relate some supersymmetric
properties of the ground-state manifold
when open boundary conditions are chosen
to some values taken by the triplet
$([(\nu,\rho)],[\mu])$. 
Our conclusions can be found in Sec.\ \ref{sec:conclusion},
while Appendices \ref{appsec:Group Cohomology},
\ref{appsec:central extension review}, and \ref{appsec:classification of Gf}
review group cohomology,
central extension class,
and the detailed
definition of the triplet of boundary indices  $([(\nu,\rho)],[\mu])$,
respectively.

\section{Strategy and Summary of results}
\label{sec:IFTP in 1D}

The classification of the IFT phases in one-dimensional space is
intimately related to the classification of the projective
representations of the fermionic symmetry group $G^{\,}_{f}$,
an internal symmetry acting globally on the fermionic Fock space.
To illustrate this, we will first consider the representations
of $G^{\,}_{f}$ on a closed one-dimensional chain 
and then investigate the consequences of imposing open boundary 
conditions.

We denote by $\Lambda$ the set of points  
on a one-dimensional lattice. 
Given are the fermionic symmetry group $G^{\,}_{f}$
(Appendices \ref{appsec:Group Cohomology} and
\ref{appsec:central extension review})
and a global fermionic Fock space $\mathfrak{F}^{\,}_{\Lambda}$
defined over $\Lambda$. 
We assume that there exists a faithful trivial representation 
$\widehat{U}^{\,}_{\mathrm{bulk}}$ of the group $G^{\,}_{f}$
on the lattice $\Lambda$.
In other words, there exists an injective map 
$\widehat{U}^{\,}_{\mathrm{bulk}}:
G^{\,}_{f}\to \mathrm{Aut}\left(\mathfrak{F}^{\,}_{\Lambda}\right)$ 
where $\mathrm{Aut}\left(\mathfrak{F}^{\,}_{\Lambda}\right)$ is the set of 
automorphisms on the fermionic Fock space such that
for any $g,h\in G^{\,}_{f}$,
\begin{subequations}
\label{eq:def faithful rep on bulk}
\begin{align}
\widehat{U}^{\,}_{\mathrm{bulk}}(g)\,
\widehat{U}^{\,}_{\mathrm{bulk}}(h)=
\widehat{U}^{\,}_{\mathrm{bulk}}(g\,h),
\end{align}
where $g\,h$ denotes the 
composition of the elements $g,h\in G^{\,}_{f}$.
We define a group homomorphism
\begin{equation}
\mathfrak{c}:G^{\,}_{f}\to\left\{0,1\right\}
\end{equation}
that specifies if an element $g\in G^{\,}_{f}$
is to be represented by a unitary operator,
in which case $\left[\mathfrak{c}(g)=0\right]$,
or by an antiunitary operator,
in which case $\left[\mathfrak{c}(g)=1\right]$. 
For any $g\in G^{\,}_{f}$, its representation 
$\widehat{U}^{\,}_{\mathrm{bulk}}(g)$ can be written as
\begin{align}
\widehat{U}^{\,}_{\mathrm{bulk}}(g)=
\widehat{V}^{\,}_{\mathrm{bulk}}(g)\,
\mathsf{K}^{\mathfrak{c}(g)},
\end{align}
\end{subequations}
where $\widehat{V}^{\,}_{\mathrm{bulk}}(g)$ is a 
unitary operator acting on $\mathfrak{F}^{\,}_{\Lambda}$
and $\mathsf{K}$ is the complex conjugation map. 

For each point $j\in \Lambda$, we associate 
a set of Hermitian Majorana operators
\begin{subequations}
\label{eq:def local degrees of freedom summary}
\begin{align}
\mathfrak{O}^{\,}_{j}
\df
\left\{
\hat{\gamma}^{(j)}_{1},
\hat{\gamma}^{(j)}_{2},
\cdots,
\hat{\gamma}^{(j)}_{n^{\,}_{j}}
\right\},
\label{eq:def local degrees of freedom summary a}
\end{align}
that realizes the Clifford algebra
\begin{align}
\mathrm{C\ell}^{\,}_{n^{\,}_{j}}\df
\mathrm{span}
\Bigg\{
&\left.
\prod_{\iota=1}^{n^{\,}_{j}}
\left(\hat{\gamma}^{(j)}_{\iota}\right)^{m^{\,}_{\iota}}
\ \right|\
\big\{
\hat{\gamma}^{(j)}_{\iota},
\hat{\gamma}^{(j')}_{\iota'}
\big\}=
2\delta^{\,}_{\iota\iota'},
\nonumber\\
&
\,\,\,
m^{\,}_{i}=0,1,
\quad
\iota,\iota'=1,\cdots,n^{\,}_{j}
\Bigg\}.
\label{eq:def local degrees of freedom summary b}
\end{align}
\end{subequations}
The $n^{\,}_{j}$ Majorana
operators \eqref{eq:def local degrees of freedom summary a}
span a local fermionic Fock space $\mathfrak{F}^{\,}_{j}$ 
if $n^{\,}_{j}$ is an even integer.
If $n^{\,}_{j}$ is odd, the Clifford algebra 
$\mathrm{C\ell}^{\,}_{n^{\,}_{j}}$ contains
a two-dimensional center, reason for which
the $n^{\,}_{j}$ Majorana operators 
\eqref{eq:def local degrees of freedom summary a}
span a Hilbert space that cannot be interpreted
as a fermionic Fock space%
~\footnote{The Hilbert space on which we choose to define
a representation of the Clifford algebra
$\mathrm{C}\ell^{\,}_{2n}$
spanned by $2n$ Majorana operators is the fermionic
Fock space of dimension $2^{n}$
defined as follows. First, $n$ pairs of Majorana operators
are chosen. Second, any one of these $n$ pairs of Majorana operators
defines a conjugate pair of creation and annihilation fermion operators.
Third, the vacuum state that is annihilated by all fermion operators
is the highest weight state from which $2^{n}-1$
orthonormal states with the fermion number
$n^{\,}_{\mathrm{f}}=1,2,\cdots,n-1,n$
descend by acting on the vacuum state with the product of
$n^{\,}_{\mathrm{f}}$ distinct fermionic creation operators.
By construction, $\mathrm{C}\ell^{\,}_{2n}$
has a nontrivial complex 
irreducible representation of dimension $2^{n}$ 
for which each element of the basis that defines 
the fermionic Fock space
has a well-defined fermion parity. Because the center
of the Clifford algebra $\mathrm{C}\ell^{\,}_{2n}$
is spanned by the identity alone,
a redefinition of any one of the $2n$ Majorana operators by multiplication
with an element of the center of $\mathrm{C}\ell^{\,}_{2n}$
changes any basis element of the fermionic Fock space by at most a
multiplicative $\mathbb{C}$ number.
A redefinition of any one of the $2n$ Majorana operators by multiplication
with an element of the center of $\mathrm{C}\ell^{\,}_{2n}$
thus leaves the fermion parity of each element of the
fermionic Fock basis unchanged.
This is not so any more for a representation of the Clifford algebra
$\mathrm{C}\ell^{\,}_{2n+1}$
spanned by $2n+1$ Majorana operators.
Even though it is still possible to define 
a Hilbert space of dimension $2^{n+1}$ on which 
the Clifford algebra $\mathrm{C}\ell^{\,}_{2n+1}$
has a nontrivial irreducible representation%
~\cite{Jackiw2012},
the center of $\mathrm{C}\ell^{\,}_{2n+1}$
is a two-dimensional subalgebra for the product of all $2n+1$
Majorana operators commutes with any one of these $2n+1$
Majorana operators.
It follows that there is no element in $\mathrm{C}\ell^{\,}_{2n+1}$,
which anticommutes with 
all the Majorana generators of $\mathrm{C}\ell^{\,}_{2n+1}$, i.e., 
it is not possible to distinguish an element in $\mathrm{C}\ell^{\,}_{2n+1}$,
which assigns odd fermion parity to all $2n+1$ Majorana generators.
The best one can do is to construct a $2^{n}$-dimensional
fermionic Fock space 
using the generators of a $\mathrm{C}\ell^{\,}_{2n}$ subalgebra of
$\mathrm{C}\ell^{\,}_{2n+1}$ and a two-dimensional Hilbert space 
in which states do not have any assigned fermion parity or 
fermion number.  Consequently,
the action on $\mathrm{C}\ell^{\,}_{2n+1}$ of the symmetries
inherited from the bulk does not determine uniquely,
i.e., up to a phase factor,
the action on the representation of $\mathrm{C}\ell^{\,}_{2n+1}$
of this symmetry.}. 
The consistent definition of a global fermionic Fock space
thus requires the total number of 
Majorana degrees of freedom to be even, i.e.,
\begin{align}
\sum_{j}
n^{\,}_{j}
=
0
\text{ mod 2}.
\end{align}

We define a \textit{local}%
~\footnote{
When $n^{\,}_{j}$ is an odd integer, it is not always
possible to construct a local representation 
${\tiny\widehat{U}^{\,}_{j}(g)}$ for any $g\in G^{\,}_{f}$
only out of the Majorana degrees of freedom in the set
$\mathfrak{O}^{\,}_{j}$ defined in
\eqref{eq:def local degrees of freedom summary a}.
However, we still call ${\tiny\widehat{U}^{\,}_{j}(g)}$ a local representation
in the sense that it can always be constructed by supplementing 
the Clifford algebra \eqref{eq:def local degrees of freedom summary b}
by an additional Majorana degree of freedom ${\tiny\hat{\gamma}^{\,}_{j,\infty}}$
that is localized at some other site $j'$ with the number
of Majorana operators $n^{\,}_{j'}$ being an odd integer.
          } 
representation $\widehat{U}^{\,}_{j}$
of the symmetry group $G^{\,}_{f}$ by demanding that
on the degrees of freedom localized at site $j\in \Lambda$,
$\widehat{U}^{\,}_{j}$ acts in the same way as
the global bulk representation 
$\widehat{U}^{\,}_{\mathrm{bulk}}$ does, i.e., 
the consistency condition
\begin{align}
\widehat{U}^{\,}_{j}(g)\,
\hat{\gamma}^{(j)}_{\iota}\,
\widehat{U}^{\dagger}_{j}(g)
=
\widehat{U}^{\,}_{\mathrm{bulk}}(g)\,
\hat{\gamma}^{(j)}_{\iota}\,
\widehat{U}^{\dagger}_{\mathrm{bulk}}(g),
\label{eq:projection to local rep}
\end{align}
for any $g\in G^{\,}_{f}$
and $\iota=1,\cdots,n^{\,}_{j}$ must hold.
Hereby, we assume that the bulk representation
$\widehat{U}^{\,}_{\mathrm{bulk}}$ 
is not anomalous in the sense that 
there are no obstructions that prevent
decomposing $\widehat{U}^{\,}_{\mathrm{bulk}}$
into the product of
local representations $\widehat{U}^{\,}_{j}$
(see Refs.\ \onlinecite{Chen2011b,Else2014} 
for examples when this is not possible).
The definition \eqref{eq:projection to local rep}
implies that the representation $\widehat{U}^{\,}_{j}$
satisfies for any $g,h\in G^{\,}_{f}$
\begin{align}
\widehat{U}^{\,}_{j}(g)\,
\widehat{U}^{\,}_{j}(h)=
e^{\mathrm{i}\phi^{\,}_{j}(g,h)}\,
\widehat{U}^{\,}_{j}(g\,h).
\end{align}
The phase factor $\phi^{\,}_{j}(g,h)\in C^{2}(G^{\,}_{f},U(1))$
defines a 2-cochain
(Appendix \ref{appsec:Group Cohomology}). 
Its equivalence classes $[\phi^{\,}_{j}]$
takes values in the second cohomology group 
$H^{2}(G^{\,}_{f},U(1)^{\,}_{\mathfrak{c}})$
(Appendix \ref{appsec:Group Cohomology}).
The equivalence class $[\phi^{\,}_{j}]$
characterizes the projective nature of the representation 
$\widehat{U}^{\,}_{j}$. The value $[\phi^{\,}_{j}]=0$ is
assigned to the trivial projective representation 
for which the vanishing phase $\phi(g,h)=0$ for any 
$g,h\in G^{\,}_{f}$ is a representative.

By definition, local Hamiltonians with the symmetry group $G^{\,}_{f}$
that realize IFT phases of matter must necessarily have nondegenerate
and gapped ground states that transform as singlets under the symmetry group
$G^{\,}_{f}$ with any closed boundary conditions. 
We restrict our attention to one-dimensional space and to
IFT phases of matter with translation symmetry
$G^{\,}_{\mathrm{trsl}}$ in addition to the internal fermionic
symmetry group $G^{\,}_{f}$. In other words, the total symmetry group
$G^{\,}_{\mathrm{tot}}$
is by hypothesis the direct product
\begin{equation}
G^{\,}_{\mathrm{tot}}\equiv
G^{\,}_{\mathrm{trsl}}\times G^{\,}_{f}.
\end{equation}

Imposing translation symmetry $G^{\,}_{\mathrm{trsl}}$
requires the number $n^{\,}_{j}$ 
of Majorana degrees of freedom at
each site to be independent of $j$
with the same local representation $\widehat{U}^{\,}_{j}(g)$ 
for any element $g\in G^{\,}_{f}$.
If so, the Lieb-Schulz-Mattis (LSM) theorems from 
Ref.\ \onlinecite{Aksoy2021} apply (see also Ref.\ \onlinecite{Cheng2019}). 
A nondegenerate and gapped ground state that transforms as a 
singlet under the symmetry group $G^{\,}_{\mathrm{tot}}$ 
is permissible if and only if:
\begin{enumerate}
\item
The number $n^{\,}_{j}$ 
of Majorana degrees of freedom
at each site $j\in \Lambda$ is even, i.e.,
$n^{\,}_{j}\equiv 2n$
\item
The local representation 
$\widehat{U}^{\,}_{j}(g)$ realizes a trivial projective representation, i.e., $[\phi^{\,}_{j}]\equiv [\phi] = 0$.
\end{enumerate}
The first condition requires that there exist a \textit{local}
fermionic Fock space $\mathfrak{F}^{\,}_{j}$ spanned by the 
even number of local Majorana degrees of freedom 
\eqref{eq:def local degrees of freedom summary a}.
Therefore, the global fermionic Fock space $\mathfrak{F}^{\,}_{\Lambda}$
decomposes as a $\mathbb{Z}^{\,}_{2}$-graded tensor product
$\otimes^{\,}_{\mathfrak{g}}$ of the local Fock spaces $\mathfrak{F}^{\,}_{j}$,
i.e.,
\begin{align}
\mathfrak{F}^{\,}_{\Lambda}
=
\bigotimes\limits_{j\in\Lambda}{}^{\,}_{\mathfrak{g}}\,
\mathfrak{F}^{\,}_{j}.
\end{align} 
The second condition requires that the local representation
$\widehat{U}^{\,}_{j}\in 
\mathrm{Aut}\left(\mathfrak{F}^{\,}_{j}\right)$
is a representation in the trivial 
equivalence class $[\phi]=0$.
This implies that the global bulk representation 
$\widehat{U}^{\,}_{\mathrm{bulk}}$ decomposes as
the product of local representations $\widehat{U}^{\,}_{j}$,
i.e., for any $g\in G^{\,}_{f}$ 
\begin{align}
\widehat{U}^{\,}_{\mathrm{bulk}}(g)
=
\left[
\prod_{j\in\Lambda}
\widehat{V}^{\,}_{j}(g)
\right]
\mathsf{K}^{\mathfrak{c}(g)}.
\end{align}

\begin{figure}[t]
\begin{center}
\includegraphics[angle=0,width=0.49\textwidth]{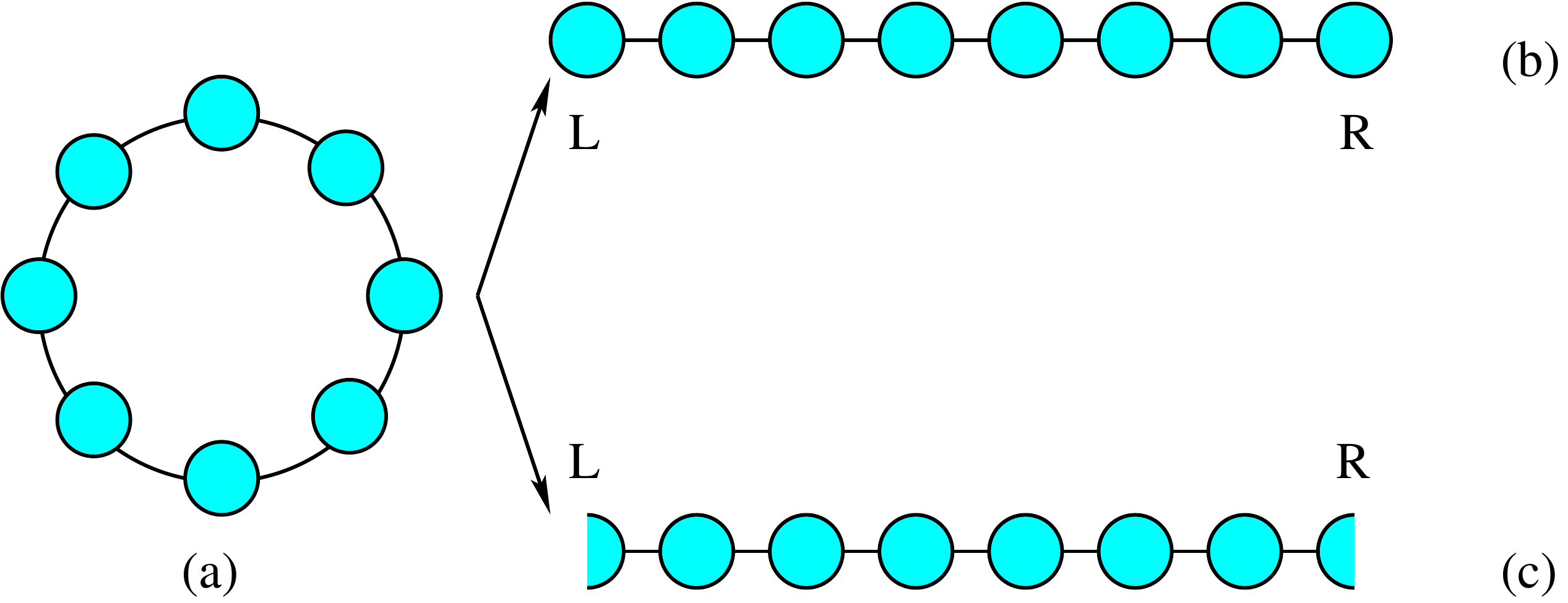}
\end{center}
\caption{
The repeat unit cells of a one-dimensional lattices are pictured
by colored discs. Each repeat unit cell hosts an even number $2n$
of Majorana degrees of freedom. Without loss of generality,
the range of the couplings between Majorana degrees of freedom is
one lattice spacing (the thick line between the repeat unit cells).
Translation symmetry is imposed by choosing periodic boundary conditions,
in which case the one-dimensional lattice is the discretization of
a ring. Open boundary conditions break the translation symmetry.
This can be achieved by cutting a thick line connecting two
repeat unit cell or by cutting open a repeat unit cell.
In the former case, the number of Majorana degrees of freedom on
any one of the upmost left or right cells is the same even
number $2n$ of Majorana degrees of freedom as that
in a single repeat unit cell.
In the latter case, the number of Majorana degrees of freedom on the
upmost left cell is any integer $1<n^{\,}_{\mathrm{L}}<2n$
while that on the upmost cell is $n^{\,}_{\mathrm{R}}=2n-n^{\,}_{\mathrm{L}}$.
        }
\label{Fig: cutting open a ring in two ways}
\end{figure}

Open boundary conditions break the hypothesis of
translation symmetry in the LSM
theorem from Ref.\ \onlinecite{Aksoy2021}.
When a closed chain is opened up, the 
degrees of freedom localized in  one
or multiple repeat unit cells may be split into 
two disconnected components 
$\Lambda^{\,}_{\mathrm{L}}$ and $\Lambda^{\,}_{\mathrm{R}}$
of the boundary 
$\Lambda^{\,}_{\mathrm{bd}}\df\partial \Lambda\equiv
\Lambda^{\,}_{\mathrm{L}}\cup\Lambda^{\,}_{\mathrm{R}}$,
as is illustrated in
Fig.\ \ref{Fig: cutting open a ring in two ways}.
If so, the two requirements of the LSM theorem need
no longer hold at each disconnected component.

Any one-dimensional IFT phase of matter 
is thus characterized by the following data:
\begin{enumerate}
\item
There is a $\mathbb{Z}^{\,}_{2}$-valued index
$[\mu^{\,}_{\mathrm{B}}]=\left\{0,1\right\}$ (B = L, R) 
that measures the parity of the number of Majorana degrees of freedom
that are localized on either one of the left (L)
or right (R) boundaries (B)
of the open chain
$\Lambda^{\,}_{\mathrm{bd}}=
\Lambda^{\,}_{\mathrm{L}}\cup\Lambda^{\,}_{\mathrm{R}}$.
The index $[\mu^{\,}_{\mathrm{B}}]$ can be viewed
as an element of the zero-th cohomology group
$H^{0}\big(G^{\,}_{f},\mathbb{Z}^{\,}_{2}\big)=\mathbb{Z}^{\,}_{2}$.
\item
There is an equivalence class
$[\phi^{\,}_{\mathrm{B}}]\in 
H^{2}\big(G^{\,}_{f},\mathrm{U}(1)^{\,}_{\mathfrak{c}}\big)$
of the second cohomology group
(Appendix \ref{appsec:Group Cohomology})
that characterizes the projective representation of
the internal symmetry group $G^{\,}_{f}$ at either one
of the left or right boudaries of an open chain
$\Lambda^{\,}_{\mathrm{bd}}=\Lambda^{\,}_{\mathrm{L}}\cup\Lambda^{\,}_{\mathrm{R}}$.
\end{enumerate}
Given a disconnected component $\Lambda^{\,}_{\mathrm{B}}$ 
of the boundary $\Lambda^{\,}_{\mathrm{bd}}$,
we assume the existence of
a set of boundary Majorana degrees of freedom
\begin{subequations}
\label{eq:projection to boundary rep}
\begin{align}
\mathfrak{O}^{\,}_{\mathrm{B}}
\df
\left\{
\hat{\gamma}^{(\mathrm{B})}_{1},
\hat{\gamma}^{(\mathrm{B})}_{2},
\cdots,
\hat{\gamma}^{(\mathrm{B})}_{n^{\,}_{\mathrm{B}}}
\right\}
\label{eq:projection to boundary rep a}
\end{align}
that are associated with states 
exponentially localized in space at the boundary $\mathrm{B}$.
The pair of data 
$([\phi^{\,}_{\mathrm{B}}],[\mu^{\,}_{\mathrm{B}}])\in 
H^{2}\big(G^{\,}_{f},\mathrm{U}(1)^{\,}_{\mathfrak{c}}\big)
\times 
H^{0}\big(G^{\,}_{f},\mathbb{Z}^{\,}_{2}\big)$ 
are assigned as follows. 
The index $[\mu^{\,}_{\mathrm{B}}]$ is nothing but the parity 
of the number of Majorana degrees of freedom at 
$\Lambda^{\,}_{\mathrm{B}}$, i.e., 
$[\mu^{\,}_{\mathrm{B}}] = n^{\,}_{\mathrm{B}}$ mod 2.
The equivalence class $[\phi^{\,}_{\mathrm{B}}]$ of
the projective phase $\phi^{\,}_{\mathrm{B}}(g,h)$ 
is computed by constructing a boundary representation 
$\widehat{U}^{\,}_{\mathrm{B}}$. This is done by 
demanding the consistency condition 
\begin{align}
\widehat{U}^{\,}_{\mathrm{B}}(g)\,
\hat{\gamma}^{(\mathrm{B})}_{\iota}\,
\widehat{U}^{\dagger}_{\mathrm{B}}(g)
=
\widehat{U}^{\,}_{\mathrm{bulk}}(g)\,
\hat{\gamma}^{(\mathrm{B})}_{\iota}\,
\widehat{U}^{\dagger}_{\mathrm{bulk}}(g),
\label{eq:projection to boundary rep b}
\end{align}
\end{subequations}
for any  $g\in G^{\,}_{f}$ and 
$\iota= 1, 2, \cdots ,n^{\,}_{\mathrm{B}}$.

The index
$[\phi^{\,}_{\mathrm{B}}]
\in H^{2}\big(G^{\,}_{f},\mathrm{U}(1)^{\,}_{\mathfrak{c}}\big)$
depends both on $[\mu^{\,}_{\mathrm{B}}]=0,1$ 
and the fermionic symmetry group
$G^{\,}_{f}$. This is so because $G^{\,}_{f}$ is the central
extension of the internal symmetry group $G$
by the fermion-parity symmetry group $\mathbb{Z}^{\mathrm{F}}_{2}$
with extension class $[\gamma]\in H^{2}(G,\mathbb{Z}^{\mathrm{F}}_{2})$ 
(Appendix \ref{appsec:central extension review}),
i.e., $G$ is isomorphic to the group
$G^{\,}_{f}/\mathbb{Z}^{\mathrm{F}}_{2}$%
~\footnote{
The coset $G^{\,}_{f}/\mathbb{Z}^{\mathrm{F}}_{2}$
is a group since $\mathbb{Z}^{\mathrm{F}}_{2}$
is in the center of $G^{\,}_{f}$ and therefore a normal 
subgroup.
          }
As the center of the fermionic symmetry group $G^{\,}_{f}$
is the fermion-parity subgroup $\mathbb{Z}^{\mathrm{F}}_{2}$,
its projective representations
are sensitive to the values of $[\mu^{\,}_{\mathrm{B}}]$. 
This sensitivity can be made
explicit if, following Turzillo and You in Ref.\
\onlinecite{Turzillo2019},
one trades the equivalence classes
$
[\phi^{\,}_{\mathrm{B}}]
\in H^{2}\big(G^{\,}_{f},\mathrm{U}(1)^{\,}_{\mathfrak{c}}\big)
$
for the equivalence classes
$
[(\nu^{\,}_{\mathrm{B}},\rho^{\,}_{\mathrm{B}})]\in
\mathrm{ker}\, \mathcal{D}^{2}_{\gamma,\mathfrak{c}}
\big/
\mathrm{im}\, \mathcal{D}^{1}_{\gamma,\mathfrak{c}}
$
where $\mathcal{D}^{2}_{\gamma,\mathfrak{c}}$ and 
$\mathcal{D}^{1}_{\gamma,\mathfrak{c}}$ are modified
coboundary operators
(Appendix \ref{appsec:classification of Gf}).

The 2-cochain
$\nu^{\,}_{\mathrm{B}}\in C^{2}[G,\mathrm{U}(1)]$
encodes the projective representations of the 
group. The interpretation of the 1-cochain
$\rho^{\,}_{\mathrm{B}}\in C^{1}[G,\mathbb{Z}^{\,}_{2}]$
depends on the value of $[\mu^{\,}_{\mathrm{B}}]=0,1$.
When $[\mu^{\,}_{\mathrm{B}}]=0$,
the 1-cochain
$\rho^{\,}_{\mathrm{B}}\in C^{1}[G,\mathbb{Z}^{\,}_{2}]$
encodes the relation between the representations of the elements of
the group $G$ and the representation of the
fermion-parity from $\mathbb{Z}^{\mathrm{F}}_{2}$.
When $[\mu^{\,}_{\mathrm{B}}]=1$,
the 1-cochain
$\rho^{\,}_{\mathrm{B}}\in C^{1}[G,\mathbb{Z}^{\,}_{2}]$
encodes the relation between the representations of the elements of
the group $G$ and the representation of the
nontrivial central element of a Clifford algebra C$\ell^{\,}_{2k+1}$
with an odd number of generators.

There are two possible scenarios for the fate
of the set \eqref{eq:projection to boundary rep a}
of boundary degrees of freedom 
on the boundary $\Lambda^{\,}_{\mathrm{B}}$
that realize the triplet of boundary data
$([(\nu^{\,}_{\mathrm{B}},\rho^{\,}_{\mathrm{B}})],[\mu^{\,}_{\mathrm{B}}])$
when the bulk is perturbed by
local and continuous interactions
that break neither explicitly nor spontaneously the $G^{\,}_{f}$ symmetry.
In scenario I, the set \eqref{eq:projection to boundary rep a}
is unchanged by the bulk perturbation.
If so, the triplet of boundary data
$([(\nu^{\,}_{\mathrm{B}},\rho^{\,}_{\mathrm{B}})],[\mu^{\,}_{\mathrm{B}}])$
does not change.
In scenario II, the bulk perturbation changes
the set \eqref{eq:projection to boundary rep a}
by either the addition or removal of boundary degrees of freedom.	
If the degrees of freedom added to or removed from the boundary 
$\Lambda^{\,}_{\mathrm{B}}$ 
realize the trivial triplet of data,
then the resulting triplet of boundary data is unchanged
according to the fermionic stacking rules, i.e.,
$([(\nu^{\,}_{\mathrm{B}},\rho^{\,}_{\mathrm{B}})],[\mu^{\,}_{\mathrm{B}}])$.
If the degrees of freedom added to or removed from the boundary 
$\Lambda^{\,}_{\mathrm{B}}$ 
realize a nontrivial triplet of data,
then the triplet of boundary data is changed to
$([(\nu^{\prime}_{\mathrm{B}},\rho^{\prime}_{\mathrm{B}})],[\mu^{\prime}_{\mathrm{B}}])
\neq ([(\nu^{\,}_{\mathrm{B}},\rho^{\,}_{\mathrm{B}})],[\mu^{\,}_{\mathrm{B}}])$
according to the fermionic stacking rules.
If the bulk-boundary correspondence were to hold, then
a gap-closing transition in the bulk that is induced by the bulk perturbations
is required to change the boundary triplet of data.
This hypothesis is plausible because
Bourne and Ogata have shown rigorously in Ref.\ \onlinecite{Bourne2021}
the existence of triplets of bulk data that take values in the same cohomology
groups as the triplets of boundary data
$([(\nu^{\,}_{\mathrm{B}},\rho^{\,}_{\mathrm{B}})],[\mu^{\,}_{\mathrm{B}}])$,
obey the same stacking rules, and offer a bulk classification of
IFT phases of matter. 
In this paper, we shall assume without proof this bulk-boundary correspondence. 

There is no need to specify
the triplets 
$([(\nu^{\,}_{\mathrm{L}},\rho^{\,}_{\mathrm{L}})],[\mu^{\,}_{\mathrm{L}}])$
and 
$([(\nu^{\,}_{\mathrm{R}},\rho^{\,}_{\mathrm{R}})],[\mu^{\,}_{\mathrm{R}}])$
associated with the disconnected components $\Lambda^{\,}_{\mathrm{L}}$
and  $\Lambda^{\,}_{\mathrm{R}}$ independently. 
The triplet of data 
on the left boundary $\Lambda^{\,}_{\mathrm{L}}$ fixes
their counterparts on the right boundary
$\Lambda^{\,}_{\mathrm{R}}$, owing to the condition that
the ground state of a Hamiltonian realizing 
an IFT phase of matter must be nondegenerate and
$G^{\,}_{f}$-symmetric when periodic boundary conditions are selected.
Thus, we drop the subscripts when denoting the triplet of data 
$([(\nu,\rho)],[\mu])$ that characterize the IFT phases.

For any $G^{\,}_{f}$ that splits, i.e., $G^{\,}_{f}$ is
isomorphic to the product $G\times\mathbb{Z}^{F}_{2}$,
the index $[\mu]$ can take the values $0$ or $1$.
If the group $G^{\,}_{f}$ is a nonsplit group, then
$[\mu]=0$ is the only possibility. 
When $[\mu]=1$, the minimal degeneracy of the eigenspace for
the ground states when open boundary conditions are selected is
two for any split fermionic symmetry group $G^{\,}_{f}$,
including the smallest possible fermionic symmetry group
$G^{\,}_{f}=\mathbb{Z}^{\mathrm{F}}_{2}$. Hence, 
one-dimensional Hamiltonians realizing
IFT phases of matter with $[\mu]=1$ 
cannot be deformed adiabatically to
a Hamiltonian realizing the
trivial IFT phase of matter at the expense of breaking explicitly
any of the protecting symmetries in $G^{\,}_{f}$ other than
$\mathbb{Z}^{\mathrm{F}}_{2}$. \textit{A forteriori}, these phases of matter
are distinct from the fermionic SPT (FSPT)
phases of matter in one-dimensional space.
In one-dimensional space, FSPT phases of matter are only possible when
$[\mu]=0$.

Once the IFT phases in one-dimension are
characterized by the triplet $\left([(\nu,\rho)],[\mu]\right)$, it is
imperative to derive the stacking rules, i.e., the group composition
rules of the triplets $\left([(\nu,\rho)],[\mu]\right)$
that are compatible with the $\mathbb{Z}^{\,}_{2}$-graded tensor
product between fermionic Fock spaces (in physics terminology,
antisymmetrization).
Stacking rules can be derived by considering the topological indices
$([(\nu^{\,}_{\wedge},\rho^{\,}_{\wedge})],[\mu^{\,}_{\wedge}])$
of an IFT phase of matter that is constructed
by combining the boundary degrees of freedoms of
any representatives of
two other IFT phases with topological indices
$\left([(\nu^{\,}_{1},\rho^{\,}_{1})],[\mu^{\,}_{1}]\right)$ and
$\left([(\nu^{\,}_{2},\rho^{\,}_{2})],[\mu^{\,}_{2}]\right)$,
respectively. 
The stacking rules are essential properties of IFT phases of matter.
They enforce a group composition law
between IFT phases of matter sharing the
same fermionic symmetry group $G^{\,}_{f}$.
This group composition law can be interpreted as the physical operation
by which two blocks of matter,
each realizing IFT phases of matter sharing the
same fermionic symmetry group $G^{\,}_{f}$,
are brought into contact so as to form a single larger block of matter
sharing the same fermionic symmetry group $G^{\,}_{f}$.
This group composition law is also needed to implement a consistency
condition corresponding to changing from open to closed boundary conditions.
Topological data associated with the left and the right
disconnected components of the one-dimensional boundary 
must be the inverse of each other with respect to the stacking
rules, i.e., one should obtain the
trivial data $([(0,0)],0)$ if the change from open to periodic boundary
conditions is interpreted as the stacking of opposite boundaries.

The main result of this paper is the derivation of the stacking rules
of any IFT phase of matter in one-dimensional space from the
perspective of the boundaries. Working on the boundaries allows
to use elementary tools of quantum mechanics and mathematics.

To achieve this goal, we first
define the set of boundary degrees of freedom $\mathfrak{O}^{\,}_{\mathrm{B}}$
and the corresponding representation $\widehat{U}^{\,}_{\mathrm{B}}$
that satisfy the consistency condition \eqref{eq:projection to boundary rep b}
for the cases of $[\mu]=0$ and $[\mu]=1$ separately
in Secs.\ \ref{sec:boundary rep gen} and
\ref{sec: Boundary projective representation of Gf}, respectively.
In doing so, we give the explicit representation for the 
center $\mathbb{Z}^{\mathrm{F}}_{2}\subset G^{\,}_{f}$ of the fermion 
parity symmetry. 
In Sec.\ \ref{sec: Definition of indices}, we define the 2-cochains
$\phi(g,h)\in C^{2}(G^{\,}_{f},U(1))$
and  
$\nu(g,h)\in C^{2}(G,U(1))$,
and the 1-cochain $\rho(g)\in H^{1}(G,\mathbb{Z}^{\,}_{2})$
for the cases of $[\mu]=0$ and $[\mu]=1$ separately.

Second, we derive the fermionic stacking rules using elementary means
in Sec.\ \ref{sec: Fermionic Stacking Rules}. 
To this end, we consider two boundary representations
$\widehat{U}^{\,}_{1}$
and
$\widehat{U}^{\,}_{2}$ 
together with their associated triplets
$\left([(\nu^{\,}_{1},\rho^{\,}_{1})],[\mu^{\,}_{1}]\right)$
and
$\left([(\nu^{\,}_{2},\rho^{\,}_{2})],[\mu^{\,}_{2}]\right)$,
respectively.
We then explicitly construct the \textit{stacked}
representation $\widehat{U}^{\,}_{\wedge}$ that acts 
on the combined degrees of freedom of the two representations.
This is done by demanding that the stacked representation 
$\widehat{U}^{\,}_{\wedge}$ satisfies the two conditions 
that are the counterparts to
the consistency condition 
\eqref{eq:projection to boundary rep b}.
\begin{widetext}
When constructing the stacked representation $\widehat{U}^{\,}_{\wedge}$,
we shall consider the following
four cases separately: 
(i) even-even 
($[\mu^{\,}_{1}]=[\mu^{\,}_{2}]=0$) stacking,
(ii) even-odd 
($[\mu^{\,}_{1}]=0$, $[\mu^{\,}_{2}]=1$) stacking,
(iii) odd-even 
($[\mu^{\,}_{1}]=1$, $[\mu^{\,}_{2}]=0$) stacking,
(iv) and 
odd-odd ($[\mu^{\,}_{1}]=[\mu^{\,}_{2}]=1$) stacking.
We find the four fermionic stacking rules
\begin{subequations}
\label{eq:gen stacking rules}
\begin{align}
&
\left(
[(\nu^{\,}_{1},\,\rho^{\,}_{1})],\,
0
\right)
\wedge
\left(
[(\nu^{\,}_{2},\,\rho^{\,}_{2})],\,
0
\right)
=
\left(
[
(\nu^{\,}_{1}+\nu^{\,}_{2}
+
\pi\left(\rho^{\,}_{1}\smile\rho^{\,}_{2}\right),\,
\rho^{\,}_{1}
+
\rho^{\,}_{2})],\,
0
\right),
\label{eq:gen stacking rules a}
\\
&
\left(
[(\nu^{\,}_{1},\,\rho^{\,}_{1})],\,
0
\right)
\wedge
\left(
[(\nu^{\,}_{2},\,\rho^{\,}_{2})],\,
1
\right)
=
\left(
[
(\nu^{\,}_{1}+\nu^{\,}_{2}
+
\pi\left(
\rho^{\,}_{1}\smile\rho^{\,}_{2}
+
\rho^{\,}_{1}\smile\mathfrak{c}
\right),\,
\rho^{\,}_{1}
+
\rho^{\,}_{2})],\,
1
\right),
\label{eq:gen stacking rules b}
\\
&
\left(
[(\nu^{\,}_{1},\,\rho^{\,}_{1})],\,
1
\right)
\wedge
\left(
[(\nu^{\,}_{2},\,\rho^{\,}_{2})],\,
0
\right)
=
\left(
[
(\nu^{\,}_{1}+\nu^{\,}_{2}
+
\pi\left(
\rho^{\,}_{1}\smile\rho^{\,}_{2}
+
\rho^{\,}_{2}\smile\mathfrak{c}
\right),\,
\rho^{\,}_{1}
+
\rho^{\,}_{2})],\,
1
\right),
\label{eq:gen stacking rules c}
\\
&
\left(
[(\nu^{\,}_{1},\,\rho^{\,}_{1})],\,
1
\right)
\wedge
\left(
[(\nu^{\,}_{2},\,\rho^{\,}_{2})],\,
1
\right)
=
\left(
[
(\nu^{\,}_{1}+\nu^{\,}_{2}
+
\pi\left(\rho^{\,}_{1}\smile\rho^{\,}_{2}\right),\,
\rho^{\,}_{1}
+
\rho^{\,}_{2}
+
\mathfrak{c})],\,
0
\right),
\label{eq:gen stacking rules d}
\end{align}
\end{subequations}
respectively. The derivation of the stacking rules
(\ref{eq:gen stacking rules})
is  the main result of this paper. 
Here, we denote the stacking operation with the symbol $\wedge$.
We also made use of the cup product $\smile$
defined in Appendix \ref{appsec:Group Cohomology}
to construct a 2-cochains out of two 1-cochains.
If
the group $G^{\,}_{f}$ only contains unitary symmetries,
i.e.,
$\mathfrak{c}(g)=0$ for any $g\in G^{\,}_{f}$, then the stacking rules
\eqref{eq:gen stacking rules} reduce to
\begin{align}
\left(
[(\nu^{\,}_{1},\,\rho^{\,}_{1})],\,
[\mu^{\,}_{1}]
\right)
\wedge
\left(
[(\nu^{\,}_{2},\,\rho^{\,}_{2})],\,
[\mu^{\,}_{2}]
\right)
=
\left(
[
(\nu^{\,}_{1}+\nu^{\,}_{2}
+
\pi\left(\rho^{\,}_{1}\smile\rho^{\,}_{2}\right),\,
\rho^{\,}_{1}
+
\rho^{\,}_{2})],\,
[\mu^{\,}_{1}]+[\mu^{\,}_{2}]
\right).
\end{align}
\end{widetext}
Finally, we discuss the protected ground-state degeneracy
of representatives of an IFT phase
with the index $([(\nu,\rho)],[\mu])$
when open boundary conditions are imposed
in Sec.\ \ref{sec: Ground-state degeneracies}.
This exercise allows us to give the generic conditions on the index
$([(\nu,\rho)],[\mu])$
that imply an emergent boundary supersymmetry,
as was recently explored in Refs.\
\onlinecite{Prakash2021,Turzillo2021}.

As an illustration of this general result,
we turn our attention to the symmetry class BDI 
from the tenfold way 
(the protecting symmetries are fermion parity and
spinless time reversal)
in one-dimensional space~\cite{Fidkowski2010}.
In the supplemental material~\footnote{See Supplemental Material 
for an application of the toolkit we developed to the
one-dimensional IFT phases in symmetry class BDI from the tenfold way 
and closely related spin-1/2 cluster chains that realize bosonic
SPT phases.},
we derive in detail (i) the explicit values of
$([(\nu^{\,}_{\mathrm{B}},\rho^{\,}_{\mathrm{B}})],[\mu^{\,}_{\mathrm{B}}])$
for the left (B = L) and the right (B = R) boundaries,
(ii) their stacking rules,
(iii) and the protected ground-state degeneracies
for a class of Hamiltonians that we call Majorana $c$ chains
with $c\in\mathbb{Z}$.
The asymmetry between the left and right boundaries is
made explicit.
We then use the Jordan-Wigner transformation
to bosonize the Majorana $c$ chains  into a family of
spin-1/2 cluster $c$ chains~\cite{Suzuki1971,Verresen2017}. 
These spin-1/2 cluster $c$ chains
are shown to realize bosonic symmetry-protected topological
phases of matter characterized by a doublet
$([\nu^{\,}_{\mathrm{B}}],[\rho^{\,}_{\mathrm{B}}])$ of indices that quantify
which projective representations of the protecting symmetries
(a global rotation in Pauli space by the angle $\pi$
around some given direction in Pauli space and spinless time-reversal symmetry)
is realized on the boundaries. Hereto, we derive
(i) the explicit values of
$([\nu^{\,}_{\mathrm{B}}],[\rho^{\,}_{\mathrm{B}}])$
for the left (B=L) and the right (B=R) boundaries,
(ii) their stacking rules,
(iii) and the protected ground-state degeneracies.
The differences with the Majorana $c$ chains are explained.
In particular, it is shown that the left and right boundaries
share the same projective representations of the protecting symmetries
for the spin-1/2 cluster $c$ chains.

\section{Assumptions and definitions}
\label{sec:boundary rep gen}

We start from a given internal symmetry group $G^{\,}_{f}$
acting on Majorana degrees of freedom. The subscript $f$ is attached 
to emphasize the ``fermionic'' nature of the group $G^{\,}_{f}$ 
as we shall now explain. 
For quantum systems built out of Majorana degrees of freedom, 
any Hamiltonian that dictates the quantum dynamics is built from
products of even numbers of Majorana operators.
In other words, the fermion parity operator $(-1)^{\widehat{F}}$
necessarily commutes with the Hamiltonian, where $\widehat{F}$ 
denotes the total fermion number operator. 
We denote by $\mathbb{Z}^{\mathrm{F}}_{2} \df \left\{e,p\right\}$
the cyclic group generated by the abstract element $p\in \mathbb{Z}^{\mathrm{F}}_{2}$
that we shall interpret as the fermion parity.
The superscript F is attached
to the cyclic group $\mathbb{Z}^{\mathrm{F}}_{2}$
to distinguish its role as the fermion parity symmetry.
We assume that the fermion parity symmetry 
$\mathbb{Z}^{\mathrm{F}}_{2}$ can be neither
explicitly nor spontaneously broken and is a subgroup of
the center of $G^{\,}_{f}$. We denote by 
$G\cong G^{\,}_{f}/\mathbb{Z}^{\mathrm{F}}_{2}$ the group consisting
of all symmetries other than the fermion parity symmetry 
$\mathbb{Z}^{\mathrm{F}}_{2}$.

The internal symmetry group $G^{\,}_{f}$ is specified by
two pieces of data. The first piece is the central extension class 
$[\gamma]\in H^{2}(G,\mathbb{Z}^{\mathrm{F}}_{2})$ that characterize 
how the group $G$ and the fermion parity symmetry group $\mathbb{Z}^{\mathrm{F}}_{2}$
are glued together to produce the group $G^{\,}_{f}$. 
This is to say that, $G^{\,}_{f}$ is not restricted to be
the direct product
$G^{\,}_{f}=G\times\mathbb{Z}^{\mathrm{F}}_{2}$.
The group $G^{\,}_{f}$ is such that
(i) $\mathbb{Z}^{\mathrm{F}}_{2}$ is a subgroup of the
center of $G^{\,}_{f}$
(ii) and $G$ is isomorphic to
$G^{\,}_{f}/\mathbb{Z}^{\mathrm{F}}_{2}$.
We assign the equivalence class $[\gamma]=0$
to the case of $G^{\,}_{f}$ being 
isomorphic to the direct product $G \times \mathbb{Z}^{\mathrm{F}}_{2}$
and say that $G^{\,}_{f}$ splits
(Appendices \ref{appsec:Group Cohomology} and
\ref{appsec:central extension review}).
The second piece is the group homomorphism 
$\mathfrak{c}: G^{\,}_{f} \to\left\{0,1\right\}$ that specifies if
an element $g\in G^{\,}_{f}$ is to be represented
by a unitary $\left[\mathfrak{c}(g)=0\right]$ operator or by an
antiunitary $\left[\mathfrak{c}(g)=1\right]$ operator 
[by definition, $\mathfrak{c}(p)=0$].

We denote by $\Lambda$ the set of points on a one-dimensional lattice
that we shall call the bulk. We assume that there exists a
nonvanishing boundary 
\begin{subequations}
\begin{equation}
\Lambda^{\,}_{\mathrm{bd}}\equiv\partial\Lambda
\end{equation}
of the bulk $\Lambda$. The boundary
$\Lambda^{\,}_{\mathrm{bd}}$
is the union of two disconnected components
$\Lambda^{\,}_{\mathrm{L}}$
or
$\Lambda^{\,}_{\mathrm{R}}$
of the one-dimensional universe
$\Lambda$,
\begin{equation}
\Lambda^{\,}_{\mathrm{bd}}=
\Lambda^{\,}_{\mathrm{L}}
\cup
\Lambda^{\,}_{\mathrm{R}},
\qquad
\Lambda^{\,}_{\mathrm{L}}
\cap
\Lambda^{\,}_{\mathrm{R}}=\emptyset.
\label{eq:def Lambda bd}
\end{equation}
\end{subequations}
The hypothesis that states bound to
$\Lambda^{\,}_{\mathrm{L}}$
or
$\Lambda^{\,}_{\mathrm{R}}$
do not overlap in space only holds 
for all fermionic invertible topological phases
after the thermodynamic limit has been taken.

We assume that there exists
a faithful representation 
of the group $G^{\,}_{f}$ acting on the bulk $\Lambda$, 
i.e., an injective map $\widehat{U}^{\,}_{\mathrm{bulk}}:
G^{\,}_{f}\to \mathrm{Aut}\left(\mathfrak{F}^{\,}_{\Lambda}\right)$ 
where $\mathrm{Aut}\left(\mathfrak{F}^{\,}_{\Lambda}\right)$ is the set of 
automorphisms on the fermionic Fock space $\mathfrak{F}^{\,}_{\Lambda}$ of
the one-dimensional universe.
We impose that the map $\widehat{U}^{\,}_{\mathrm{bulk}}$ forms 
an ordinary representation of the group $G^{\,}_{f}$
on the bulk $\Lambda$, 
i.e., it satisfies Eq.\ \eqref{eq:def faithful rep on bulk}. 
This might not be so anymore
when restricting the action of $G^{\,}_{f}$
to any one of the disconnected components
$\Lambda^{\,}_{\mathrm{L}}$
or
$\Lambda^{\,}_{\mathrm{R}}$
on the boundary
$\Lambda^{\,}_{\mathrm{bd}}$,
in which case the existence of degenerate ground states must follow
when open boundary conditions are selected.

Without loss of generality, 
we consider any one of
$\Lambda^{\,}_{\mathrm{L}}$
and
$\Lambda^{\,}_{\mathrm{R}}$,
which we denote $\Lambda^{\,}_{\mathrm{B}}$.
We are going to construct a projective representation
of the symmetry group $G^{\,}_{f}$ on this component
$\Lambda^{\,}_{\mathrm{B}}$
of the boundary
$\Lambda^{\,}_{\mathrm{bd}}$,
while the opposite component of the boundary
must then always be represented by the ``inverse'' projective representation.

On the boundary $\Lambda^{\,}_{\mathrm{B}}$,
we assume the existence of a set of $n$ Hermitian Majorana operators
\begin{subequations}
\label{eq:def mathfrak O}
\begin{equation}
\mathfrak{O}^{\,}_{n}
\df
\left\{
\hat{\gamma}^{\,}_{1},
\hat{\gamma}^{\,}_{2},
\cdots,
\hat{\gamma}^{\,}_{n}
\right\}
\label{eq:def mathfrak O a}
\end{equation}
that realizes the Clifford algebra
\begin{align}
\mathrm{C\ell}^{\,}_{n}\df
\mathrm{span}
\Bigg\{
&\left.
\prod_{i=1}^{n}
\left(\hat{\gamma}^{\,}_{i}\right)^{m^{\,}_{i}}
\ \right|\
\big\{
\hat{\gamma}^{\,}_{i},
\hat{\gamma}^{\,}_{j}
\big\}=
2\delta^{\,}_{ij},
\nonumber\\
&
\qquad
m^{\,}_{i}=0,1,
\quad
i,j=1,\cdots,n
\Bigg\}.
\label{eq:def mathfrak O b}
\end{align}
We call these operators Majorana operators.
We assign the index $[\mu]\in \left\{0,1\right\}$ to 
the parity of $n$, i.e.,
\begin{align}
[\mu]=
\hbox{$n$ mod $2$}.
\label{eq:def mathfrak O c}
\end{align}
\end{subequations}
We consider the cases of even and odd $n$ separately. 

When $[\mu]=0$, the even number $n$ of Majorana operators
from the set (\ref{eq:def mathfrak O a})
span the fermionic Fock space
\begin{subequations}
\begin{align}
\mathfrak{F}^{\,}_{\Lambda^{\,}_{\mathrm{B}},0}\df
&
\mathrm{span}
\Bigg\{
\left.
\prod_{\alpha=1}^{n/2}
\left(
\frac{\hat{\gamma}^{\,}_{2\alpha-1}-\mathrm{i}\hat{\gamma}^{\,}_{2\alpha}}{2}
\right)^{m^{\,}_{\alpha}}\,\ket{0}
\ \right|\
\nonumber\\
&
\left(
\frac{\hat{\gamma}^{\,}_{2\alpha-1}+\mathrm{i}\hat{\gamma}^{\,}_{2\alpha}}{2}
\right)
\ket{0}=0,
\quad
m^{\,}_{\alpha}=0,1
\Bigg\}
\label{eq:def F[mu]=0}
\end{align}
of dimension%
~\footnote{The partition of a set of $n$ labels into two pairs of $n/2$ labels
is here arbitrary.
}
\begin{equation}
\mathrm{dim}\,\mathfrak{F}^{\,}_{\Lambda^{\,}_{\mathrm{B}},0}=2^{n/2}.
\end{equation}
\end{subequations}

When $[\mu]=1$, the odd number $n$ of Majorana operators
from the set (\ref{eq:def mathfrak O a})
span a vector space that is not a fermionic Fock space.
In order to recover a fermionic Fock space,
we add to the set (\ref{eq:def mathfrak O a})
made of an odd number $n$ of Majorana operators
the Majorana operator $\hat{\gamma}^{\,}_{\infty}$~\cite{Fidkowski2011},
\begin{equation}
\mathfrak{O}^{\,}_{n,\infty}
\df
\left\{
\hat{\gamma}^{\,}_{1},
\hat{\gamma}^{\,}_{2},
\cdots,
\hat{\gamma}^{\,}_{2\lfloor n/2\rfloor},
\hat{\gamma}^{\,}_{n},
\hat{\gamma}^{\,}_{\infty}
\right\},
\label{eq:def mathfrak O infty}
\end{equation}
thereby defining the Clifford algebra $\mathrm{C\ell}^{\,}_{n+1}$.
Here, the lower floor function $\lfloor\cdot\rfloor$ returns the
largest integer $\lfloor x\rfloor$ smaller than the positive real number $x$.
We may then define the fermionic Fock space
\begin{subequations}
\begin{align}
\mathfrak{F}^{\,}_{\Lambda^{\,}_{\mathrm{B}},1}\df
&
\mathrm{span}
\Bigg\{
\left.
\prod_{\alpha=1}^{(n+1)/2}
\left(
\frac{\hat{\gamma}^{\,}_{2\alpha-1}-\mathrm{i}\hat{\gamma}^{\,}_{2\alpha}}{2}
\right)^{m^{\,}_{\alpha}}\,\ket{0}
\ \right|\
\nonumber\\
&
\left(
\frac{\hat{\gamma}^{\,}_{2\alpha-1}+\mathrm{i}\hat{\gamma}^{\,}_{2\alpha}}{2}
\right)
\ket{0}=0,
\
m^{\,}_{\alpha}=0,1
\Bigg\}
\label{eq:def F[mu]=1}
\end{align}
of dimension 
\begin{equation}
\mathrm{dim}\,\mathfrak{F}^{\,}_{\Lambda^{\,}_{\mathrm{B}},1}=2^{(n+1)/2},
\end{equation}
\end{subequations}
where it is understood that
$\hat{\gamma}^{\,}_{n+1}\equiv\hat{\gamma}^{\,}_{\infty}$.
In this fermionic Fock space, all 
creation and annihilation fermion operators are local, except for one pair.
The pair of creation and annihilation
operator built out of the pair
$\hat{\gamma}^{\,}_{n}$
and
$\hat{\gamma}^{\,}_{\infty}$
of Majorana operators
is nonlocal as $\hat{\gamma}^{\,}_{\infty}$ originates from
the opposite component of the boundary of one-dimensional space
owing to the open boundary conditions,
a distance infinitely far away after the thermodynamic limit has been taken.
The same is true of the two-dimensional fermionic Fock space
\begin{align}
\mathfrak{F}^{\,}_{\mathrm{LR}}\df
&
\mathrm{span}
\Big\{
\left.
\left(
\frac{\hat{\gamma}^{\,}_{n}-\mathrm{i}\hat{\gamma}^{\,}_{\infty}}{2}
\right)^{m^{\,}_{\alpha}}\,|0\rangle
\ \right|\
\nonumber\\
&\qquad\qquad\qquad\quad
\left(
\frac{\hat{\gamma}^{\,}_{n}+\mathrm{i}\hat{\gamma}^{\,}_{\infty}}{2}
\right)
|0\rangle=0
\Big\}
\label{eq:def FLR}
\end{align}
spanned by the pair
$\hat{\gamma}^{\,}_{n}$
and
$\hat{\gamma}^{\,}_{\infty}$.

Finally, it is assumed that the component $\Lambda^{\,}_{\mathrm{B}}$
of the boundary $\Lambda^{\,}_{\mathrm{bd}}$
defined in Eq.\ (\ref{eq:def Lambda bd})
is symmetric under the action of $G^{\,}_{f}$ in the sense that
\begin{align}
\widehat{U}^{\,}_{\mathrm{bulk}}(g)\,
\mathrm{C}\ell^{\,}_{n}\,
\widehat{U}^{\dag}_{\mathrm{bulk}}(g)\,
\subset
\mathrm{C}\ell^{\,}_{n},
\qquad
\forall g\in G^{\,}_{f}.
\label{eq: boundary modes are closed under Gf}
\end{align}

\section{Boundary projective representation of $G^{\,}_{f}$}
\label{sec: Boundary projective representation of Gf}

We assume that, for any $g\in G^{\,}_{f}$, there exists
a norm-preserving operator
$\widehat{U}^{\,}_{\mathrm{B}}(g)$
acting on the Fock space
$\mathfrak{F}^{\,}_{\Lambda^{\,}_{\mathrm{B}},[\mu]}$
as domain of definition
such that 
\begin{align}
\widehat{U}^{\,}_{\mathrm{B}}(g)\,
\hat{\gamma}^{\,}_{i}\,
\widehat{U}^{\dag}_{\mathrm{B}}(g)
=
\widehat{U}^{\,}_{\mathrm{bulk}}(g)\,
\hat{\gamma}^{\,}_{i}\,
\widehat{U}^{\dag}_{\mathrm{bulk}}(g),
\label{eq:consistency cond general}
\end{align}
for $i=1,2,\cdots,n$.
The boundary representation $\widehat{U}^{\,}_{\mathrm{B}}(g)$
of any element $g\neq e,\,p$ is not unique since
Eq.\ (\ref{eq:consistency cond general})
is left invariant by the multiplication from the right
of $\widehat{U}^{\,}_{\mathrm{B}}(g)$ with any norm-preserving element 
from the center of the Clifford algebra 
$\mathrm{C}\ell^{\,}_{n}$.
When $n$ is even this center is 
trivial and one-dimensional.
When $n$ is odd ($[\mu]=1$) this center
is nontrivial and two-dimensional.
In contrast, irrespective of $[\mu]$ the representations 
$\widehat{U}^{\,}_{\mathrm{B}}(e)$ and $\widehat{U}^{\,}_{\mathrm{B}}(p)$
of the identity
and fermion parity acting on the 
fermionic Fock space $\mathfrak{F}^{\,}_{\Lambda^{\,}_{\mathrm{B}},[\mu]}$
are uniquely determined up to 
a multiplicative phase factor.

Finally, we observe two consequences of
Eq.\ \eqref{eq:consistency cond general}.
First, the boundary representation $\widehat{U}^{\,}_{\mathrm{B}}$
inherits the injectivity
of the bulk representation $\widehat{U}^{\,}_{\mathrm{bulk}}$
of the fermionic symmetry group $G^{\,}_{f}$.
Second, for any element $g \in G^{\,}_{f}$,
the boundary representation $\widehat{U}^{\,}_{\mathrm{B}}(g)$
has a definite fermion parity.
However, unlike the representation
$\widehat{U}^{\,}_{\mathrm{bulk}}$,
the representation
$\widehat{U}^{\,}_{\mathrm{B}}$
can be projective as we shall explain.

\subsection{The case of $[\mu]=0$} 

When the number $n$ of Majorana operators
on the boundary $\Lambda^{\,}_{\mathrm{B}}$ is even, $[\mu]=0$.
We denote the identity on the local fermionic Fock space
(\ref{eq:def F[mu]=0})
by $\hat{\mathbb{1}}^{\,}_{\mathrm{B},0}$.
The boundary representation of element $p\in G^{\,}_{f}$
that generates the fermion parity group $\mathbb{Z}^{\mathrm{F}}_{2}$
is chosen to be
\begin{subequations}
\label{eq:boundary rep parity gen mu0}
\begin{align}
\label{eq:boundary rep parity gen mu0 a}
\widehat{U}^{\,}_{\mathrm{B}}(p)\df
\prod_{\alpha=1}^{n/2}
\widehat{P}^{\,}_{\alpha},
\quad
\widehat{P}^{\,}_{\alpha}\df
\mathrm{i}
\hat{\gamma}^{\,}_{2\alpha-1}\,
\hat{\gamma}^{\,}_{2\alpha}.
\end{align}
The parity operators $\widehat{P}^{\,}_{1},\cdots,\widehat{P}^{\,}_{n/2}$  
are Hermitian, square to the identity,
and are pairwise commuting. Hence, 
$\widehat{U}^{\,}_{\mathrm{B}}(p)$ is Hermitian and squares to
the identity.
Since operators $\widehat{P}^{\,}_{1},\cdots,\widehat{P}^{\,}_{n/2}$  
are pairwise commuting,
we can simultaneously diagonalize them and 
choose any one of them to be even
under complex conjugation $\mathsf{K}$,
\begin{equation}
\mathsf{K}\,\widehat{P}^{\,}_{\alpha}\,\mathsf{K}=\widehat{P}^{\,}_{\alpha},
\label{eq:boundary rep parity gen mu0 c}
\end{equation}
\end{subequations}
for $\alpha=1,\cdots,n/2$.
The most general form of a representation of element $g\in G^{\,}_{f}$ is
\begin{align}
\widehat{U}^{\,}_{\mathrm{B}}(g)\df
\widehat{V}^{\,}_{\mathrm{B}}(g)\,
\mathsf{K}^{\mathfrak{c}(g)},
\label{eq:boundary rep gen mu0}
\end{align}
where $\widehat{V}^{\,}_{\mathrm{B}}(g)$ is a unitary operator
that belongs to $\mathrm{C}\ell^{\,}_{n}$ defined in
Eq.\ (\ref{eq:def mathfrak O}).

\subsection{The case of $[\mu]=1$}

When the number $n$ of Majorana operators
on the boundary $\Lambda^{\,}_{\mathrm{B}}$ is odd, $[\mu]=1$.
We denote the identity on the nonlocal fermionic Fock space
(\ref{eq:def F[mu]=1})
by $\hat{\mathbb{1}}^{\,}_{\mathrm{B},1}$.
The boundary representation of element $p\in G^{\,}_{f}$
that generates the fermion parity group $\mathbb{Z}^{\mathrm{F}}_{2}$
is chosen to be
\begin{subequations}
\label{eq:boundary rep parity gen mu1}
\begin{align}
&
\widehat{U}^{\,}_{\mathrm{B}}(p)\df
\widehat{P}\,
\widehat{P}^{\,}_{\mathrm{nonloc}},
\label{eq:boundary rep parity gen mu1 a}
\\
&
\widehat{P}\df
\prod\limits_{\alpha=1}^{(n-1)/2}
\widehat{P}^{\,}_{\alpha},
\quad
\widehat{P}^{\,}_{\alpha}\df
\mathrm{i}
\hat{\gamma}^{\,}_{2\alpha-1}\,
\hat{\gamma}^{\,}_{2\alpha},
\label{eq:boundary rep parity gen mu1 b}
\\
&
\widehat{P}^{\,}_{\mathrm{nonloc}}\df
\mathrm{i}
\hat{\gamma}^{\,}_{n}\,
\hat{\gamma}^{\,}_{\infty},
\label{eq:boundary rep parity gen mu1 c}
\end{align}
for $\widehat{U}^{\,}_{\mathrm{B}}(p)$
is proportional to the product
$\hat{\gamma}^{\,}_{1}\,\cdots\,\hat{\gamma}^{\,}_{n}\,\hat{\gamma}^{\,}_{\infty}$
of all the generators in $\mathrm{C}\ell^{\,}_{n+1}$. As such,
$\widehat{U}^{\,}_{\mathrm{B}}(p)$ 
anticommutes with all the Majorana operators that span 
the nonlocal fermionic Fock space (\ref{eq:def F[mu]=1}). 
The parity operators $\widehat{P}^{\,}_{1},\cdots,\widehat{P}^{\,}_{(n-1)/2},
\widehat{P}^{\,}_{\mathrm{nonloc}}$ 
are Hermitian, square to the identity, and are pairwise commuting.
We choose to diagonalize them
simultaneously and choose each of them to be even
under complex conjugation $\mathsf{K}$,
\begin{equation}
\mathsf{K}\,\widehat{P}^{\,}_{\alpha}\,\mathsf{K}=\widehat{P}^{\,}_{\alpha},
\qquad
\mathsf{K}\,\widehat{P}^{\,}_{\mathrm{nonloc}}\,\mathsf{K}=
\widehat{P}^{\,}_{\mathrm{nonloc}},
\label{eq:def parity ops mu1 b}
\end{equation}
\end{subequations}
for $\alpha,\alpha'=1,\cdots,(n-1)/2$.

In addition to defining a representation of the fermion parity $p$,
we need to account for the fact that the center of the
Clifford algebra $\mathrm{C}\ell^{\,}_{n}$ is two-dimensional when
$n$ is odd. We choose to represent the nontrivial element of
this center by
\begin{align}
\widehat{Y}^{\,}_{\mathrm{B}}\df
\widehat{P}\,
\hat{\gamma}^{\,}_{n},
\qquad
\widehat{Y}^{\dag}_{\mathrm{B}}
=
\widehat{Y}^{\,}_{\mathrm{B}},
\qquad
\widehat{Y}^{2}_{\mathrm{B}}
=
\hat{\mathbb{1}}^{\,}_{\mathrm{B},1}.
\label{eq:def re YB for mu1}
\end{align}
By construction,
$\widehat{Y}^{\,}_{\mathrm{B}}$
is proportional to the product
$\hat{\gamma}^{\,}_{1}\,\cdots\,\hat{\gamma}^{\,}_{n}
\neq\hat{\mathbb{1}}^{\,}_{\mathrm{B},1}$.
It commutes with the Majorana operators 
$\hat{\gamma}^{\,}_{1},\cdots,\hat{\gamma}^{\,}_{n}$,
while it anticommutes with the Majorana operator 
$\hat{\gamma}^{\,}_{\infty}$.
The operator
$\widehat{Y}^{\,}_{\mathrm{B}}$ is of odd fermion parity for it
anticommutes with the fermion parity operator
(\ref{eq:boundary rep parity gen mu1}).
Because $\widehat{Y}^{\,}_{\mathrm{B}}$ commutes with
all the elements of $\mathrm{C}\ell^{\,}_{n}$,
it follows that the left-hand side of Eq.\
(\ref{eq:consistency cond general})
is invariant under the $G^{\,}_{f}$-resolved transformation
\begin{equation}
\widehat{U}^{\,}_{\mathrm{B}}(g)\mapsto
\widehat{U}^{\,}_{\mathrm{B}}(g)\,
\widehat{Y}^{\,}_{\mathrm{B}}
\label{eq: symmetry of def UB(g) under right multiplicationof UB(g) by YB(g)}
\end{equation}
under which the fermion parity of $\widehat{U}^{\,}_{\mathrm{B}}(g)$
is reversed.

Since the Clifford algebra $\mathrm{C}\ell^{\,}_{n}$ is closed 
under the action of the boundary representation
$\widehat{U}^{\,}_{\mathrm{bulk}}(g)$, 
the same must be true
for the boundary representation $\widehat{U}^{\,}_{\mathrm{B}}(g)$
[recall Eqs.\ \eqref{eq: boundary modes are closed under Gf}
and \eqref{eq:consistency cond general}].
In other words, $\widehat{U}^{\,}_{\mathrm{B}}(g)$ preserves locality
in that its action on those operators whose non-trivial actions are
limited to $\Lambda^{\,}_{\mathrm{B}}$ is merely to mix them. This locality
is guaranteed only if the condition 
\begin{align}
\left[
\widehat{U}^{\,}_{\mathrm{B}}(g)\,
\hat{\gamma}^{\,}_{i}\,
\widehat{U}^{\dagger}_{\mathrm{B}}(g),
\widehat{Y}^{\,}_{\mathrm{B}}
\right]=
0,
\label{eq:cond locality preserving}
\end{align}
is satisfied
for any $g\in G^{\,}_{f}$ and 
$i=1,\cdots,n$. In turn, 
condition (\ref{eq:cond locality preserving}) implies that 
$\widehat{U}^{\,}_{\mathrm{B}}(g)$ either
commutes or anticommutes with the center $\widehat{Y}^{\,}_{\mathrm{B}}$
of $\mathrm{C}\ell^{\,}_{n}$, i.e.,
\begin{align}
\label{eq:UB decomposition proof final a}
\widehat{Y}^{\,}_{\mathrm{B}}\,
\widehat{U}^{\,}_{\mathrm{B}}(g)
=
\pm
\widehat{U}^{\,}_{\mathrm{B}}(g)\,
\widehat{Y}^{\,}_{\mathrm{B}}.
\end{align}
Furthermore, this is true only if the decomposition
\begin{align}
\widehat{U}^{\,}_{\mathrm{B}}(g)\df
\widehat{V}^{\,}_{\mathrm{B}}(g)\,
\widehat{Q}^{\,}_{\mathrm{B}}(g)\,
\mathsf{K}^{\mathfrak{c}(g)}
\label{eq:boundary rep gen mu1 pre}
\end{align}
holds. Here,
$\widehat{V}^{\,}_{\mathrm{B}}(g)\in\mathrm{C}\ell^{\,}_{n}\subset
\mathrm{C}\ell^{\,}_{n+1}$ is a unitary operator with
well-defined fermion parity and
the operator $\widehat{Q}^{\,}_{\mathrm{B}}(g)$ is either proportional
to the identity operator in $\mathrm{C}\ell^{\,}_{n+1}$ or to the operator
$\hat{\gamma}^{\,}_{\infty}$.

The invariance of Eq.\
(\ref{eq:consistency cond general})
under the $G^{\,}_{f}$-resolved transformation
(\ref{eq: symmetry of def UB(g) under right multiplicationof UB(g) by YB(g)})
allows to fix the fermion parity of $\widehat{U}^{\,}_{\mathrm{B}}(g)$
to be even for all $g\in G^{\,}_{f}$. In this ``gauge'',
\begin{align}
\widehat{U}^{\,}_{\mathrm{B}}(g)=
\widehat{V}^{\,}_{\mathrm{B}}(g)\,
\widehat{Q}^{\,}_{\mathrm{B}}(g)\,
\mathsf{K}^{\mathfrak{c}(g)},
\quad
\widehat{Q}^{\,}_{\mathrm{B}}(g)=
\left[
\hat{\gamma}^{\,}_{\infty}
\right]^{q(g)},
\label{eq:boundary rep gen mu1} 
\end{align}
where $q(g)=0,1$ denotes the fermion parity of the unitary 
operator $\widehat{V}^{\,}_{\mathrm{B}}(g)$.
Equation
(\ref{eq:boundary rep gen mu1})
together with Eqs.\
(\ref{eq:boundary rep parity gen mu1})
and
(\ref{eq:def re YB for mu1})
define the realization
of the symmetry group $G^{\,}_{f}$
on the boundary $\Lambda^{\,}_{\mathrm{B}}$ 
when $[\mu]=1$. 

\section{Definition of indices}
\label{sec: Definition of indices}

We consider a boundary representation
$\widehat{U}^{\,}_{\mathrm{B}}\colon
G^{\,}_{f}\to
\mathrm{Aut}
\big(\mathfrak{F}^{\,}_{\Lambda^{\,}_{\mathrm{B}},[\mu]}\big)$,
where $\mathrm{Aut}
\big(\mathfrak{F}^{\,}_{\Lambda^{\,}_{\mathrm{B}},[\mu]}\big)$
denotes the set of automorphisms on the fermionic Fock space
$\mathfrak{F}^{\,}_{\Lambda^{\,}_{\mathrm{B}},[\mu]}$. 
We demand that this map satisfies, for any $g,h,f\in G^{\,}_{f}$,
\begin{subequations}
\label{eq:def proj rep on one component bd}
\begin{align}
\widehat{U}^{\,}_{\mathrm{B}}(g)\,
\widehat{U}^{\,}_{\mathrm{B}}(h)=
e^{\mathrm{i}\phi(g,h)}\,
\widehat{U}^{\,}_{\mathrm{B}}(g\,h),
\label{eq:def proj rep on one component bd b}
\end{align}
where $g\,h$ denotes the 
composition of the elements $g,h\in G^{\,}_{f}$.  
The map
$\phi(\cdot,\cdot)\in
C^{2}\big(G^{\,}_{f},\mathrm{U}(1)\big)$
is a $\mathrm{U}(1)$-valued 2-cochain%
~\footnote{
Note that we denote the elements of the set of 2-cochains
$C^{2}(G^{\,}_{f},U(1))$
by the phase $\phi(g^{\,}_{1},g^{\,}_{2})$ as opposed to its exponential as is 
the usual convention.
This is because we impose an additive composition rule on the group 
$U(1)$ as opposed to a multiplicative one.
See Appendix \ref{appsec:Group Cohomology}
for the definition of the set $C^{2}(G^{\,}_{f},U(1))$
and details on the convention we adopt.
          }
(Appendix \ref{appsec:Group Cohomology}).
Furthermore, to ensure the compatibility with the
associativity of the composition law of $G^{\,}_{f}$,
we demand that, for any $g,h,f\in G^{\,}_{f}$,
\begin{align}
\phi(g,h)+\phi(gh,f)=
(-1)^{\mathfrak{c}(g)}\,\phi(h,f)
+
\phi(g,hf).
\label{eq:def proj rep on one component bd d}
\end{align}
\end{subequations}
The 2-cochains that satisfy this condition are called 2-cocycles
(Appendix \ref{appsec:Group Cohomology}).
The map (\ref{eq:def proj rep on one component bd}) defines
a projective representation of the symmetry group $G^{\,}_{f}$.

Under the gauge transformation 
\begin{subequations}
\label{eq:def gauge equiv bd}
\begin{align}
\widehat{U}^{\,}_{\mathrm{B}}(g)
\mapsto 
e^{\mathrm{i}\xi(g)}\,
\widehat{U}^{\,}_{\mathrm{B}}(g),
\label{eq:def gauge equiv bd a}
\end{align}
the phase $\phi(g,h)$ entering any
projective representation of the symmetry group $G^{\,}_{f}$
changes by
\begin{align}
\phi'(g,h)
-
\phi(g,h)
= 
\xi(g\,h)
-
\xi(g)
-
(-1)^{\mathfrak{c}(g)}\,
\xi(h)
\label{eq:def gauge equiv bd c}
\end{align} 
for any $g,h\in G^{\,}_{f}$.
Two 2-cochains $\phi$ and $\phi'$ are equivalent if they are related
by a gauge transformation.
The 2-cochains $\phi$ that vanish under a 
gauge transformation, i.e., the identity
\begin{align}
\phi(g,h)
=
\xi(g\,h)
-
\xi(g)
-
(-1)^{\mathfrak{c}(g)}\,
\xi(h)
\label{eq:def gauge equiv bd d}
\end{align}
\end{subequations}
for any $g,h\in G^{\,}_{f}$ holds,
are called 2-coboundaries.
The set of equivalence classes $[\phi]$ of
2-cocycles under the gauge transformations is the second cohomology group
$H^{2}\big(G^{\,}_{f},\mathrm{U}(1)^{\,}_{\mathfrak{c}}\big)$
(Appendix \ref{appsec:Group Cohomology}).

Elements of $G^{\,}_{f}$ were refered to, so far, by single letters $g$,
with $e$ reserved for the identity and $p$ reserved from the fermion parity.
When we want to emphasize that elements of
$G^{\,}_{f}$
are elements of the set
$G\times\mathbb{Z}^{\mathrm{F}}_{2}$
as is done in Appendix \ref{appsec:central extension review},
we will denote an element of $G^{\,}_{f}$
as $(g,h)$ with $g\in G$, $h\in\mathbb{Z}^{\mathrm{F}}_{2}$. 
With this convention,
$(\mathrm{id},e)$ is the identity, $(\mathrm{id},p)$ is the fermion parity,
and the projection $(g,e)$ of $(g,h)$ on $G$ defines the inclusion map
$G\subset G^{\,}_{f}$.
Here, the 2-cochain $\nu\in C^{2}\big(G,\mathrm{U}(1)\big)$ captures
the projective representation
(\ref{eq:def proj rep on one component bd})
for those elements of $G^{\,}_{f}$ of the form $(g,e)$. 
When $[\mu]=0$, the 1-cochain $\rho\in C^{1}\big(G,\mathbb{Z}^{\,}_{2}\big)$
measures if an operator representing an element of $G$ commutes
or anticommutes with the operator representing the fermion parity $p$.
When $[\mu]=1$, the 1-cochain $\rho\in C^{1}\big(G,\mathbb{Z}^{\,}_{2}\big)$
measures if an operator representing an element of $G$ commutes
or anticommutes 
with the central element $\widehat{Y}^{\,}_{\mathrm{B}}$ of the Clifford algebra 
$\mathrm{C}\ell^{\,}_{n}$ when $[\mu]=1$.
Indeed, it is possible to organize
$C^{2}\big(G,\mathrm{U}(1)\big)\times C^{1}\big(G,\mathbb{Z}^{\,}_{2}\big)$
into a coset of equivalence classes $\{[(\nu,\rho)]\}$ such that
there is a one-to-one correspondence between
any element $[\phi]\in H^{2}\big(G^{\,}_{f},\mathrm{U}(1)^{\,}_{\mathfrak{c}}\big)$
and $[(\nu,\rho)]$, as was shown in Ref.\ \onlinecite{Turzillo2019}
and is reviewed in Appendix \ref{appsec:classification of Gf}.
When defining the indices $(\nu,\rho)$, 
we made an implicit choice 
for which elements of $G^{\,}_{f}$ are mapped to
the pairs $(g,e)\in G\times \mathbb{Z}^{F}_{2}$.
Different choices are related to each other by group isomorphisms
on $G^{\,}_{f}$. We explain in
Appendix \ref{appsec:group isomorphism on indices}
how the pair $(\nu,\rho)$ changes under isomorphisms relating 
different representatives of the central extension class
$[\gamma]\in H^{2}(G,\mathbb{Z}^{F}_{2})$.

\subsection{The case of $[\mu]=0$}

When the number $n$ of Majorana operators
on the boundary $\Lambda^{\,}_{\mathrm{B}}$ is even, $[\mu]=0$.
The 2-cochain $\nu\in C^{2}\big(G,\mathrm{U}(1)\big)$ is defined 
by restricting the domain of definition of
the 2-cochain $\phi$ from $G^{\,}_{f}$ to $G$, 
\begin{equation}
\nu(g,h)\df
\phi\big((g,e),(h,e)\big),
\label{eq:def 2-cochain nu [mu]=0}
\end{equation}
for any $g,h\in G$.

\begin{widetext}
The 1-cochain
$\rho\in C^{1}\big(G^{\,}_{f},\mathbb{Z}^{\,}_{2}\big)$
is defined by the relation
\begin{align}
e^{\mathrm{i}\pi \rho\big((g,h)\big)}\equiv\,
(-1)^{\rho\big((g,h)\big)}
\df\,
\begin{cases}
\widehat{U}^{\,}_{\mathrm{B}}\big((g,h)\big)\,
\widehat{U}^{\,}_{\mathrm{B}}\big((\mathrm{id},p)\big)\,
\widehat{U}^{\dag}_{\mathrm{B}}\big((g,h)\big)\,
\widehat{U}^{\dag}_{\mathrm{B}}\big((\mathrm{id},p)\big),
&
\text{ if }\mathfrak{c}\big((g,h)\big)=0,
\\ &\\
\widehat{U}^{\,}_{\mathrm{B}}\big((g,h)\big)\,
\widehat{U}^{\,}_{\mathrm{B}}\big((\mathrm{id},p)\big)\,
\widehat{U}^{\dag}_{\mathrm{B}}\big((g,h)\big)\,
\widehat{U}^{\,}_{\mathrm{B}}\big((\mathrm{id},p)\big),
&
\text{ if }\mathfrak{c}\big((g,h)\big)=1,
\end{cases}
\label{eq:def rho gen mu0 on Gf}
\end{align}
for any $(g,h)\in G^{\,}_{f}$.
The 1-cochain
$\rho\in C^{1}\big(G^{\,}_{f},\mathbb{Z}^{\,}_{2}\big)$
takes the values $0$ or $1$.
The 1-cochain $\rho\in C^{1}\big(G^{\,}_{f},\mathbb{Z}^{\,}_{2}\big)$
is a group homomorphism from
$G^{\,}_{f}$ to $\mathbb{Z}^{\,}_{2}=\left\{0,1\right\}$,
since it has a vanishing coboundary and, hence, is a 1-cocycle%
~\footnote{
A $\mathbb{Z}^{\,}_{2}$ valued 1-cocycle
$\rho\in Z^{1}(G^{\,}_{f},\mathbb{Z}^{\,}_{2})$ satisfies by definition 
$(\delta^{1}_{\mathfrak{c}}\rho)(g,h)=\rho(g)+\mathfrak{c}(g)\rho(h)-\rho(g\,h)=0$
for any $g,h\in G^{\,}_{f}$. Since in the group $\mathbb{Z}^{\,}_{2}$,
$\rho(h)=\pm\rho(h)$ for any $h\in G^{\,}_{f}$, the 1-cocycle
$\rho$ is a group homomorphism.
          }
(Appendix \ref{appsec:Group Cohomology}).
It measures the fermion parity of the operator
$\widehat{U}^{\,}_{\mathrm{B}}\big((g,h)\big)$.  As expected we have
$\rho\big((\mathrm{id},p)\big)=0$.

The 1-cochain $\rho\in C^{1}\big(G,\mathbb{Z}^{\,}_{2}\big)$
is defined by restricting the domain of definition of
$\rho\in C^{1}\big(G^{\,}_{f},\mathbb{Z}^{\,}_{2}\big)$
from $G^{\,}_{f}$ to $G$, i.e.,
\begin{subequations}
\label{eq:def rho gen mu0}
\begin{align}
e^{\mathrm{i}\pi \rho(g)}\equiv\,
(-1)^{\rho(g)}
\df\,
\begin{cases}
\widehat{U}^{\,}_{\mathrm{B}}\big((g,e)\big)\,
\widehat{U}^{\,}_{\mathrm{B}}\big((\mathrm{id},p)\big)\,
\widehat{U}^{\dag}_{\mathrm{B}}\big((g,e)\big)\,
\widehat{U}^{\dag}_{\mathrm{B}}\big((\mathrm{id},p)\big),
&
\text{ if }\mathfrak{c}\big((g,e)\big)=0,
\\ &\\
\widehat{U}^{\,}_{\mathrm{B}}\big((g,e)\big)\,
\widehat{U}^{\,}_{\mathrm{B}}\big((\mathrm{id},p)\big)\,
\widehat{U}^{\dag}_{\mathrm{B}}\big((g,e)\big)\,
\widehat{U}^{\,}_{\mathrm{B}}\big((\mathrm{id},p)\big),
&
\text{ if }\mathfrak{c}\big((g,e)\big)=1,
\end{cases}
\end{align}
for any $g\in G$.
In terms of the 2-cocycle $\phi$,
$\rho\in C^{1}\big(G,\mathbb{Z}^{\,}_{2}\big)$
is, for any $g\in G$, given by
\begin{align}
\rho(g)=
\frac{1}{\pi}
\left[
\phi\big((g,e),(\mathrm{id},p)\big)
-
\phi\big((\mathrm{id},p),(g,e)\big)
+
\mathfrak{c}(g,e)\,
\phi\big((\mathrm{id},p),(\mathrm{id},p)\big)
\right].
\end{align}
\end{subequations} 
\end{widetext}
The definitions \eqref{eq:def rho gen mu0 on Gf} 
and \eqref{eq:def rho gen mu0} are made so that
the 1-cochain $\rho$ is invariant under the gauge transformation
\eqref{eq:def gauge equiv bd a}. We note that when a
gauge choice is made by choosing the representation 
$\widehat{U}\big((\mathrm{id},p)\big)$ to be Hermitian,
the two cases in the definitions \eqref{eq:def rho gen mu0 on Gf} 
and \eqref{eq:def rho gen mu0} are equivalent.

\subsection{The case of $[\mu]=1$} 
\label{subsec: The case of [mu]=1.} 

When the number $n$ of Majorana operators
on the boundary $\Lambda^{\,}_{\mathrm{B}}$ is odd, $[\mu]=1$.
The 2-cochain $\nu\in C^{2}\big(G,\mathrm{U}(1)\big)$ is defined 
by restricting the domain of definition of
the 2-cochain $\phi$ from $G^{\,}_{f}$ to $G$, 
\begin{equation}
\nu(g,h)\df
\phi\big((g,e),(h,e)\big),
\label{eq:def 2-cochain nu [mu]=1}
\end{equation}  
for any $g,h\in G$.

When $[\mu]=1$, the Clifford algebra $\mathrm{C}\ell^{\,}_{n}$ 
spanned by the Majorana operators (\ref{eq:def mathfrak O})
has a two-dimensional center, in which case
the fermion parity of the boundary representation
$\widehat{U}^{\,}_{\mathrm{B}}\big((g,h)\big)$
for any element $(g,h)\in G^{\,}_{f}$
can be reversed by multiplying it with
the generator $\widehat{Y}^{\,}_{\mathrm{B}}$
of the two-dimensional center of
the Clifford algebra $\mathrm{C}\ell^{\,}_{n}$.
Moreover,
any $\widehat{U}^{\,}_{\mathrm{B}}\big((g,h)\big)$ 
must either commute or anticommute with 
$\widehat{Y}^{\,}_{\mathrm{B}}$
according to Eq.\ (\ref{eq:UB decomposition proof final a}).

\begin{widetext}
For this reason, we define the 1-cochain
$\rho\in C^{1}\big(G^{\,}_{f},\mathbb{Z}^{\,}_{2}\big)$
through 
\begin{align}
e^{\mathrm{i}\pi \rho\big((g,h)\big)}\equiv\,
(-1)^{\rho\big((g,h)\big)}\df\,
\begin{cases}
\widehat{U}^{\,}_{\mathrm{B}}\big((g,h)\big)\,
\widehat{Y}^{\,}_{\mathrm{B}}\,
\widehat{U}^{\dag}_{\mathrm{B}}\big((g,h)\big)\,
\widehat{Y}^{\dag}_{\mathrm{B}},
&
\text{ if }\mathfrak{c}\big((g,h)\big)=0,
\\&\\
\widehat{U}^{\,}_{\mathrm{B}}\big((g,h)\big)\,
\widehat{Y}^{\,}_{\mathrm{B}}\,
\widehat{U}^{\dag}_{\mathrm{B}}\big((g,h)\big)\,
\widehat{Y}^{\,}_{\mathrm{B}},
&
\text{ if }\mathfrak{c}\big((g,h)\big)=1,
\end{cases}
\label{eq:def rho gen mu1 on Gf}
\end{align}
for any $(g,h)\in G^{\,}_{f}$. 
The 1-cochain $\rho\in C^{1}\big(G^{\,}_{f},\mathbb{Z}^{\,}_{2}\big)$
takes the value 0 and 1.
The 1-cochain $\rho\in C^{1}\big(G^{\,}_{f},\mathbb{Z}^{\,}_{2}\big)$
is a group homomorphism from
$G^{\,}_{f}$ to $\mathbb{Z}^{\,}_{2}=\left\{0,1\right\}$
since it has a vanishing coboundary and, hence, is a 1-cocycle
(Appendix \ref{appsec:Group Cohomology}).
Since $\widehat{Y}^{\,}_{\mathrm{B}}$ is of odd fermion parity 
by definition \eqref{eq:def re YB for mu1}, it anticommutes with the 
representation $\widehat{U}^{\,}_{\mathrm{B}}\big((\mathrm{id},p)\big)$.
This implies that $\rho(\mathrm{id},p)=1$.
More generally,
the 1-cochain $\rho\in C^{1}\big(G^{\,}_{f},\mathbb{Z}^{\,}_{2}\big)$
measures if the representation
$\widehat{U}^{\,}_{\mathrm{B}}(g,h)$
of $(g,h)\in G^{\,}_{f}$ commutes
or anticommutes with 
$\widehat{Y}^{\,}_{\mathrm{B}}$.

The 1-cochain $\rho\in C^{1}\big(G,\mathbb{Z}^{\,}_{2}\big)$
is defined by restricting the domain of definition of
$\rho\in C^{1}\big(G^{\,}_{f},\mathbb{Z}^{\,}_{2}\big)$
from $G^{\,}_{f}$ to $G$, i.e.,
\begin{align}
e^{\mathrm{i}\pi \rho(g)}
\equiv\,
(-1)^{\rho(g)}
\df\,
\begin{cases}
\widehat{U}^{\,}_{\mathrm{B}}\big((g,e)\big)\,
\widehat{Y}^{\,}_{\mathrm{B}}\,
\widehat{U}^{\dag}_{\mathrm{B}}\big((g,e)\big)\,
\widehat{Y}^{\dag}_{\mathrm{B}},
&
\text{ if }\mathfrak{c}\big((g,e)\big)=0,
\\&\\
\widehat{U}^{\,}_{\mathrm{B}}\big((g,e)\big)\,
\widehat{Y}^{\,}_{\mathrm{B}}\,
\widehat{U}^{\dag}_{\mathrm{B}}\big((g,e)\big)\,
\widehat{Y}^{\,}_{\mathrm{B}},
&
\text{ if }\mathfrak{c}\big((g,e)\big)=1,
\end{cases}
\label{eq:def rho gen mu1}
\end{align}
for any $g\in G$.
\end{widetext}

The definitions \eqref{eq:def rho gen mu1 on Gf} 
and \eqref{eq:def rho gen mu1} are made so that
the 1-cochain $\rho$ is invariant under the gauge transformation
\eqref{eq:def gauge equiv bd a}. We note that when a
gauge choice is made by choosing the representation 
$\widehat{Y}^{\,}_{\mathrm{B}}$ to be Hermitian,
the two cases in the definitions \eqref{eq:def rho gen mu1 on Gf} 
and \eqref{eq:def rho gen mu1} are equivalent.

The fact that the 1-cochain 
$\rho\in C^{1}\big(G^{\,}_{f},\mathbb{Z}^{\,}_{2}\big)$
defined in Eq.\ \eqref{eq:def rho gen mu1 on Gf}
is a group homomorphism puts constraints on the structure of the
internal symmetry group $G^{\,}_{f}$.
Compatibility between the existence of the
group homomorphism $\rho
\in C^{1}\big(G^{\,}_{f},\mathbb{Z}^{\,}_{2}\big)$
which is
defined in Eq.\ \eqref{eq:def rho gen mu1 on Gf}
and the group composition 
rule in $G^{\,}_{f}$ [see Eq.\ (\ref{eq:def ZF2})]
requires that the central extension class
$[\gamma]\in H^{2}(G^{\,}_{f},\mathbb{Z}^{\mathrm{F}}_{2})$
is trivial, i.e., $[\gamma]=0$.
This is so because, when restricted to the center
$\mathbb{Z}^{\mathrm{F}}_{2}\subset G^{\,}_{f}$,
the homomorphism $\rho\in C^{1}\big(G^{\,}_{f},\mathbb{Z}^{\,}_{2}\big)$ 
is a group isomorphism%
~\footnote{
This is not true when $[\mu]=0$. The group homomorphism 
$\rho\in C^{1}\big(G^{\,}_{f},\mathbb{Z}^{\,}_{2}\big)$
defined in Eq.\ \eqref{eq:def rho gen mu0 on Gf} takes the values
$\rho\big((\mathrm{id},e)\big)=\rho\big((\mathrm{id},p)\big)=0$. 
Therefore, when restricted to 
the center $\mathbb{Z}^{\mathrm{F}}_{2}\subset G^{\,}_{f}$, it is not 
an isomorphism.
          }.
It can then be used to 
construct a group isomorphism from $G^{\,}_{f}$ to the direct product
$G\times\mathbb{Z}^{\mathrm{F}}_{2}$.
In other words, the only internal symmetry groups $G^{\,}_{f}$
compatible with boundaries supporting an odd number 
of Majorana degrees of 
freedom ($[\mu]=1$) 
are those that split, i.e, 
$G^{\,}_{f}=G\times\mathbb{Z}^{\mathrm{F}}_{2}$.

For simplicity, 
we revert back to the single letters (e.g., $g$ or $h$) 
to denote elements of the group $G^{\,}_{f}$ from now on.   
We will use this notation as long as there are no ambiguities. 
Whenever used, the appropriate definition for 1-cochain $\rho$ 
[definitions \eqref{eq:def rho gen mu0 on Gf} or \eqref{eq:def rho gen mu0}
and 
definitions \eqref{eq:def rho gen mu1 on Gf} or \eqref{eq:def rho gen mu1}]
should be understood from the context.

We close Sec.\
\ref{subsec: The case of [mu]=1.}
by spelling out two identities that will be convenient 
when deriving the stacking rules
in Sec.\ \ref{sec: Fermionic Stacking Rules}.
We note that definition \eqref{eq:def rho gen mu1 on Gf}
involves conjugation of the central element $\widehat{Y}^{\,}_{\mathrm{B}}$
by the boundary representation 
$\widehat{U}^{\,}_{\mathrm{B}}(g)$ 
of some element $g\in G^{\,}_{f}$. 
By definitions
\eqref{eq:boundary rep parity gen mu1}
and
\eqref{eq:def re YB for mu1},
$\widehat{Y}^{\,}_{\mathrm{B}}$ can be written as
\begin{subequations}
\label{eq:useful identity gen gamm inf conjugation}
\begin{align}
\widehat{Y}^{\,}_{\mathrm{B}}
=
-
\mathrm{i}\,
\widehat{U}^{\,}_{\mathrm{B}}(p)\,
\hat{\gamma}^{\,}_{\infty}.
\end{align}
Using this identity in definition \eqref{eq:def rho gen mu1 on Gf}
allows one to express the complex conjugation of $\hat{\gamma}^{\,}_{\infty}$
in terms of group homomorphisms $\mathfrak{c}$, $q$, and $\rho$.
Since Eq.\ \eqref{eq:def parity ops mu1 b} implies
that the Majorana operators $\hat{\gamma}^{\,}_{\infty}$ 
and $\hat{\gamma}^{\,}_{n}$ transform oppositely under complex conjugation,
one finds the pair of identities
\begin{align}
&
\mathsf{K}^{\mathfrak{c}(g)}\,
\hat{\gamma}^{\,}_{\infty}\,
\mathsf{K}^{\mathfrak{c}(g)}=
(-1)^{\mathfrak{c}(g)+q(g)+\rho(g)}
\hat{\gamma}^{\,}_{\infty},
\label{eq:useful identity gen gamm inf conjugation a}
\\
&
\mathsf{K}^{\mathfrak{c}(g)}\,
\hat{\gamma}^{\,}_{n}\,
\mathsf{K}^{\mathfrak{c}(g)}=
(-1)^{q(g)+\rho(g)}
\hat{\gamma}^{\,}_{n},
\label{eq:useful identity gen gamm inf conjugation b}
\end{align}
\end{subequations}
for any $g\in G^{\,}_{f}$

\section{Fermionic stacking rules}
\label{sec: Fermionic Stacking Rules}

Given the two triplets
$\left((\nu^{\,}_{1},\rho^{\,}_{1}),[\mu^{\,}_{1}]\right)$
and
$\left((\nu^{\,}_{2},\rho^{\,}_{2}),[\mu^{\,}_{2}]\right)$
associated to the pair 
$\widehat{U}^{\,}_{1}$
and
$\widehat{U}^{\,}_{2}$
of boundary representations, respectively, 
we shall construct the triplet
$\left((\nu^{\,}_{\wedge},
\rho^{\,}_{\wedge}),[\mu^{\,}_{\wedge}]\right)$
that is associated with the representation
$\widehat{U}^{\,}_{\wedge}$,
whereby $\widehat{U}^{\,}_{\wedge}$ must be
compatible with the symmetry group $G^{\,}_{f}$
and is obtained from taking the tensor product of
the two set of boundary degrees of freedom.
We call this operation stacking.

Since the number of boundary Majorana degrees of freedom
on which $\widehat{U}^{\,}_{\wedge}$ acts is obtained by adding
the boundary Majorana degrees of freedom
\begin{equation}
\mathfrak{O}^{\,}_{1}\df
\left\{
\hat{\gamma}^{(1)}_{1},\,
\hat{\gamma}^{(1)}_{2},\,
\cdots,\,
\hat{\gamma}^{(1)}_{n^{\,}_{1}}
\right\}
\end{equation}
on which $\widehat{U}^{\,}_{1}$ acts
to the boundary Majorana degrees of freedom
\begin{equation}
\mathfrak{O}^{\,}_{2}\df
\left\{
\hat{\gamma}^{(2)}_{1},\,
\hat{\gamma}^{(2)}_{2},\,
\cdots,\,
\hat{\gamma}^{(2)}_{n^{\,}_{2}}
\right\}
\end{equation}
on which $\widehat{U}^{\,}_{2}$ acts,
we define the index $[\mu^{\,}_{\mathrm{\wedge}}]$
of the stacked representation to be
\begin{align}
[\mu^{\,}_{\wedge}]\df
[\mu^{\,}_{1}]
+
[\mu^{\,}_{2}]
\text{ mod }2.
\end{align}
For any $g\in G^{\,}_{f}$, we define the
stacked representation $\widehat{U}^{\,}_{\wedge}(g)$ 
to be a norm preserving operator 
that satisfies the identities
\begin{subequations}
\label{eq:stacked consistency cond gen}
\begin{align}
&
\widehat{U}^{\,}_{\wedge}(g)\,
\hat{\gamma}^{(1)}_{i}\,
\widehat{U}^{\dag}_{\wedge}(g)\df
\widehat{U}^{\,}_{1}(g)\,
\hat{\gamma}^{(1)}_{i}\,
\widehat{U}^{\dag}_{1}(g),
\label{eq:stacked consistency cond gen a}
\\
&
\widehat{U}^{\,}_{\wedge}(g)\,
\hat{\gamma}^{(2)}_{j}\,
\widehat{U}^{\dag}_{\wedge}(g)\df
\widehat{U}^{\,}_{2}(g)\,
\hat{\gamma}^{(2)}_{j}\,
\widehat{U}^{\dag}_{2}(g),
\label{eq:stacked consistency cond gen b}
\end{align}
\end{subequations}
for $i=1,\cdots,n^{\,}_{1}$
and $j=1,\cdots,n^{\,}_{2}$.
This definition is the natural generalization of Eq.\ 
\eqref{eq:consistency cond general}. Because
$\widehat{U}^{\,}_{1}(g)$ and $\widehat{U}^{\,}_{2}(g)$ act on 
single Majorana operators in the same way as the
bulk representation of the element 
$g\in G^{\,}_{f}$ does, the same is true for the stacked representation
$\widehat{U}^{\,}_{\wedge}(g)$.
The stacked representation $\widehat{U}^{\,}_{\wedge}(g)$
is not unique since
Eqs.\
(\ref{eq:stacked consistency cond gen a})
and
(\ref{eq:stacked consistency cond gen b})
are left invariant by the multiplications from the right
of $\widehat{U}^{\,}_{\wedge}(g)$ with any norm-preserving element 
from the center of the Clifford algebra
$\mathrm{C}\ell^{\,}_{n^{\,}_{1}+n^{\,}_{2}}$.

When
constructing an explicit representation of $\widehat{U}^{\,}_{\wedge}(g)$
for any $g\in G^{\,}_{f}$,
we shall consider the three cases:
(i) even-even stacking, $[\mu^{\,}_{1}]=[\mu^{\,}_{2}]=0$,
(ii) even-odd stacking, $[\mu^{\,}_{1}]=0$, $[\mu^{\,}_{2}]=1$,
(iii) and odd-odd stacking, $[\mu^{\,}_{1}]=[\mu^{\,}_{2}]=1$.
The case of odd-even stacking is to be treated analogously to the case
of even-odd stacking.

As is done in Sec.\ \ref{sec:boundary rep gen},
we begin with the construction of a
representation of the fermion parity $p\in G^{\,}_{f}$. 
When $[\mu^{\,}_{\wedge}]=0$, the stacked representation of
$\widehat{U}^{\,}_{\wedge}(p)$ 
follows from combining
Eq.\ \eqref{eq:stacked consistency cond gen} 
with the counterpart to Eq.\
(\ref{eq:boundary rep parity gen mu0}).
When $[\mu^{\,}_{\wedge}]=1$, the stacked representation of
$\widehat{U}^{\,}_{\wedge}(p)$ 
follows from combining
Eq.\ \eqref{eq:stacked consistency cond gen} 
with the counterparts to 
Eqs.\
\eqref{eq:boundary rep parity gen mu1}.
More precisely, the stacked representation $\widehat{U}^{\,}_{\wedge}(p)$
of the fermion parity $p$ is defined to be
\begin{align}
\widehat{U}^{\,}_{\wedge}(p)\df
\begin{cases}
\widehat{U}^{\,}_{1}(p)\,
\widehat{U}^{\,}_{2}(p),
&
\text{ if } [\mu^{\,}_{1}]=[\mu^{\,}_{2}]=0,
\\
&
\\
\widehat{U}^{\,}_{1}(p)\,
\widehat{U}^{\,}_{2}(p),
&
\text{ if } [\mu^{\,}_{1}]=0, \, [\mu^{\,}_{2}]=1,
\\
&
\\
\widehat{P}^{\,}_{1}\,
\widehat{P}^{\,}_{2}\,
\mathrm{i}\hat{\gamma}^{(1)}_{n^{\,}_{1}}\,
\hat{\gamma}^{(2)}_{n^{\,}_{2}},
&
\text{ if } [\mu^{\,}_{1}]=[\mu^{\,}_{2}]=1.
\end{cases}
\label{eq:def stacked parity rep}
\end{align}
By construction, we have chosen a Hermitian representation
$\widehat{U}^{\,}_{\wedge}(p)$ of the fermion parity $p$.

Next, we fix the action 
of the stacked complex conjugation $\mathsf{K}^{\,}_{\wedge}$
on the single Majorana operators spanning the
fermionic Fock space of the stacked boundary
by demanding that some set of 
mutually commuting fermion parity operators
are left invariant under complex conjugation 
[recall Eqs.\ \eqref{eq:boundary rep parity gen mu0}
and \eqref{eq:boundary rep parity gen mu1}].
For the cases of even-even ($[\mu^{\,}_{1}]=[\mu^{\,}_{2}]=0$) 
and even-odd stacking ($[\mu^{\,}_{1}]=0$, $[\mu^{\,}_{2}]=1$),
we define $\mathsf{K}^{\,}_{\wedge}$ by
\begin{subequations}
\label{eq:stacked complex conj ee eo gen}
\begin{align}
&
\mathsf{K}^{\,}_{\wedge}\,
\hat{\gamma}^{(1)}_{i}\,
\mathsf{K}^{\,}_{\wedge}\df
\mathsf{K}^{\,}_{1}\,
\hat{\gamma}^{(1)}_{i}\,
\mathsf{K}^{\,}_{1},
\\
&
\mathsf{K}^{\,}_{\wedge}\,
\hat{\gamma}^{(2)}_{j}\,
\mathsf{K}^{\,}_{\wedge}\df
\mathsf{K}^{\,}_{2}\,
\hat{\gamma}^{(2)}_{j}\,
\mathsf{K}^{\,}_{2},
\end{align}
\end{subequations}
for $i=1,\cdots,n^{\,}_{1}$ and
$j=1,\cdots,n^{\,}_{2}$.
For the case of odd-odd stacking,
we define $\mathsf{K}^{\,}_{\wedge}$ by
\begin{subequations}
\label{eq:stacked complex conj oo gen}
\begin{align}
&
\mathsf{K}^{\,}_{\wedge}\,
\hat{\gamma}^{(1)}_{i}\,
\mathsf{K}^{\,}_{\wedge}\df
\mathsf{K}^{\,}_{1}\,
\hat{\gamma}^{(1)}_{i}\,
\mathsf{K}^{\,}_{1},
\\
&
\mathsf{K}^{\,}_{\wedge}\,
\hat{\gamma}^{(2)}_{j}\,
\mathsf{K}^{\,}_{\wedge}\df
\mathsf{K}^{\,}_{2}\,
\hat{\gamma}^{(2)}_{j}\,
\mathsf{K}^{\,}_{2},
\\
&
\mathsf{K}^{\,}_{\wedge}\,
\hat{\gamma}^{(1)}_{n^{\,}_{1}}\,
\mathsf{K}^{\,}_{\wedge}\df
+\hat{\gamma}^{(1)}_{n^{\,}_{1}},
\\
&
\mathsf{K}^{\,}_{\wedge}\,
\hat{\gamma}^{(2)}_{n^{\,}_{2}}\,
\mathsf{K}^{\,}_{\wedge}\df
-
\hat{\gamma}^{(2)}_{n^{\,}_{2}},
\end{align}
\end{subequations}
for $i=1,\cdots,n^{\,}_{1}-1$ and
$j=1,\cdots,n^{\,}_{2}-1$.
One verifies that, by construction, the fermion parity operator
$\widehat{U}^{\,}_{\wedge}(p)$ is invariant under
conjugation by $\mathsf{K}^{\,}_{\wedge}$.

For the  stacked representation with $[\mu^{\,}_{\wedge}]=1$ that is achieved
with an even-odd stacking, we define the central element
$\widehat{Y}^{\,}_{\wedge}$ by
\begin{align}
\widehat{Y}^{\,}_{\wedge}\df
\widehat{U}^{\,}_{1}(p)\,
\widehat{Y}^{\,}_{2},
\label{eq:def central element Ywedge}
\end{align}
where $\widehat{Y}^{\,}_{2}$ is the central element inherited
from the representation $\widehat{U}^{\,}_{2}$ which by assumption has
$[\mu^{\,}_{2}]=1$
for the case of even-odd stacking.

In what follows, we give explicit representations of
$\widehat{U}^{\,}_{\wedge}(g)$
in terms of the pair
$\widehat{U}^{\,}_{1}(g)$
and
$\widehat{U}^{\,}_{2}(g)$
and of
$([(\nu^{\,}_{\wedge},\,\rho^{\,}_{\wedge})],[\mu^{\,}_{\wedge}])$
in terms of the pairs
$([(\nu^{\,}_{1},\,\rho^{\,}_{1})],[\mu^{\,}_{1}])$ and
$([(\nu^{\,}_{2},\,\rho^{\,}_{2})],[\mu^{\,}_{2}])$.

\subsection{Even-even stacking}
\label{subsec:ee stacking}

For even-even stacking, we have $[\mu^{\,}_{1}]=[\mu^{\,}_{2}]=0$.
We define
\begin{equation}
[\mu^{\,}_{\wedge}]\df[\mu^{\,}_{1}]+[\mu^{\,}_{2}]=0.
\end{equation}
The representations $\widehat{U}^{\,}_{1}$ and $\widehat{U}^{\,}_{2}$ 
of the group $G^{\,}_{f}$ are of the form
\eqref{eq:boundary rep gen mu0}, i.e.,
for any $g\in G^{\,}_{f}$,
\begin{equation}
\widehat{U}^{\,}_{1}(g)=
\widehat{V}^{\,}_{1}(g)\,
\mathsf{K}^{\mathfrak{c}(g)}_{1},
\qquad
\widehat{U}^{\,}_{2}(g)=
\widehat{V}^{\,}_{2}(g)\,
\mathsf{K}^{\mathfrak{c}(g)}_{2},
\end{equation}
with the pair of unitary operators
$\widehat{V}^{\,}_{1}(g)$
and
$\widehat{V}^{\,}_{2}(g)$.
The naive guess
$\widehat{V}^{\,}_{1}(g)\,
\widehat{V}^{\,}_{2}(g)\,
\mathsf{K}^{\mathfrak{c}(g)}_{\wedge}$
is not a satisfactory definition of
$\widehat{U}^{\,}_{\wedge}(g)$,
for one verifies that it fails to satisfy
Eq.\ \eqref{eq:stacked consistency cond gen}. 
Instead,
for any $g\in G^{\,}_{f}$, we define
\begin{align}
\widehat{U}^{\,}_{\wedge}(g)\df
\widehat{V}^{\,}_{1}(g)\,
\widehat{V}^{\,}_{2}(g)\,
\left[\widehat{U}^{\,}_{1}(p)\right]^{\rho^{\,}_{2}(g)}\,
\left[\widehat{U}^{\,}_{2}(p)\right]^{\rho^{\,}_{1}(g)}\,
\mathsf{K}^{\mathfrak{c}(g)}_{\wedge}.
\label{eq:stacked gen ee}
\end{align}
One verifies that this definition
satisfies Eq.\
\eqref{eq:stacked consistency cond gen}
and, \textit{a forteriori}, Eq.\
\eqref{eq:consistency cond general}. 
The parity operators $\widehat{U}^{\,}_{1}(p)$ and
$\widehat{U}^{\,}_{2}(p)$ 
in definition \eqref{eq:stacked gen ee}
ensure that no additional minus
signs are introduced when Majorana operators 
$\hat{\gamma}^{(1)}_{i}$ and $\hat{\gamma}^{(2)}_{i}$
are conjugated  by $\widehat{U}^{\,}_{\wedge}(g)$.
This is because, by definition 
\eqref{eq:def rho gen mu0},
the values 
$\rho^{\,}_{1}(g)$
and
$\rho^{\,}_{2}(g)$
encode the fermion parity of the
unitary operators $\widehat{V}^{\,}_{1}(g)$ and
$\widehat{V}^{\,}_{2}(g)$, respectively,
and the parity operators 
$\widehat{U}^{\,}_{1}(p)$ and
$\widehat{U}^{\,}_{2}(p)$ correct
for any additional minus signs arising from 
fermionic algebra between operators from
representation $\widehat{U}^{\,}_{1}$ and
$\widehat{U}^{\,}_{2}$ in compliance with Eq.\
\eqref{eq:stacked consistency cond gen}.

As a sanity check, one verifies that 
when restricted to the center $\mathbb{Z}^{\mathrm{F}}_{2}
\subset G^{\,}_{f}$, the definition \eqref{eq:stacked gen ee} 
of the stacked representation 
together with definition
\eqref{eq:boundary rep parity gen mu0}
deliver the Hermitian representations
\begin{align}
\widehat{U}^{\,}_{\wedge}(e)
=
\hat{\mathbb{1}}^{\,}_{\wedge,0},
\quad
\widehat{U}^{\,}_{\wedge}(p)
=
\widehat{U}^{\,}_{1}(p)\,
\widehat{U}^{\,}_{2}(p),
\end{align}
that are consistent with the
definition \eqref{eq:def stacked parity rep}.

When the representations
$\widehat{U}^{\,}_{\wedge}(g)$
and
$\widehat{U}^{\,}_{\wedge}(h)$
of two elements $g$ and $h$
of $G^{\,}_{f}$
are composed, we obtain from definition (\ref{eq:stacked gen ee})
\begin{subequations}
\begin{align}
\widehat{U}^{\,}_{\wedge}(g)\,
\widehat{U}^{\,}_{\wedge}(h)
&=\,
\widehat{V}^{\,}_{1}(g)\,
\widehat{V}^{\,}_{2}(g)\,
\nonumber\\
&\times
\left[\widehat{U}^{\,}_{1}(p)\right]^{\rho^{\,}_{2}(g)}\,
\left[\widehat{U}^{\,}_{2}(p)\right]^{\rho^{\,}_{1}(g)}
\nonumber\\
&\times
\overline{\widehat{V}^{\,}_{1}(h)}^{\wedge,g}\,
\overline{\widehat{V}^{\,}_{2}(h)}^{\wedge,g}\,
\nonumber\\
&\times
\left[\widehat{U}^{\,}_{1}(p)\right]^{\rho^{\,}_{2}(h)}\,
\left[\widehat{U}^{\,}_{2}(p)\right]^{\rho^{\,}_{1}(h)}
\nonumber\\
&\times
\mathsf{K}^{\mathfrak{c}(g\,h)}_{\wedge},
\label{eq:stacking ee int step 1}
\end{align}
where we have introduced the notation
\begin{align}
&
\overline{\widehat{O}}^{1,g}\df
\mathsf{K}^{\mathfrak{c}(g)}_{1}\,
\widehat{O}\,
\mathsf{K}^{\mathfrak{c}(g)}_{1},
\\
&
\overline{\widehat{O}}^{2,g}\df
\mathsf{K}^{\mathfrak{c}(g)}_{2}\,
\widehat{O}\,
\mathsf{K}^{\mathfrak{c}(g)}_{2},
\\
&
\overline{\widehat{O}}^{\wedge,g}\df
\mathsf{K}^{\mathfrak{c}(g)}_{\wedge}\,
\widehat{O}\,
\mathsf{K}^{\mathfrak{c}(g)}_{\wedge},
\end{align}
\end{subequations}
for any operator $\widehat{O}$,
used the reality condition obeyed by
$\widehat{U}^{\,}_{1}(p)$
and
$\widehat{U}^{\,}_{2}(p)$,
and the fact that $\mathfrak{c}$ is a group 
homomorphism.
We rearrange the terms in Eq.\ 
\eqref{eq:stacking ee int step 1} to obtain
\begin{align}
\widehat{U}^{\,}_{\wedge}(g)\,
\widehat{U}^{\,}_{\wedge}(h)
=&\,
(-1)^{\rho^{\,}_{1}(g)\rho^{\,}_{2}(h)}\,
\nonumber\\
&\,\times
\widehat{V}^{\,}_{1}(g)\,
\overline{\widehat{V}^{\,}_{1}(h)}^{1,g}\,
\widehat{V}^{\,}_{2}(g)\,
\overline{\widehat{V}^{\,}_{2}(h)}^{2,g}\,
\nonumber\\
&\,\times
\left[\widehat{U}^{\,}_{1}(p)\right]^{\rho^{\,}_{2}(g\,h)}
\left[\widehat{U}^{\,}_{2}(p)\right]^{\rho^{\,}_{1}(g\,h)}
\mathsf{K}^{\mathfrak{c}(g\,h)}_{\wedge},
\label{eq:stacking ee int step 2}
\end{align}
where 
$(-1)^{\rho^{\,}_{1}(g)\rho^{\,}_{2}(h)}$
is the total multiplicative phase factor arising due to 
fermionic algebra. 
In reaching the last line, we have used definition 
\eqref{eq:stacked complex conj ee eo gen} to trade the complex conjugation
by $\mathsf{K}^{\,}_{\wedge}$ with complex conjugations by 
$\mathsf{K}^{\,}_{1}$ and $\mathsf{K}^{\,}_{2}$, and 
the fact that $\rho^{\,}_{1}$ and $\rho^{\,}_{2}$ are 
group homomorphisms.
To proceed, we observe that definition 
\eqref{eq:def proj rep on one component bd} implies
\begin{align}
\widehat{V}^{\,}_{i}(g)\,
\overline{\widehat{V}^{\,}_{i}(h)}^{i,g}
=&\,
e^{\mathrm{i}\phi^{\,}_{i}(g,h)}\,
\widehat{V}^{\,}_{i}(g\,h),
\qquad
i=1,2.
\label{eq: Vi(g) overline Vi(h) i,g}
\end{align}
Inserting these identities to Eq.\
\eqref{eq:stacking ee int step 2} delivers 
\begin{subequations}
\begin{align}
\widehat{U}^{\,}_{\wedge}(g)\,
\widehat{U}^{\,}_{\wedge}(h)
=&\,
e^{\mathrm{i}\phi^{\,}_{\wedge}(g,h)}\,
\widehat{U}^{\,}_{\wedge}(g\,h),
\end{align}
where we have defined
\begin{equation}
\phi^{\,}_{\wedge}(g,h)\df
\phi^{\,}_{1}(g,h)
+
\phi^{\,}_{2}(g,h)
+
\pi\,\rho^{\,}_{1}(g)\,\rho^{\,}_{2}(h).
\end{equation}
\end{subequations}
The construction of the indices
$([(\nu^{\,}_{\wedge},\rho^{\,}_{\wedge})],[\mu^{\,}_{\wedge}])$
in terms of the indices 
$([(\nu^{\,}_{1},\rho^{\,}_{1})],[\mu^{\,}_{1}])$ and
$([(\nu^{\,}_{2},\rho^{\,}_{2})],[\mu^{\,}_{2}])$
is achieved as follows.

According to definition \eqref{eq:def 2-cochain nu [mu]=0},
the 2-cochain $\nu^{\,}_{\wedge}$
is simply obtained by restricting $\phi^{\,}_{\wedge}$
to the elements of $G$, i.e.,
\begin{subequations}
\label{eq:ee stacking rel gen}
\begin{align}
\nu^{\,}_{\wedge}(g,h)=
\nu^{\,}_{1}(g,h)
+
\nu^{\,}_{2}(g,h)
+
\pi\left(\rho^{\,}_{1}\smile\rho^{\,}_{2}\right)(g,h).
\end{align}
In the last step we have used the cup product $\smile$ 
to construct a 2-cochain $\rho^{\,}_{1}\smile \rho^{\,}_{2}$
out of the pair of
one cochains $\rho^{\,}_{1}$ and $\rho^{\,}_{2}$.
For the 1-cochain $\rho^{\,}_{\wedge}$, definition
\eqref{eq:def rho gen mu0} delivers
\begin{align}
\rho^{\,}_{\wedge}(g)
=&\,
\rho^{\,}_{1}(g)
+
\rho^{\,}_{2}(g),
\end{align}
\end{subequations} 
which is nothing but the total fermion parity
of the stacked representation $\widehat{U}^{\,}_{\wedge}(g)$
of element $g\in G$.

\subsection{Even-odd stacking}
\label{subsec:eo stacking}

For even-odd stacking, we have $[\mu^{\,}_{1}]=0$, $[\mu^{\,}_{2}]=1$.
Hence, we define
\begin{align}
[\mu^{\,}_{\wedge}]\df[\mu^{\,}_{1}]+[\mu^{\,}_{2}]=1.
\end{align}
The representations $\widehat{U}^{\,}_{1}$ and $\widehat{U}^{\,}_{2}$
of the group $G^{\,}_{f}$ are of the form
\eqref{eq:boundary rep gen mu0}
and
\eqref{eq:boundary rep gen mu1}, respectively, 
i.e., for any $g\in G^{\,}_{f}$,
\begin{subequations}
\begin{align}
&
\widehat{U}^{\,}_{1}(g)=
\widehat{V}^{\,}_{1}(g)\,
\mathsf{K}^{\mathfrak{c}(g)}_{1},
\\
&
\widehat{U}^{\,}_{2}(g)=
\widehat{V}^{\,}_{2}(g)\,
\widehat{Q}^{\,}_{2}(g)\,
\mathsf{K}^{\mathfrak{c}(g)}_{2},
\\
&
\widehat{Q}^{\,}_{2}(g)=
\left[\hat{\gamma}^{(2)}_{\infty}\right]^{q^{\,}_{2}(g)}.
\end{align}
\end{subequations}
The naive guess
$\widehat{V}^{\,}_{1}(g)\,
\widehat{V}^{\,}_{2}(g)\,
\widehat{Q}^{\,}_{2}(g)\,
\mathsf{K}^{\mathfrak{c}(g)}_{\wedge}$
is not a satisfactory definition of
$\widehat{U}^{\,}_{\wedge}(g)$,
for one verifies that it fails to satisfy
Eq.\ \eqref{eq:stacked consistency cond gen}
and to be of even fermion parity. Instead,
we define the stacked representation to be 
\begin{subequations}
\label{eq:stacked gen eo}
\begin{align}
&
\widehat{U}^{\,}_{\wedge}(g)
\df
\widehat{V}^{\,}_{1}(g)\,
\widehat{V}^{\,}_{2}(g)\,
\widehat{Q}^{\,}_{2}(g)\,
\left[
\widehat{U}^{\,}_{1}(p)\,
\hat{\gamma}^{(2)}_{\infty}
\right]^{\rho^{\,}_{1}(g)}\,
\mathsf{K}^{\mathfrak{c}(g)}_{\wedge}
\nonumber\\
&
\qquad\,\,\,
\equiv
\widehat{V}^{\,}_{\wedge}(g)\,
\widehat{Q}^{\,}_{\wedge}(g)\,
\mathsf{K}^{\mathfrak{c}(g)}_{\wedge},
\\
&
\widehat{V}^{\,}_{\wedge}(g)
\df
\widehat{V}^{\,}_{1}(g)\,
\widehat{V}^{\,}_{2}(g)\,
\left[
\widehat{U}^{\,}_{1}(p)\,
\right]^{\rho^{\,}_{1}(g)},
\\
&
\widehat{Q}^{\,}_{\wedge}(g)
\df
\widehat{Q}^{\,}_{2}(g)\,
\left[
\hat{\gamma}^{(2)}_{\infty}
\right]^{\rho^{\,}_{1}(g)}
=
\left[
\hat{\gamma}^{(2)}_{\infty}
\right]^{q^{\,}_{2}(g)+\rho^{\,}_{1}(g)}.
\end{align}
\end{subequations}
One verifies that this definition
satisfies Eq.\
\eqref{eq:stacked consistency cond gen}
and, \textit{a forteriori}, Eq.\
\eqref{eq:consistency cond general}. 
For any $g\in G^{\,}_{f}$, the definition
(\ref{eq:stacked gen eo}) guarantees that  
$\widehat{U}^{\,}_{\wedge}(g)$ is of even fermion parity.
This property is inherited from the fact that
$\widehat{U}^{\,}_{2}(g)$
is of even fermion parity according to
Eq.\ \eqref{eq:boundary rep gen mu1}
and the factor
$\widehat{U}^{\,}_{1}(p)\,\hat{\gamma}^{(2)}_{\infty}$
compensates for the fermion parity of the operator
$\widehat{V}^{\,}_{1}(g)$. 
The product $\widehat{U}^{\,}_{1}(p)\,\hat{\gamma}^{(2)}_{\infty}$
also compensates for additional minus signs 
arising from fermionic algebra between the operators 
from representations $\widehat{U}^{\,}_{1}$ and 
$\widehat{U}^{\,}_{2}$ in compliance with Eq.\
\eqref{eq:stacked consistency cond gen}.

As a sanity check, one verifies that, 
when restricted to the center $\mathbb{Z}^{\mathrm{F}}_{2}
\subset G^{\,}_{f}$, the definition \eqref{eq:stacked gen eo} 
of the stacked representation 
together with definitions
\eqref{eq:boundary rep parity gen mu0}
and \eqref{eq:boundary rep parity gen mu1}
deliver the Hermitian representations
\begin{align}
\widehat{U}^{\,}_{\wedge}(e)
=
\hat{\mathbb{1}}^{\,}_{\wedge,1},
\quad
\widehat{U}^{\,}_{\wedge}(p)
=
\widehat{U}^{\,}_{1}(p)\,
\widehat{U}^{\,}_{2}(p),
\end{align}
that are  consistent with the
definition \eqref{eq:def stacked parity rep}.

When representations $\widehat{U}^{\,}_{\wedge}(g)$ and
$\widehat{U}^{\,}_{\wedge}(h)$ of two elements $g,h\in G^{\,}_{f}$ are
composed, we obtain from definition \eqref{eq:stacked gen eo}
\begin{align}
\widehat{U}^{\,}_{\wedge}(g)\,
\widehat{U}^{\,}_{\wedge}(h)
=&\,
\widehat{V}^{\,}_{1}(g)\,
\widehat{V}^{\,}_{2}(g)\,
\widehat{Q}^{\,}_{2}(g)\,
\left[
\widehat{U}^{\,}_{1}(p)\,
\hat{\gamma}^{(2)}_{\infty}
\right]^{\rho^{\,}_{1}(g)}
\nonumber\\
\times&
\overline{\widehat{V}^{\,}_{1}(h)}^{1,g}\,
\overline{\widehat{V}^{\,}_{2}(h)}^{2,g}\,
\overline{\widehat{Q}^{\,}_{2}(h)}^{2,g}
\nonumber\\
\times&
\left[
\widehat{U}^{\,}_{1}(p)\,
\overline{\hat{\gamma}^{(2)}_{\infty}}^{2,g}
\right]^{\rho^{\,}_{1}(h)}\,
\mathsf{K}^{\mathfrak{c}(g\,h)}_{\wedge},
\label{eq:stacking eo int step 1}
\end{align}
where we have traded the complex conjugation  
$\mathsf{K}^{\,}_{\wedge}$ by complex conjugations  
$\mathsf{K}^{\,}_{1}$ and $\mathsf{K}^{\,}_{2}$ through
Eq.\ \eqref{eq:stacked complex conj ee eo gen}.
We rearrange the terms in Eq.\ \eqref{eq:stacking eo int step 1}
to obtain
\begin{align}
\widehat{U}^{\,}_{\wedge}(g)\,
\widehat{U}^{\,}_{\wedge}(h)
=&\,
(-1)^{\rho^{\,}_{1}(g)\,q^{\,}_{2}(h)}\,
\widehat{V}^{\,}_{1}(g)\,
\overline{\widehat{V}^{\,}_{1}(h)}^{1,g}\,
\nonumber\\
&\,\times
\widehat{V}^{\,}_{2}(g)\,
\widehat{Q}^{\,}_{2}(g)\,
\overline{\widehat{V}^{\,}_{2}(h)}^{2,g}\,
\overline{\widehat{Q}^{\,}_{2}(h)}^{2,g}
\nonumber\\
&\,\times
\left[
\hat{\gamma}^{(2)}_{\infty}
\right]^{\rho^{\,}_{1}(g)}\,
\left[
\overline{\hat{\gamma}^{(2)}_{\infty}}^{2,g}
\right]^{\rho^{\,}_{1}(h)}
\nonumber\\
&\,\times
\left[
\widehat{U}^{\,}_{1}(p)
\right]^{\rho^{\,}_{1}(g\,h)}\,
\mathsf{K}^{\mathfrak{c}(g\,h)}_{\wedge}.
\label{eq:stacking eo int step 2}
\end{align}
Hereby, the multiplicative phase factor 
$(-1)^{\rho^{\,}_{1}(g)\,q^{\,}_{2}(h)}$
is induced by the fermionic algebra between 
$\hat{\gamma}^{(2)}_{\infty}$ and 
$\widehat{V}^{\,}_{2}(h)$ 
[recall that by definition \eqref{eq:boundary rep gen mu1}
$\widehat{V}^{\,}_{2}(h)$  has fermion 
parity $q^{\,}_{2}(h)$].
Using Eqs.\ \eqref{eq:def proj rep on one component bd}
and \eqref{eq:useful identity gen gamm inf conjugation a},
we obtain the identities
\begin{subequations}
\small
\begin{align}
&
\widehat{V}^{\,}_{1}(g)\,
\overline{\widehat{V}^{\,}_{1}(h)}^{1,g}
=
e^{\mathrm{i}\phi^{\,}_{1}(g,h)}
\widehat{V}^{\,}_{1}(g\,h),
\\
&
\widehat{V}^{\,}_{2}(g)\,
\widehat{Q}^{\,}_{2}(g)\,
\overline{\widehat{V}^{\,}_{2}(h)}^{2,g}\,
\overline{\widehat{Q}^{\,}_{2}(h)}^{2,g}
=
e^{\mathrm{i}\phi^{\,}_{2}(g,h)}
\widehat{V}^{\,}_{2}(g\,h)
\widehat{Q}^{\,}_{2}(g\,h),
\\
&
\left[
\overline{\hat{\gamma}^{(2)}_{\infty}}^{2,g}
\right]^{\rho^{\,}_{1}(h)}
=
(-1)^{\rho^{\,}_{1}(h)
[\mathfrak{c}(g)+q^{\,}_{2}(g)+\rho^{\,}_{2}(g)]}
\left[
\hat{\gamma}^{(2)}_{\infty}
\right]^{\rho^{\,}_{1}(h)}.
\end{align}
\normalsize
\end{subequations}
Inserting these identities to 
Eq.\ \eqref{eq:stacking eo int step 2},
one is left with
\begin{subequations}
\begin{align}
\widehat{U}^{\,}_{\wedge}(g)\,
\widehat{U}^{\,}_{\wedge}(h)
=\,
e^{\mathrm{i}\phi^{\,}_{\wedge}(g,h)}\,
\widehat{U}^{\,}_{\wedge}(g\,h),
\end{align}
where 
\begin{align}
\phi^{\,}_{\wedge}(g,h)
\df&
\phi^{\,}_{1}(g,h)+\phi^{\,}_{2}(g,h)
\nonumber\\
&
+\pi \rho^{\,}_{1}(g)\,q^{\,}_{2}(h)
\nonumber\\
&
+
\pi \rho^{\,}_{1}(h)
\left[\mathfrak{c}(g)+q^{\,}_{2}(g)+\rho^{\,}_{2}(g)\right].
\label{eq:phi gen stacking eo}
\end{align}
\end{subequations}
The projective phase \eqref{eq:phi gen stacking eo}
can be simplified by noting that terms
that contain the 1-cochain $q^{\,}_{2}$
can be gauged away under the transformation 
\eqref{eq:def gauge equiv bd}. 
More concretely, for any two $\mathbb{Z}^{\,}_{2}$
valued 1-cochains $\alpha,\beta\in 
C^{1}(G^{\,}_{f},\mathbb{Z}^{\,}_{2})$, the equivalence relations
\begin{align}
&
\alpha(g)\beta(h)
\sim 
\alpha(h)\beta(g),
\nonumber\\
&
\alpha(g)\,\beta(h)
+
\alpha(h)\,\beta(g)
\sim
0\text{ mod 2},
\label{eq:Z2 cochain gauge equiv}
\end{align}
hold. Therefore, the
2-cochain $\phi^{\,}_{\wedge}(g,h)$
defined in \eqref{eq:phi gen stacking eo}
is gauge equivalent to
\begin{align}
\phi^{\prime}_{\wedge}(g,h)
\df&
\phi^{\,}_{1}(g,h)+\phi^{\,}_{2}(g,h)
\nonumber\\
&
+
\pi \rho^{\,}_{1}(g)\,\rho^{\,}_{2}(h)
+
\pi \rho^{\,}_{1}(g)\,\mathfrak{c}(h),
\label{eq:phi gen stacking eo gauge equiv}
\end{align}
where in reaching the last line we have used the 
equivalence \eqref{eq:Z2 cochain gauge equiv}
to trade $\rho^{\,}_{1}(h)\,\rho^{\,}_{2}(g)$
and $\rho^{\,}_{1}(h)\,\mathfrak{c}(g)$ 
with $\rho^{\,}_{1}(g)\,\rho^{\,}_{2}(h)$
and $\rho^{\,}_{1}(g)\,\mathfrak{c}(h)$,
respectively. 

The construction of the indices
$([(\nu^{\,}_{\wedge},\rho^{\,}_{\wedge})],[\mu^{\,}_{\wedge}])$
in terms of the indices $([(\nu^{\,}_{1},\rho^{\,}_{1})],[\mu^{\,}_{1}])$ and
$([(\nu^{\,}_{2},\rho^{\,}_{2})],[\mu^{\,}_{2}])$
is achieved as follows.

According to definition \eqref{eq:def 2-cochain nu [mu]=1},
the 2-cochain $\nu^{\,}_{\wedge}$
is simply obtained by restricting $\phi^{\,}_{\wedge}$
to the elements of $G$, i.e.,
\begin{subequations}
\label{eq:eo stacking rel gen}
\begin{align}
&
\nu^{\,}_{\wedge}(g,h)
\df
\nu^{\,}_{1}(g,h)
+
\nu^{\,}_{2}(g,h)
\nonumber\\
&\qquad\quad
+
\pi\left(\rho^{\,}_{1}\smile\rho^{\,}_{2}\right)(g,h)
+
\pi\left(\rho^{\,}_{1}\smile\mathfrak{c}\right)(g,h),
\end{align}
where we introduced the cup product $\smile$ to construct
a 2-cochain out of 1-cochains.

Since the stacked representation has index $[\mu^{\,}_{\wedge}]=1$,
the 1-cochain $\rho^{\,}_{\wedge}$ can be either
determined by the definition \eqref{eq:def rho gen mu1} 
or by the identity
\eqref{eq:useful identity gen gamm inf conjugation a}. 
The definition \eqref{eq:stacked complex conj ee eo gen}
implies
\begin{align}
\overline{\hat{\gamma}^{(2)}_{\infty}}^{\wedge,g}\,
\hat{\gamma}^{(2)}_{\infty}
=
\overline{\hat{\gamma}^{(2)}_{\infty}}^{2,g}\,
\hat{\gamma}^{(2)}_{\infty}.
\end{align}
Using identity \eqref{eq:useful identity gen gamm inf conjugation a}
for the left and right hand sides separately, and comparing 
the two we find
\begin{align}
\rho^{\,}_{\wedge}(g)
&=
q^{\,}_{2}(g) + q^{\,}_{\wedge}(g) + \rho^{\,}_{2}(g)
\nonumber\\
&=
\rho^{\,}_{1}(g)
+ \rho^{\,}_{2}(g)
\hbox{ mod 2},
\end{align}
\end{subequations}
where the value of the 1-cochain 
$q^{\,}_{\wedge}(g) = 
\rho^{\,}_{1}(g) + q^{\,}_{2}(g)$ 
is read off from the fermion parity of 
the unitary operator $\widehat{V}^{\,}_{\wedge}(g)$
defined in Eq.\ \eqref{eq:stacked gen eo}.

\subsection{Odd-odd stacking}
\label{subsec:oo stacking}

For odd-odd stacking, we have $[\mu^{\,}_{1}]=[\mu^{\,}_{2}]=1$.
Hence,
we define
\begin{align}
[\mu^{\,}_{\wedge}]
\df
[\mu^{\,}_{1}]+[\mu^{\,}_{2}]=0.
\end{align}
The representations $\widehat{U}^{\,}_{1}$ and $\widehat{U}^{\,}_{2}$
of the group $G^{\,}_{f}$ are of the form
\eqref{eq:boundary rep gen mu1}, i.e., for any $g\in G^{\,}_{f}$,
\begin{subequations}
\begin{align}
&
\widehat{U}^{\,}_{1}(g)=
\widehat{V}^{\,}_{1}(g)\,
\widehat{Q}^{\,}_{1}(g)\,
\mathsf{K}^{\mathfrak{c}(g)}_{1},
\\
&
\widehat{Q}^{\,}_{1}(g)
=
\left[\hat{\gamma}^{(1)}_{\infty}\right]^{q^{\,}_{1}(g)},
\\
&
\widehat{U}^{\,}_{2}(g)=
\widehat{V}^{\,}_{2}(g)\,
\widehat{Q}^{\,}_{2}(g)\,
\mathsf{K}^{\mathfrak{c}(g)}_{2},
\\
&
\widehat{Q}^{\,}_{2}(g)
=
\left[\hat{\gamma}^{(2)}_{\infty}\right]^{q^{\,}_{2}(g)}.
\end{align}
\end{subequations}
The naive guess
$\widehat{V}^{\,}_{1}(g)\,
\widehat{Q}^{\,}_{1}(g)\,
\widehat{V}^{\,}_{2}(g)\,
\widehat{Q}^{\,}_{2}(g)\,
\mathsf{K}^{\mathfrak{c}(g)}_{\wedge}$
is not a satisfactory definition of
$\widehat{U}^{\,}_{\wedge}(g)$,
for one verifies that it fails to satisfy
Eq.\ \eqref{eq:stacked consistency cond gen}. 
Instead,
we define the stacked representation to be 
\begin{subequations}
\label{eq:stacked gen oo}
\begin{align}
&
\widehat{U}^{\,}_{\wedge}(g)
\df
(-\mathrm{i})^{\delta^{\,}_{g,p}}
\widehat{V}^{\,}_{1}(g)\,
\widehat{V}^{\,}_{2}(g)\,
\left[
\widehat{U}^{\,}_{\wedge}(p)
\right]^{\mathfrak{c}(g)+\rho^{\,}_{1}(g)+\rho^{\,}_{2}(g)}\,
\nonumber\\
&\qquad\qquad
\times
\left[
\hat{\gamma}^{(1)}_{n^{\,}_{1}}
\right]^{q^{\,}_{1}(g)+\rho^{\,}_{1}(g)}\,
\left[
\hat{\gamma}^{(2)}_{n^{\,}_{2}}
\right]^{\mathfrak{c}(g)+q^{\,}_{2}(g)+\rho^{\,}_{2}(g)}
\nonumber\\
&\qquad\qquad
\times
\mathsf{K}^{\mathfrak{c}(g)}_{\wedge},
\label{eq:stacked gen oo a}
\\
&
\widehat{U}^{\,}_{\wedge}(p)
\df
\widehat{P}^{\,}_{1}\,
\widehat{P}^{\,}_{2}\,
\mathrm{i}
\hat{\gamma}^{(1)}_{n^{\,}_{1}}\,
\hat{\gamma}^{(2)}_{n^{\,}_{2}},
\label{eq:stacked gen oo b}
\end{align}
\end{subequations}
where $\widehat{P}^{\,}_{1}$ and $\widehat{P}^{\,}_{2}$
are the fermion parity operators
constructed out of the Majorana operators 
$\hat{\gamma}^{(1)}_{1},\cdots,
\hat{\gamma}^{(1)}_{n^{\,}_{1}-1}$
and 
$\hat{\gamma}^{(2)}_{1},\cdots,
\hat{\gamma}^{(2)}_{n^{\,}_{2}-1}$,
respectively 
[recall definitions \eqref{eq:boundary rep parity gen mu1} 
and \eqref{eq:def stacked parity rep}].
The exponent $\delta^{\,}_{g,p}$ of the multiplicative phase
factor $(-\mathrm{i})^{\delta^{\,}_{g,p}}$ is the Kronecker
delta defined over the group $G^{\,}_{f}$.

As a sanity check, one verifies that, 
when restricted to the center $\mathbb{Z}^{\mathrm{F}}_{2}
\subset G^{\,}_{f}$, the definition \eqref{eq:stacked gen oo} 
of the stacked representation 
together with the definition
\eqref{eq:boundary rep parity gen mu1}
deliver the Hermitian representations
\begin{align}
\widehat{U}^{\,}_{\wedge}(e)
=
\hat{\mathbb{1}}^{\,}_{\wedge,1},
\quad
\widehat{U}^{\,}_{\wedge}(p)
=
\widehat{P}^{\,}_{1}\,
\widehat{P}^{\,}_{2}\,
\mathrm{i}
\hat{\gamma}^{(1)}_{n^{\,}_{1}}\,
\hat{\gamma}^{(2)}_{n^{\,}_{2}},
\end{align}
that are consistent with the
definition \eqref{eq:def stacked parity rep}.
The choice of the multiplicative phase factor 
$(-\mathrm{i})^{\delta^{\,}_{g,p}}$
in Eq.\ \eqref{eq:stacked gen oo} is not unique since 
representation $\widehat{U}(g)$ of any element $g\in G^{\,}_{f}$ 
is defined up to a multiplicative $U(1)$ phase. 
We observe that the multiplicative factor 
$(-\mathrm{i})^{\delta^{\,}_{g,p}}$
in Eq.\ \eqref{eq:stacked gen oo}
ensures that the stacked representation $\widehat{U}^{\,}_{\wedge}(p)$ is
Hermitian in compliance with the ``gauge'' choice made in definition 
\eqref{eq:boundary rep parity gen mu1}.

Several comments are due. 
First, one verifies that the definition 
(\ref{eq:stacked gen oo})
satisfies Eq.\
\eqref{eq:stacked consistency cond gen}
and, \textit{a forteriori}, Eq.\
\eqref{eq:consistency cond general}.
Second, 
the Majorana operators 
$\hat{\gamma}^{(1)}_{\infty}$ and $\hat{\gamma}^{(2)}_{\infty}$
do not enter the definition \eqref{eq:stacked gen oo} 
of the stacked representation $\widehat{U}^{\,}_{\wedge}$. 
This is expected as the stacked representation $\widehat{U}^{\,}_{\wedge}$
has $[\mu^{\,}_{\wedge}]=0$. Accordingly,
$\widehat{U}^{\,}_{\wedge}$ is constructed solely out of the even number
$n^{\,}_{1}+n^{\,}_{2}$ of 
Majorana operators spanning the fermionic Fock space of
the stacked boundary
[recall definition \eqref{eq:boundary rep gen mu0}].
Third, 
the definition
(\ref{eq:stacked gen oo})
is not symmetric under exchange of the labels 1 and 2,
as is to be expected by inspection of
Eq.\ \eqref{eq:stacked complex conj oo gen}.

Before computing the stacked
2-cochain $\phi^{\,}_{\wedge}$, we 
shall derive two useful identities 
that relate complex conjugation by $\mathsf{K}^{\,}_{\wedge}$
to complex conjugation by $\mathsf{K}^{\,}_{1}$
and $\mathsf{K}^{\,}_{2}$ for any pair $g,h\in G^{\,}_{f}$.
This is needed as definition \eqref{eq:stacked complex conj oo gen}
of $\mathsf{K}^{\,}_{\wedge}$ is not completely fixed
from the definitions of $\mathsf{K}^{\,}_{1}$ and $\mathsf{K}^{\,}_{2}$
as in the case of even-even and even-odd stacking, recall Eq.\ 
\eqref{eq:stacked complex conj ee eo gen}. 
The consistency conditions
\eqref{eq:consistency cond general}
and
\eqref{eq:stacked consistency cond gen}
imply the identities
\begin{subequations}
\label{eq:stacking oo wedge conj}
\begin{align}
&
\widehat{U}^{\,}_{\wedge}(g)\,
\widehat{V}^{\,}_{1}(h)\,
\widehat{U}^{\dag}_{\wedge}(g)
=
\widehat{U}^{\,}_{1}(g)\,
\widehat{V}^{\,}_{1}(h)\,
\widehat{U}^{\dag}_{1}(g),
\label{eq:stacking oo wedge conj a}
\\
&
\widehat{U}^{\,}_{\wedge}(g)\,
\widehat{V}^{\,}_{2}(h)\,
\widehat{U}^{\dag}_{\wedge}(g)
=
\widehat{U}^{\,}_{2}(g)\,
\widehat{V}^{\,}_{2}(h)\,
\widehat{U}^{\dag}_{2}(g),
\label{eq:stacking oo wedge conj b}
\end{align}
\end{subequations}
for any $g,h\in G^{\,}_{f}$.
If $g$ is to be represented antiunitarily,
then complex conjugation is denoted by
$\mathsf{K}^{\,}_{\wedge}$,
$\mathsf{K}^{\,}_{1}$,
and
$\mathsf{K}^{\,}_{2}$
for
$\widehat{U}^{\,}_{\wedge}(g)$,
$\widehat{U}^{\,}_{1}(g)$,
and
$\widehat{U}^{\,}_{2}(g)$,
respectively.
Comparing the two sides delivers the pair of identities
\begin{subequations}
\label{eq:stacking oo exchanging wedge conj}
\begin{align}
&
\overline{\widehat{V}^{\,}_{1}(h)}^{\wedge,g}=
(-1)^{q^{\,}_{1}(h)\big[\rho^{\,}_{1}(g)+q^{\,}_{1}(g)\big]}\,
\nonumber\\
&\qquad\qquad\times
\left[
\hat{\gamma}^{(1)}_{n^{\,}_{1}}
\right]^{q^{\,}_{1}(g)+\rho^{\,}_{1}(g)}\,
\overline{\widehat{V}^{\,}_{1}(h)}^{1,g}\,
\nonumber\\
&\qquad\qquad\times
\left[
\hat{\gamma}^{(1)}_{n^{\,}_{1}}
\right]^{q^{\,}_{1}(g)+\rho^{\,}_{1}(g)},
\label{eq:stacking oo exchanging wedge conj a}
\\
&
\overline{\widehat{V}^{\,}_{2}(h)}^{\wedge,g}
=
(-1)^{q^{\,}_{2}(h)\big[\mathfrak{c}(g)+\rho^{\,}_{2}(g)+q^{\,}_{2}(g)\big]}\,
\nonumber\\
&\qquad\qquad\times
\left[
\hat{\gamma}^{(2)}_{n^{\,}_{2}}
\right]^{\mathfrak{c}(g)+q^{\,}_{2}(g)+\rho^{\,}_{2}(g)}\,
\overline{\widehat{V}^{\,}_{2}(h)}^{2,g}\,
\nonumber\\
&\qquad\qquad\times
\left[
\hat{\gamma}^{(2)}_{n^{\,}_{2}}
\right]^{\mathfrak{c}(g)+q^{\,}_{2}(g)+\rho^{\,}_{2}(g)}.
\label{eq:stacking oo exchanging wedge conj b}
\end{align}
\end{subequations}
These pair of identities 
are not symmetric under exchange of 
the labels 1 and 2 as 
the definitions 
\eqref{eq:stacked complex conj oo gen}
and \eqref{eq:stacked complex conj oo gen} 
are not symmetric as well. 
The phase factors multiplying the right-hand side
are due to the fermionic algebra between the operators.

We are now ready to compute the 2-cochain $\phi^{\,}_{\wedge}(g,h)$
associated with the stacked representation $\widehat{U}^{\,}_{\wedge}$.
Composing the representations $\widehat{U}^{\,}_{\wedge}(g)$ and 
$\widehat{U}^{\,}_{\wedge}(h)$
of any pair $g,h\in G^{\,}_{f}$ delivers
\begin{widetext}
\small
\begin{align}
\widehat{U}^{\,}_{\wedge}(g)\,
\widehat{U}^{\,}_{\wedge}(h)
&=\,
(-\mathrm{i})^{
\delta^{\,}_{g,p}
+
(-1)^{\mathfrak{c}(g)}\delta^{\,}_{h,p}
              }
(-1)^{
\mathfrak{c}(g)\big[\mathfrak{c}(h)+q^{\,}_{2}(h)+\rho^{\,}_{2}(h)\big]
}
\widehat{V}^{\,}_{1}(g)\,
\widehat{V}^{\,}_{2}(g)\,
\left[
\widehat{U}^{\,}_{\wedge}(p)
\right]^{\mathfrak{c}(g)+\rho^{\,}_{1}(g)+\rho^{\,}_{2}(g)}
\left[
\hat{\gamma}^{(1)}_{n^{\,}_{1}}
\right]^{q^{\,}_{1}(g)+\rho^{\,}_{1}(g)}\,
\nonumber\\
&\times
\left[
\hat{\gamma}^{(2)}_{n^{\,}_{2}}
\right]^{\mathfrak{c}(g)+q^{\,}_{2}(g)+\rho^{\,}_{2}(g)}\,
\overline{\widehat{V}^{\,}_{1}(h)}^{\wedge,g}
\overline{\widehat{V}^{\,}_{2}(h)}^{\wedge,g}
\left[
\widehat{U}^{\,}_{\wedge}(p)
\right]^{\mathfrak{c}(h)+\rho^{\,}_{1}(h)+\rho^{\,}_{2}(h)}
\left[
\hat{\gamma}^{(1)}_{n^{\,}_{1}}
\right]^{q^{\,}_{1}(h)+\rho^{\,}_{1}(h)}\,
\left[
\hat{\gamma}^{(2)}_{n^{\,}_{2}}
\right]^{\mathfrak{c}(h)+q^{\,}_{2}(h)+\rho^{\,}_{2}(h)}\,
\mathsf{K}^{\mathfrak{c}(g\,h)}_{\wedge},
\label{eq:stacking oo inter step 1}
\end{align}
\normalsize
where the multiplier $(-1)^{\mathfrak{c}(g)}$ in the phase factor 
$(-\mathrm{i})^{
\delta^{\,}_{g,p}
+		
(-1)^{\mathfrak{c}(g)}\delta^{\,}_{h,p}}$ arises when 
the complex conjugation $\mathsf{K}^{\mathfrak{c}(g)}_{\wedge}$
is passed
through $(-\mathrm{i})^{\delta^{\,}_{h,p}}$.
The multiplicative phase factor 
$(-1)^{
\mathfrak{c}(g)\big[\mathfrak{c}(h)+q^{\,}_{2}(h)+\rho^{\,}_{2}(h)\big]}$
is due to complex conjugation of Majorana operators 
$\hat{\gamma}^{(1)}_{n^{\,}_{1}}$ and 
$\hat{\gamma}^{(2)}_{n^{\,}_{2}}$, through Eq.\ 
\eqref{eq:stacked complex conj oo gen}.

We rearrange the terms in Eq.\ \eqref{eq:stacking oo inter step 1}
to obtain
\begin{subequations}
\label{eq:stacking oo inter step final}
\begin{align}
\widehat{U}^{\,}_{\wedge}(g)\,
\widehat{U}^{\,}_{\wedge}(h)
&=
(-\mathrm{i})^{
\delta^{\,}_{g,p}
+
(-1)^{\mathfrak{c}(g)}\delta^{\,}_{h,p}}\,
(-1)^{\chi^{\,}_{1}(g,h)}\,
\widehat{V}^{\,}_{1}(g)\,
\left[
\hat{\gamma}^{(1)}_{n^{\,}_{1}}
\right]^{q^{\,}_{1}(g)+\rho^{\,}_{1}(g)}\,
\overline{\widehat{V}^{\,}_{1}(h)}^{\wedge,g}\,
\widehat{V}^{\,}_{2}(g)\,
\left[
\hat{\gamma}^{(2)}_{n^{\,}_{2}}
\right]^{\mathfrak{c}(g)+q^{\,}_{2}(g)+\rho^{\,}_{2}(g)}\,
\overline{\widehat{V}^{\,}_{2}(h)}^{\wedge,g}
\nonumber\\
&\times
\left[
\widehat{U}^{\,}_{\wedge}(p)
\right]^{\mathfrak{c}(g\,h)+\rho^{\,}_{1}(g\,h)+\rho^{\,}_{2}(g\,h)}\,
\left[
\hat{\gamma}^{(1)}_{n^{\,}_{1}}
\right]^{q^{\,}_{1}(h)+\rho^{\,}_{1}(h)}\,
\left[
\hat{\gamma}^{(2)}_{n^{\,}_{2}}
\right]^{\mathfrak{c}(h)+q^{\,}_{2}(h)+\rho^{\,}_{2}(h)}\,
\mathsf{K}^{\mathfrak{c}(g\,h)}_{\wedge},
\label{eq:stacking oo inter step final a}
\end{align}
where
\begin{align}
\chi^{\,}_{1}(g,h)
&\df\,
\mathfrak{c}(g)
\big[
1+\mathfrak{c}(h)+q^{\,}_{1}(g)+q^{\,}_{2}(g)+\rho^{\,}_{2}(h)
\big]
+
\rho^{\,}_{1}(g)
\big[
1+
q^{\,}_{1}(g)
+
q^{\,}_{1}(h)
+
q^{\,}_{2}(h)
\big]
\nonumber\\
&\,
+
\rho^{\,}_{2}(g)
\big[
1+
q^{\,}_{1}(g)
+
q^{\,}_{2}(g)
+
q^{\,}_{2}(h)
\big]
+
q^{\,}_{1}(g)
q^{\,}_{2}(g),
\label{eq:stacking oo inter step final b}
\end{align}
\end{subequations}
is the total multiple phase factor
that is due to the fermionic algebra when rearranging 
the terms together with the multiplicative phase factor
in Eq.\ \eqref{eq:stacking oo inter step 1}.

We use the identity \eqref{eq:stacking oo exchanging wedge conj}
in Eq.\ \eqref{eq:stacking oo inter step 1}
to obtain
\begin{subequations}
\label{eq:stacking oo before grp comp}
\small
\begin{align}
\widehat{U}^{\,}_{\wedge}(g)\,
\widehat{U}^{\,}_{\wedge}(h)
&=\,
(-1)^{\chi^{\,}_{1}(g,h)}\,
(-1)^{\chi^{\,}_{\mathrm{conj}}(g,h)}\,
(-\mathrm{i})^{
\delta^{\,}_{g,p}
+
(-1)^{\mathfrak{c}(g)}\delta^{\,}_{h,p}}\,
\widehat{V}^{\,}_{1}(g)\,
\overline{\widehat{V}^{\,}_{1}(h)}^{1,g}\,
\left[
\hat{\gamma}^{(1)}_{n^{\,}_{1}}
\right]^{q^{\,}_{1}(g)+\rho^{\,}_{1}(g)}
\widehat{V}^{\,}_{2}(g)\,
\overline{\widehat{V}^{\,}_{2}(h)}^{2,g}\,
\left[
\hat{\gamma}^{(2)}_{n^{\,}_{2}}
\right]^{\mathfrak{c}(g)+q^{\,}_{2}(g)+\rho^{\,}_{2}(g)}
\nonumber\\
&\times
\left[
\widehat{U}^{\,}_{\wedge}(p)
\right]^{\mathfrak{c}(g\,h)+\rho^{\,}_{1}(g\,h)+\rho^{\,}_{2}(g\,h)}\,
\left[
\hat{\gamma}^{(1)}_{n^{\,}_{1}}
\right]^{q^{\,}_{1}(h)+\rho^{\,}_{1}(h)}\,
\left[
\hat{\gamma}^{(2)}_{n^{\,}_{2}}
\right]^{\mathfrak{c}(h)+q^{\,}_{2}(h)+\rho^{\,}_{2}(h)}\,
\mathsf{K}^{\mathfrak{c}(g\,h)}_{\wedge},
\end{align}
\normalsize
where we have consolidated the two multiplicative phase factors
on the right-hand sides of
Eqs.\
(\ref{eq:stacking oo exchanging wedge conj a})
and
(\ref{eq:stacking oo exchanging wedge conj b})
into the multiplicative phase factor
\begin{align}
\chi^{\,}_{\mathrm{conj}}(g,h)
&\df
q^{\,}_{1}(h)[\rho^{\,}_{1}(g)+q^{\,}_{1}(g)]
+
q^{\,}_{2}(h)[\mathfrak{c}(g)+\rho^{\,}_{2}(g)+q^{\,}_{2}(g)].
\label{eq:stacking oo chi conj}
\end{align}
\end{subequations}
To proceed, we observe 
that the definitions \eqref{eq:def proj rep on one component bd}
and \eqref{eq:boundary rep gen mu1} can be cast into 
\begin{align}
\widehat{V}^{\,}_{i}(g)
\overline{\widehat{V}^{\,}_{i}(h)}^{i,g}
=
e^{\mathrm{i}\phi^{\,}_{i}(g,h)
+
\mathrm{i}\pi
q^{\,}_{i}(h)[\mathfrak{c}(g)+\rho^{\,}_{i}(g)]
}\,
\widehat{V}^{\,}_{i}(g\,h),
\label{eq:stacking oo unitary composition rule}
\end{align}
for $i=1,2$.
Inserting Eq.\
\eqref{eq:stacking oo unitary composition rule} into 
Eq.\ \eqref{eq:stacking oo before grp comp} delivers
\begin{subequations}
\label{eq:stacking oo before final ordering}
\begin{align}
\widehat{U}^{\,}_{\wedge}(g)\,
\widehat{U}^{\,}_{\wedge}(h)
&=\,
e^{
\mathrm{i}\phi^{\,}_{\mathrm{comp}}(g,h)
+
\mathrm{i}\pi\chi^{\,}_{1}(g,h)
+
\mathrm{i}\pi\chi^{\,}_{\mathrm{conj}}(g,h)
+
\mathrm{i}\frac{3\pi}{2}
\left(
\delta^{\,}_{g,p}
+
(-1)^{\mathfrak{c}(g)}\delta^{\,}_{h,p}
\right)
}
\widehat{V}^{\,}_{1}(g\,h)\,
\left[
\hat{\gamma}^{(1)}_{n^{\,}_{1}}
\right]^{q^{\,}_{1}(g)+\rho^{\,}_{1}(g)}\,
\widehat{V}^{\,}_{2}(g\,h)\,
\nonumber\\
&\times
\left[
\hat{\gamma}^{(2)}_{n^{\,}_{2}}
\right]^{\mathfrak{c}(g)+q^{\,}_{2}(g)+\rho^{\,}_{2}(g)}\,
\left[
\widehat{U}^{\,}_{\wedge}(p)
\right]^{\mathfrak{c}(g\,h)+\rho^{\,}_{1}(g\,h)+\rho^{\,}_{2}(g\,h)}\,
\left[
\hat{\gamma}^{(1)}_{n^{\,}_{1}}
\right]^{q^{\,}_{1}(h)+\rho^{\,}_{1}(h)}\,
\left[
\hat{\gamma}^{(2)}_{n^{\,}_{2}}
\right]^{\mathfrak{c}(h)+q^{\,}_{2}(h)+\rho^{\,}_{2}(h)}\,
\mathsf{K}^{\mathfrak{c}(g\,h)}_{\wedge},
\label{eq:stacking oo before final ordering a}
\end{align}
where we have defined the phase factor accumulated from the 
group composition rule \eqref{eq:stacking oo unitary composition rule}
\begin{align}
\phi^{\,}_{\mathrm{comp}}(g,h)
&\df
\phi^{\,}_{1}(g,h)
+
\phi^{\,}_{2}(g,h)
+
\pi\,
q^{\,}_{1}(h)[\mathfrak{c}(g)+\rho^{\,}_{1}(g)]
+
\pi\,
q^{\,}_{2}(h)[\mathfrak{c}(g)+\rho^{\,}_{2}(g)].
\label{eq:stacking oo before final ordering b}
\end{align}
\end{subequations}
It remains to reorder operators
on the right-hand side of Eq.\
(\ref{eq:stacking oo before final ordering a})
with the goal to isolate the operator
$\widehat{U}^{\,}_{\wedge}(g\,h)$,
whose definition is given by
Eq.\
\eqref{eq:stacked gen oo}.
Doing so, one finds 
\begin{subequations}
\begin{align}
\widehat{U}^{\,}_{\wedge}(g)\,
\widehat{U}^{\,}_{\wedge}(h)
&=
e^{
\mathrm{i}\phi^{\,}_{\mathrm{comp}}(g,h)
+
\mathrm{i}\pi\chi^{\,}_{1}(g,h)
+
\mathrm{i}\pi\chi^{\,}_{\mathrm{conj}}(g,h)
+
\mathrm{i}\pi\chi^{\,}_{\mathrm{ord}}(g,h)
+
\mathrm{i}
\chi^{\,}_{\mathrm{gag}}(g,h)
}
\nonumber\\
&\,\times
(-\mathrm{i})^{\delta^{\,}_{gh,p}}\,
\widehat{V}^{\,}_{1}(g\,h)\,
\widehat{V}^{\,}_{2}(g\,h)\,
\left[
\widehat{U}^{\,}_{\wedge}(p)
\right]^{\mathfrak{c}(g\,h)+\rho^{\,}_{1}(g\,h)+\rho^{\,}_{2}(g\,h)}
\left[
\hat{\gamma}^{(1)}_{n^{\,}_{1}}
\right]^{q^{\,}_{1}(g\,h)+\rho^{\,}_{1}(g\,h)}\,
\left[
\hat{\gamma}^{(2)}_{n^{\,}_{2}}
\right]^{\mathfrak{c}(g\,h)+q^{\,}_{2}(g\,h)+\rho^{\,}_{2}(g\,h)}\,
\mathsf{K}^{\mathfrak{c}(g\,h)}_{\wedge},
\nonumber\\
&\equiv
e^{\mathrm{i}\phi^{\,}_{\wedge}(g,h)}\,
\widehat{U}^{\,}_{\wedge}(g\,h),
\label{eq: final composition rule o-o case a}
\end{align}
where we have defined the phase factors
\begin{align}
&
\chi^{\,}_{\mathrm{ord}}(g,h)
\df
[
\mathfrak{c}(g\,h)+
\rho^{\,}_{1}(g)+
\rho^{\,}_{2}(g\,h)+
q^{\,}_{1}(h)
]
[\mathfrak{c}(g)+q^{\,}_{2}(g)+\rho^{\,}_{2}(g)]
\nonumber\\
&\qquad\qquad\qquad
+
[
\mathfrak{c}(g\,h)+
\rho^{\,}_{1}(g\,h)+
\rho^{\,}_{2}(g\,h)+
q^{\,}_{2}(g\,h)
]
[q^{\,}_{1}(g)+\rho^{\,}_{1}(g)],
\label{eq:stacking oo chi ord}
\\
&
\chi^{\,}_{\mathrm{gag}}(g,h)
\df
\frac{3\pi}{2}
\left(
\delta^{\,}_{g,p}
+
(-1)^{\mathfrak{c}(g)}\delta^{\,}_{h,p}
-
\delta^{\,}_{gh,p}
\right).
\\
&
\phi^{\,}_{\wedge}(g,h)
\df\,
\phi^{\,}_{\mathrm{comp}}(g,h)
+
\pi\,\chi^{\,}_{1}(g,h)
+
\pi\,\chi^{\,}_{\mathrm{conj}}(g,h)
+
\pi\,\chi^{\,}_{\mathrm{ord}}(g,h)
+
\chi^{\,}_{\mathrm{gag}}(g,h),
\label{eq: final composition rule o-o case b}
\end{align}
\end{subequations}
and used the definition
\eqref{eq:stacked gen oo} 
for $\widehat{U}^{\,}_{\wedge}(g\,h)$.
The phase factor $\chi^{\,}_{\mathrm{gag}}(g,h)$
that appears
in Eq.\ \eqref{eq: final composition rule o-o case b}
is an artifact of 
the particular gauge choice we have made 
when defining an Hermitian representation
for the fermion parity operator in Eq.\ 
\eqref{eq:boundary rep parity gen mu0}. 
Indeed, we observe that $\chi^{\,}_{\mathrm{gag}}(g,h)$
is nothing but a pure gauge under the gauge transformation
\eqref{eq:def gauge equiv bd a}, i.e.,
the right-hand side of Eq.\ \eqref{eq:def gauge equiv bd c}
is identical to $\chi^{\,}_{\mathrm{gag}}(g,h)$ if we choose
$\xi(g) = -3\pi\delta^{\,}_{g,h}/2$.
Under such a gauge transformation, the representation 
$\widehat{U}^{\,}_{\wedge}(p)$ of fermion parity $p$ is
no longer Hermitian. 
However, by definition, the equivalence classes
$[\phi^{\,}_{\wedge}]$ of 
the stacked 2-cochain $\phi^{\,}_{\wedge}(g,h)$
are invariant under the gauge transformations
\eqref{eq:def gauge equiv bd a}. 
Therefore, the stacked 2-cochain $\phi^{\,}_{\wedge}(g,h)$
is gauge equivalent to 
\begin{align}
\phi^{\,}_{\wedge}(g,h)
\sim 
\phi^{\,}_{1}(g,h)
+
\phi^{\,}_{2}(g,h)
+
\pi\,\chi(g,h).
\end{align}
We have reserved
the phase $\chi(g,h)$ for all phases other than the 
2-cochains $\phi^{\,}_{1}(g,h)$, $\phi^{\,}_{2}(g,h)$, and 
$\chi^{\,}_{\mathrm{gag}}(g,h)$
in Eq.\ (\ref{eq: final composition rule o-o case b}), i.e.,
\begin{align}
\chi(g,h)
&\df\,
\frac{1}{\pi}
\left(
\phi^{\,}_{\mathrm{comp}}(g,h)
-
\phi^{\,}_{1}(g,h)
-
\phi^{\,}_{2}(g,h)
\right)
+
\chi^{\,}_{1}(g,h)
+
\chi^{\,}_{\mathrm{conj}}(g,h)
+
\chi^{\,}_{\mathrm{ord}}(g,h)
\nonumber\\
&=\,
\rho^{\,}_{1}(g)
\big[
q^{\,}_{1}(h)
+
\mathfrak{c}(h)
+
\rho^{\,}_{1}(h)
+
\rho^{\,}_{2}(h)
\big]
+
\rho^{\,}_{2}(g)
\big[
q^{\,}_{1}(h)
+
q^{\,}_{2}(h)
+
\mathfrak{c}(h)
+
\rho^{\,}_{2}(h)
\big]
\nonumber\\
&\quad
+
q^{\,}_{1}(h)\,q^{\,}_{1}(g)
+
q^{\,}_{2}(h)\,q^{\,}_{2}(g)
+
q^{\,}_{2}(g)
\big[
\mathfrak{c}(g)
+
\rho^{\,}_{2}(h)
+
q^{\,}_{1}(h)
\big]
+
q^{\,}_{1}(g)
\big[
\mathfrak{c}(h)
+
\rho^{\,}_{1}(h)
+
\rho^{\,}_{2}(h)
+
q^{\,}_{2}(h)
\big].
\label{eq: final composition rule o-o case c}
\end{align}
\end{widetext}

In defining $\chi(g,h)$ we have simplified all
the contributions
from Eqs.\ \eqref{eq:stacking oo inter step final b},
\eqref{eq:stacking oo chi conj},
\eqref{eq:stacking oo before final ordering b},
and \eqref{eq:stacking oo chi ord}
by using the facts that 
1-cochains $\mathfrak{c}$, $\rho^{\,}_{i}$, and $q^{\,}_{i}$ for
$i=1,2$ are all group homomorphisms and $\chi(g,h)$
is only defined modulo 2.
Further simplifications 
in Eq.\ \eqref{eq: final composition rule o-o case c}
can be made by using the 
gauge equivalence \eqref{eq:Z2 cochain gauge equiv}
of products of $\mathbb{Z}^{\,}_{2}$-valued 1-cochains.
Consequently,
one is left with
\begin{subequations}
\begin{align}
&
\chi(g,h)
\sim\,
\rho^{\,}_{1}(g)\,
\rho^{\,}_{2}(h)
+
\sum_{i=1}^{2}
\varphi^{\,}_{i}(g,h),
\label{eq:oo chi after gag equiv}
\\
&
\varphi^{\,}_{i}(g,h)
\df
\rho^{\,}_{i}(g)
[
\mathfrak{c}(h)
+
\rho^{\,}_{i}(h)
]
+
q^{\,}_{i}(g)\,
[
\mathfrak{c}(h)
+
q^{\,}_{i}(h)
],
\label{eq:oo stacking phase term after gag equiv}
\end{align}
\end{subequations}
where $\varphi^{\,}_{i}(g,h)$ is a 2-cochain 
for $i=1,2$. 

Finally, we will show that $\varphi^{\,}_{i}$ defined by
Eq.\ \eqref{eq:oo stacking phase term after gag equiv}
vanishes for $i=1,2$. 
To this end, we define
\begin{subequations}
\begin{align}
\overline{\hat{\gamma}^{(i)}_{\infty}}^{i}
=
(-1)^{\zeta^{\,}_{i}}\,
\hat{\gamma}^{(i)}_{\infty},
\end{align}
for $\zeta^{\,}_{i},
=
0,1$ and $i=1,2$
which, together with 
Eq.\ \eqref{eq:useful identity gen gamm inf conjugation}, 
deliver the following two identities
\begin{align}
&
q^{\,}_{i}(g)
=
\mathfrak{c}(g)(1+\zeta^{\,}_{i})+\rho^{\,}_{i}(g)
\text{ mod }2,
\label{eq:oo stacking gen key identities}
\end{align}
\end{subequations}
for any $g\in G^{\,}_{f}$ and $i=1,2$.
Inserting identity \eqref{eq:oo stacking gen key identities}
to Eq.\ \eqref{eq:oo stacking phase term after gag equiv}
delivers the desired result
\begin{align}
\varphi^{\,}_{i}(g,h)
=
\mathfrak{c}(g)\,
\mathfrak{c}(h)\,
\zeta^{\,}_{i}
\big(
1
+
\zeta^{\,}_{i}
\big)=
0
\text{ mod 2}.
\end{align}
We therefore obtained the stacked 2-cochain $\phi^{\,}_{\wedge}(g,h)$
\begin{align}
\phi^{\,}_{\wedge}(g,h)
\df
\phi^{\,}_{1}(g,h)
+
\phi^{\,}_{2}(g,h)
+
\pi\,\rho^{\,}_{1}(g)\,\rho^{\,}_{2}(h).
\end{align}

The construction of the indices
$([(\nu^{\,}_{\wedge},\rho^{\,}_{\wedge})],[\mu^{\,}_{\wedge}])$
in terms of the indices $([(\nu^{\,}_{1},\rho^{\,}_{1})],[\mu^{\,}_{1}])$ and
$([(\nu^{\,}_{2},\rho^{\,}_{2})],[\mu^{\,}_{2}])$
is achieved as follows.

The 2-cochain $\nu^{\,}_{\wedge}$
is obtained by restricting $\phi^{\,}_{\wedge}$
to the elements of $G$, i.e.,
\begin{subequations}
\label{eq:oo stacking rel gen}
\begin{align}
\nu^{\,}_{\wedge}(g,h)
\df
\nu^{\,}_{1}(g,h)
+
\nu^{\,}_{2}(g,h)
+
\pi\left(\rho^{\,}_{1}\smile\rho^{\,}_{2}\right)(g,h),
\end{align}
where we introduced the cup product $\smile$ to construct
a 2-cochain out of 1-cochains.

Since $[\mu^{\,}_{\wedge}]=0$, we
identify the 1-cochain $\rho^{\,}_{\wedge}(g)$ as
the total fermion parity of the representation of element
$g\in G^{\,}_{f}$
[recall definition \eqref{eq:def rho gen mu0}]. 
From the definition \eqref{eq:stacked gen oo}, 
we thus find
\begin{align}
\rho^{\,}_{\wedge}(g)
=
\rho^{\,}_{1}(g)
+
\rho^{\,}_{2}(g)
+
\mathfrak{c}(g),
\end{align}
where the first two terms originate from
$\widehat{V}^{\,}_{1}(g)$ and $\widehat{V}^{\,}_{2}(g)$,
the next two terms originate from
$\hat{\gamma}^{(1)}_{n^{\,}_{1}}$,
and the last three terms originate from
$\hat{\gamma}^{(2)}_{n^{\,}_{2}}$. 
\end{subequations}

\subsection{Summary of the fermionic stacking rules}

In Secs.\ \ref{subsec:ee stacking}, \ref{subsec:eo stacking},
and \ref{subsec:oo stacking}, 
we have explicitly constructed the stacked representation
$\widehat{U}^{\,}_{\wedge}$ given two representations $\widehat{U}^{\,}_{1}$
and $\widehat{U}^{\,}_{2}$ in 
Eqs.\ \eqref{eq:stacked gen ee}, \eqref{eq:stacked gen eo},
and \eqref{eq:stacked gen oo}. 
This was achieved by defining for any $g\in G^{\,}_{f}$
\begin{widetext}
\small
\begin{align}
\label{eq:stacked gen summ}
\widehat{U}^{\,}_{\wedge}(g)
\df
\begin{cases}
\widehat{V}^{\,}_{1}(g)\,
\widehat{V}^{\,}_{2}(g)\,
\left[\widehat{U}^{\,}_{1}(p)\right]^{\rho^{\,}_{2}(g)}\,
\left[\widehat{U}^{\,}_{2}(p)\right]^{\rho^{\,}_{1}(g)}\,
\mathsf{K}^{\mathfrak{c}(g)}_{\wedge},
&\text{ if $[\mu^{\,}_{1}]=[\mu^{\,}_{2}]=0$,}
\\
\widehat{V}^{\,}_{1}(g)\,
\widehat{V}^{\,}_{2}(g)\,
\widehat{Q}^{\,}_{2}(g)\,
\left[
\widehat{U}^{\,}_{1}(p)\,
\hat{\gamma}^{(2)}_{\infty}
\right]^{\rho^{\,}_{1}(g)}\,
\mathsf{K}^{\mathfrak{c}(g)}_{\wedge},
&\text{ if $[\mu^{\,}_{1}]=0$, $[\mu^{\,}_{2}]=1$,}
\\
(-\mathrm{i})^{\delta^{\,}_{g,p}}
\widehat{V}^{\,}_{1}(g)
\widehat{V}^{\,}_{2}(g)
\left[
\widehat{U}^{\,}_{\wedge}(p)
\right]^{\mathfrak{c}(g)+\rho^{\,}_{1}(g)+\rho^{\,}_{2}(g)}
\left[
\hat{\gamma}^{(1)}_{n^{\,}_{1}}
\right]^{q^{\,}_{1}(g)+\rho^{\,}_{1}(g)}
\left[
\hat{\gamma}^{(2)}_{n^{\,}_{2}}
\right]^{\mathfrak{c}(g)+q^{\,}_{2}(g)+\rho^{\,}_{2}(g)}
\mathsf{K}^{\mathfrak{c}(g)}_{\wedge},
&\text{ if $[\mu^{\,}_{1}]=[\mu^{\,}_{2}]=1$,}
\end{cases}
\end{align}
\normalsize
and deriving Eqs.\
\eqref{eq:ee stacking rel gen},
\eqref{eq:eo stacking rel gen},
and 
\eqref{eq:oo stacking rel gen}
by comparing
$\widehat{U}^{\,}_{\wedge}(g)\,
\widehat{U}^{\,}_{\wedge}(h)$
to
$\widehat{U}^{\,}_{\wedge}(g\,h)$
for any pair $g,h\in G^{\,}_{f}$.
We collect these equations
into the fermionic stacking rules 
of one-dimensional IFT phases
\begin{subequations}
\label{eq:stacking rules gen summ}
\begin{align}
&
\left(
[(\nu^{\,}_{1},\,\rho^{\,}_{1})],\,
0
\right)
\wedge
\left(
[(\nu^{\,}_{2},\,\rho^{\,}_{2})],\,
0
\right)
=
\left(
[
(\nu^{\,}_{1}+\nu^{\,}_{2}+\pi\left(\rho^{\,}_{1}\smile\rho^{\,}_{2}\right),\,
\rho^{\,}_{1}
+
\rho^{\,}_{2})],\,
0
\right),
\label{eq:stacking rules gen summ a}
\\
&
\left(
[(\nu^{\,}_{1},\,\rho^{\,}_{1})],\,
0
\right)
\wedge
\left(
[(\nu^{\,}_{2},\,\rho^{\,}_{2})],\,
1
\right)
=
\left(
[
(\nu^{\,}_{1}
+\nu^{\,}_{2}
+\pi\left(\rho^{\,}_{1}\smile\rho^{\,}_{2}
+ \rho^{\,}_{1}\smile\mathfrak{c}
\right),\,
\rho^{\,}_{1}
+
\rho^{\,}_{2})],\,
1
\right),
\label{eq:stacking rules gen summ b}
\\
&
\left(
[(\nu^{\,}_{1},\,\rho^{\,}_{1})],\,
1
\right)
\wedge
\left(
[(\nu^{\,}_{2},\,\rho^{\,}_{2})],\,
0
\right)
=
\left(
[
(\nu^{\,}_{1}
+\nu^{\,}_{2}
+\pi\left(\rho^{\,}_{1}\smile\rho^{\,}_{2}
+ \rho^{\,}_{2}\smile\mathfrak{c}
\right),\,
\rho^{\,}_{1}
+
\rho^{\,}_{2})],\,
1
\right),
\label{eq:stacking rules gen summ c}
\\
&
\left(
[(\nu^{\,}_{1},\,\rho^{\,}_{1})],\,
1
\right)
\wedge
\left(
[
(\nu^{\,}_{2},\,\rho^{\,}_{2})],\,
1
\right)
=
\left(
[
(
\nu^{\,}_{1}
+
\nu^{\,}_{2}
+
\pi\,\rho^{\,}_{1}
\smile
\rho^{\,}_{2},
\rho^{\,}_{1}
+
\rho^{\,}_{2}
+
\mathfrak{c})],\,
0
\right).
\label{eq:stacking rules gen summ d}
\end{align}
\end{subequations}
They correspond to the even-even,
even-odd, odd-even, and odd-odd stacking, respectively.
The stacking rules \eqref{eq:stacking rules gen summ} 
agree with the ones derived in Refs.\ \onlinecite{Turzillo2019} and
\onlinecite{Bourne2021}. We note that 
the even-odd stacking rule derived
in Ref.\ \onlinecite{Turzillo2019} contains
the term $\rho^{\,}_{1}\smile\rho^{\,}_{1}$
instead of the term $\rho^{\,}_{1}\smile\mathfrak{c}$. 
These two terms are gauge equivalent to each other under the transformation 
\eqref{eq:def gauge equiv bd c} with 
$\xi=\pi\,\rho^{\,}_{1}\smile\mathfrak{c}
-\frac{\pi}{2}\,\rho^{\,}_{1}\smile\rho^{\,}_{1}$. 
The presentation in Eq.\ \eqref{eq:stacking rules gen summ}
makes the role of antiunitary symmetries in the stacking rules
explicit. If the group $G^{\,}_{f}$ consist of 
only unitary symmetries, i.e., 
$\mathfrak{c}(g)=0$
for any $g\in G^{\,}_{f}$, the stacking rules \eqref{eq:stacking rules gen summ}
reduce to
\begin{align}
\left(
[(\nu^{\,}_{1},\,\rho^{\,}_{1})],\,
[\mu^{\,}_{1}]
\right)
\wedge
\left(
[(\nu^{\,}_{2},\,\rho^{\,}_{2})],\,
[\mu^{\,}_{2}]
\right)
=
\left(
[
(\nu^{\,}_{1}+\nu^{\,}_{2}+\pi\left(\rho^{\,}_{1}\smile\rho^{\,}_{2}\right),\,
\rho^{\,}_{1}
+
\rho^{\,}_{2})],\,
[\mu^{\,}_{1}] + [\mu^{\,}_{2}]
\right).
\label{eq:stacking rules gen simplifed for unitary}
\end{align}
\end{widetext}

The stacking rules
\eqref{eq:stacking rules gen summ}
dictate the group structure of
IFT phases that are symmetric under the group $G^{\,}_{f}$.
This group structure encodes the physical operation by which two
open chains realizing IFT phases that are symmetric under group $G^{\,}_{f}$
are brought adiabatically into contact so as to realize
an IFT phases that is symmetric under group
$G^{\,}_{f}$.
The stacking rules
\eqref{eq:stacking rules gen summ a}
and
\eqref{eq:stacking rules gen summ d}
each encodes how the left and right boundaries of
an open chain realizing an
IFT phase that is symmetric under the group $G^{\,}_{f}$
are glued back together in such a way that the resulting chain obeying
periodic boundary conditions supports
a nondegenerate gapped ground state. 

\section{Bulk Representations derived from stacking rules}
\label{sec:bulk rep from stacking}

Invertible fermionic topological phases of matter
in one-dimensional space have an internal
symmetry group $G^{\,}_{f}$ that is represented in the bulk
by the faithful representation
$\widehat{U}^{\,}_{\mathrm{bulk}}$
given in Eq.\ (\ref{eq:def faithful rep on bulk}).
Because these symmetries are internal,
they induce for any site $j$ of any
one-dimensional lattice $\Lambda$
a faithful representation
$\widehat{U}^{\,}_{j}$.
However, representatives of IFT phases
can also accomodate projective representations of
the internal symmetry group $G^{\,}_{f}$
on the left and right boundaries of $\Lambda$
provided the stacking of these two boundary representations
is gauge-equivalent to a faithful representation of $G^{\,}_{f}$,
as is captured by Fig.\ \ref{Fig: cutting open a ring in two ways}.

Instead of deducing the existence of local projective representations for
the internal symmetry group $G^{\,}_{f}$ from a faithful bulk representation
$\widehat{U}^{\,}_{\mathrm{bulk}}$,
we are going to construct a bulk representation
$\widehat{U}^{\,}_{\mathrm{bulk}}$ of the symmetry group $G^{\,}_{f}$ 
out of a given set of projective representations
$\widehat{U}^{\,}_{\bm{\bm{j}}}$
acting on the Clifford algebra 
\begin{align}
\mathrm{C}\ell^{\,}_{n^{\,}_{\bm{j}}}\df
\mathrm{span}\,
\left\{
\hat{\gamma}^{(\bm{j})}_{1},\,
\hat{\gamma}^{(\bm{j})}_{2},\,
\cdots,
\hat{\gamma}^{(\bm{j})}_{n^{\,}_{\bm{j}}}
\right\}
\label{eq:set of majoranas reconstr}
\end{align}
spanned by
$n^{\,}_{\bm{j}}$ Majorana degrees of freedom
for any site $\bm{\bm{j}}$ from a $d$-dimensional lattice $\Lambda$
provided
\begin{align}
\label{eq:even total majorana reconstr}
\sum_{j\in \Lambda}
n^{\,}_{j}=
0\hbox{ mod 2}.
\end{align}

To this end, we use the fact that the definition \eqref{eq:stacked gen summ}
and the stacking rules
(\ref{eq:stacking rules gen summ})
are associative%
~\footnote{
Stacking three representations ${\tiny\widehat{U}^{\,}_{1}}$, 
${\tiny\widehat{U}^{\,}_{2}}$,
and ${\tiny\widehat{U}^{\,}_{3}}$ using the definition 
\protect{\eqref{eq:stacked gen summ}
          }
is associative, i.e., it is independent of which two of the three representations
are first stacked.
This associativity 
follows from the consistency condition 
\protect{\eqref{eq:stacked consistency cond gen}}.
This is because, for a Clifford algebra ${\tiny\mathrm{C}\ell^{\,}_{2n}}$
with an even number of generators,
specifying the transformation rules on its generators together with the action 
of the complex conjugation uniquely (up to a phase factor) determines the representation 
${\tiny\widehat{U}(g)}$ of any element ${\tiny g\in G^{\,}_{f}}$. 
For a Clifford algebra ${\tiny\mathrm{C}\ell^{\,}_{2n+1}}$ with an odd number of 
generators, this is no longer true since ${\tiny\mathrm{C}\ell^{\,}_{2n+1}}$ has a
two-dimensional center spanned by $\hat{\mathbb{1}}$ and ${\tiny\widehat{Y}}$. 
We removed the ambiguity consisting in multiplying ${\tiny\widehat{U}(g)}$ by
the central element ${\tiny\widehat{Y}}$
by demanding that ${\tiny\widehat{U}(g)}$ is of even fermion parity.}.
If so, for any $g\in G^{\,}_{f}$ and any labeling
$\bm{j}^{\,}_{1}$,
$\bm{j}^{\,}_{2}$,
$\cdots$,
$\bm{j}^{\,}_{|\Lambda|}$
with $|\Lambda|$ the cardinality of $\Lambda$,
we can define $\widehat{U}^{\,}_{\mathrm{bulk}}(g)$
by stacking
$\widehat{U}^{\,}_{\bm{j}^{\,}_{1}}(g)$
with
$\widehat{U}^{\,}_{\bm{j}^{\,}_{2}}(g)$,
which we then stack with
$\widehat{U}^{\,}_{\bm{j}^{\,}_{3}}(g)$,
and so on. By construction, it follows that
\begin{align}
\widehat{U}^{\,}_{\bm{j}}(g)\,
\hat{\gamma}^{(\bm{j})}_{\iota}\,
\widehat{U}^{\dagger}_{\bm{j}}(g)
=
\widehat{U}^{\,}_{\mathrm{bulk}}(g)\,
\hat{\gamma}^{(\bm{j})}_{\iota}\,
\widehat{U}^{\dagger}_{\mathrm{bulk}}(g),
\label{eq: counterpart to consistency cond general} 
\end{align}
for any $\iota=1,\cdots,n^{\,}_{\bm{j}}$, $\bm{j}\in\Lambda$,
and $g\in G^{\,}_{f}$. Equation
(\ref{eq: counterpart to consistency cond general})
is the counterpart to the
consistency condition \eqref{eq:consistency cond general} 
that we used to construct boundary representations.
It also follows that the representation
\begin{align}
\widehat{U}^{\,}_{\mathrm{bulk}}(g)=
\left[
\prod_{\bm{j}\in\Lambda}
\widehat{V}^{\,}_{\bm{j}}(g)
\right]\,
\mathsf{K}^{\mathfrak{c}(g)},
\label{eq:naive bulk rep reconstr}
\end{align}
for any $g\in G^{\,}_{f}$
holds if and only if
the local representation $\widehat{U}^{\,}_{\bm{j}}(g)$
has the indices $\rho^{\,}_{\bm{j}}(g)=0$ and $[\mu^{\,}_{\bm{j}}]=0$
for any $\bm{j}\in\Lambda$.
This implies that the local representation $\widehat{U}^{\,}_{\bm{j}}(g)$
of any element $g\in G^{\,}_{f}$ is of even fermion parity and the number of
Majorana degrees of freedom $n^{\,}_{\bm{j}}$ is an even integer for any site
$\bm{j}\in\Lambda$.
It is then appropriate to call a local representation $\widehat{U}^{\,}_{\bm{j}}$
that has nontrivial indices $\rho^{\,}_{\bm{j}}$ and $[\mu^{\,}_{\bm{j}}]$
an \textit{intrinsically fermionic} representation.
In other words, the decomposition \eqref{eq:naive bulk rep reconstr}
is possible if and only if the local representation $\widehat{U}^{\,}_{\bm{j}}$
for any site $\bm{j}\in\Lambda$ is not intrinsically fermionic. In particular,
if all local degrees of freedom are bosonic,
then the decomposition \eqref{eq:naive bulk rep reconstr} is always valid. 
However, instead of the decomposition
\eqref{eq:naive bulk rep reconstr},
$\widehat{U}^{\,}_{\mathrm{bulk}}(g)$
is obtained in all generality by iterating
Eq.\ (\ref{eq:stacked gen summ})
for any $g\in G^{\,}_{f}$.

\section{Ground-state degeneracies}
\label{sec: Ground-state degeneracies}

In Secs.\
\ref{sec:boundary rep gen},
\ref{sec: Boundary projective representation of Gf},
and
\ref{sec: Definition of indices}
we have shown that the distinct IFT phases are characterized by
the projective character of the
boundary representation $\widehat{U}^{\,}_{\mathrm{B}}$.
In turn this projective character is
captured by the triplet of indices
$([(\nu,\rho)],[\mu])$. Let us now consider the implications of this
triplet being nontrivial, i.e.,
$([(\nu,\rho)],[\mu])\neq([(0,0)],[0])$,
for the spectral degeneracy of the boundary states.

The foremost consequence of the nontrivial indices $([(\nu,\rho)],[\mu])$
is the robustness of the boundary degeneracy
that is protected by 
a combination of the symmetry group $G^{\,}_{f}$
being represented projectively
and the existence of a nonlocal boundary Fock space, denoted
$\mathfrak{F}^{\,}_{\mathrm{LR}}$
in Eq.\ (\ref{eq:def FLR}),
whenever opposite boundaries host
odd numbers of Majorana degrees of freedom.  

Recently, a robust quantum mechanical supersymmetry
~\cite{Witten1982}
was shown in
Refs.\ \onlinecite{Prakash2021,Turzillo2021,Behrends2020}
to be generically present in nontrivial IFT phases.
We are going to recast these results by showing how
the quantum mechanical supersymmetry
present at the boundaries can be deduced from the
indices $([(\nu,\rho)],[\mu])$.

In what follows, we consider the two cases $[\mu]=0$ and $[\mu]=1$ 
separately. For each case, we first discuss the degeneracies associated with 
nontrivial pair $[(\nu,\rho)]$ on general grounds.

\subsection{The case of $[\mu]=0$}
\label{subsec: The case [mu]=0} 

When $[\mu]=0$, there always are even numbers
of Majorana degrees of freedom 
localized  on each disconnected component
$\Lambda^{\,}_{\mathrm{L}}$
and
$\Lambda^{\,}_{\mathrm{R}}$
of the boundary $\Lambda^{\,}_{\mathrm{bd}}$
[recall definition \eqref{eq:def Lambda bd}].
In this case, the boundary Fock space
$\mathfrak{F}^{\,}_{\Lambda^{\,}_{\mathrm{bd}}}$ spanned by the Majorana
degrees of freedom supported on $\Lambda^{\,}_{\mathrm{bd}}$
decomposes as
\begin{align}
\mathfrak{F}^{\,}_{\Lambda^{\,}_{\mathrm{bd}}}
=
\mathfrak{F}^{\,}_{\Lambda^{\,}_{\mathrm{L}}}
\otimes^{\,}_{\mathfrak{g}}
\mathfrak{F}^{\,}_{\Lambda^{\,}_{\mathrm{R}}},
\label{eq:Fock space decomp mu0}
\end{align}
where $\otimes^{\,}_{\mathfrak{g}}$ denotes a $\mathbb{Z}^{\,}_{2}$
graded tensor product, while $\mathfrak{F}^{\,}_{\Lambda^{\,}_{\mathrm{L}}}$
and $\mathfrak{F}^{\,}_{\Lambda^{\,}_{\mathrm{R}}}$ are the Fock spaces spanned
by the Majorana degrees of freedom localized at the disconnected components
$\Lambda^{\,}_{\mathrm{L}}$ and $\Lambda^{\,}_{\mathrm{R}}$, respectively.
The Fock spaces $\mathfrak{F}^{\,}_{\Lambda^{\,}_{\mathrm{L}}}$
and $\mathfrak{F}^{\,}_{\Lambda^{\,}_{\mathrm{R}}}$ are defined by Eq.\
\eqref{eq:def F[mu]=0}.
We denote with
$\widehat{H}^{\,}_{\mathrm{L}}$
and
$\widehat{H}^{\,}_{\mathrm{R}}$
the Hamiltonians
that act on Fock spaces $\mathfrak{F}^{\,}_{\Lambda^{\,}_{\mathrm{L}}}$
and $\mathfrak{F}^{\,}_{\Lambda^{\,}_{\mathrm{R}}}$
and govern the dynamics of the local
Majorana degrees of freedom localized at boundaries 
$\Lambda^{\,}_{\mathrm{L}}$ and $\Lambda^{\,}_{\mathrm{R}}$, respectively.
By assumption, the Hamiltonians $\widehat{H}^{\,}_{\mathrm{L}}$ and
$\widehat{H}^{\,}_{\mathrm{R}}$
are invariant under the representations (possibly projective)
$\widehat{U}^{\,}_{\mathrm{L}}$ and 
$\widehat{U}^{\,}_{\mathrm{R}}$ of the
given symmetry group $G^{\,}_{f}$,
respectively.

Since $[\mu]=0$, the only nontrivial IFT phases are those with 
nontrivial equivalence classes $[(\nu,\rho)]\neq [(0,0)]$, i.e., the
FSPT phases.
By definition, 
the indices $([\nu^{\,}_{\mathrm{L}},\rho^{\,}_{\mathrm{L}}],0)$
and $([\nu^{\,}_{\mathrm{R}},\rho^{\,}_{\mathrm{R}}],0)$
associated with the representations (possibly projective)
$\widehat{U}^{\,}_{\mathrm{L}}$ and $\widehat{U}^{\,}_{\mathrm{R}}$,
respectively, satisfy
\begin{align}
([(\nu^{\,}_{\mathrm{L}},\rho^{\,}_{\mathrm{L}})],0)
\wedge
([(\nu^{\,}_{\mathrm{R}},\rho^{\,}_{\mathrm{R}})],0)
=
([(0,0)],0)
\end{align}
under the stacking rule \eqref{eq:stacking rules gen summ a}.

If we focus on a single boundary (denoted by B), 
the equivalence class $[(\nu^{\,}_{\mathrm{B}},\rho^{\,}_{\mathrm{B}})]$
characterizes the nontrivial projective nature of the boundary representation
$\widehat{U}^{\,}_{\mathrm{B}}$.
Whenever $[(\nu^{\,}_{\mathrm{B}},\rho^{\,}_{\mathrm{B}})]\neq [(0,0)]$,
it is guaranteed that there is
no state that is invariant under the action of $\widehat{U}^{\,}_{\mathrm{B}}(g)$
for all $g\in G^{\,}_{f}$. In other words, there is no state in the Fock space 
$\mathfrak{F}^{\,}_{\Lambda^{\,}_{\mathrm{B}}}$ that transforms as a singlet under 
the representation $\widehat{U}^{\,}_{\mathrm{B}}$.
Any eigenenergy
of a $G^{\,}_{f}$-symmetric boundary Hamiltonian
$\widehat{H}^{\,}_{\mathrm{B}}$
must be degenerate. The degeneracy is protected by the particular 
representation $\widehat{U}^{\,}_{\mathrm{B}}$ of the symmetry group $G^{\,}_{f}$ 
and cannot be lifted without breaking the $G^{\,}_{f}$ symmetry.
The minimal degeneracy that is protected
by the $G^{\,}_{f}$ symmetry depends 
on the explicit structure
of the group $G^{\,}_{f}$
and the equivalence class
$[(\nu^{\,}_{\mathrm{B}},\rho^{\,}_{\mathrm{B}})]$
of the boundary representation
$\widehat{U}^{\,}_{\mathrm{B}}$.

Since for $[\mu]=0$, the boundary representations $\widehat{U}^{\,}_{\mathrm{L}}$ 
and $\widehat{U}^{\,}_{\mathrm{R}}$ act on two independent Fock spaces 
$\mathfrak{F}^{\,}_{\Lambda^{\,}_{\mathrm{L}}}$ and
$\mathfrak{F}^{\,}_{\Lambda^{\,}_{\mathrm{R}}}$,
the total protected ground-state degeneracy
$\mathrm{GSD}^{[\mu]=0}_{\mathrm{bd}}$ when open boundary conditions are
imposed is nothing but the product of the 
protected ground-state degeneracies
$\mathrm{GSD}^{[\mu]=0}_{\mathrm{L}}$ and $\mathrm{GSD}^{[\mu]=0}_{\mathrm{R}}$
of the Hamiltonians
$\widehat{H}^{\,}_{\mathrm{L}}$ 
and $\widehat{H}^{\,}_{\mathrm{R}}$, respectively, i.e.,
\begin{align}
\mathrm{GSD}^{[\mu]=0}_{\mathrm{bd}}
=
\mathrm{GSD}^{[\mu]=0}_{\mathrm{L}}
\times
\mathrm{GSD}^{[\mu]=0}_{\mathrm{R}}.
\label{eq:total degeneracy gen mu0}
\end{align} 

When $[\mu]=0$, the 1-cochain $\rho^{\,}_{\mathrm{B}}(g)=0,1$ 
encodes the commutation relation between the representations 
$\widehat{U}^{\,}_{\mathrm{B}}(g)$ of group element $g\in G^{\,}_{f}$
and $\widehat{U}^{\,}_{\mathrm{B}}(p)$ of fermion parity $p\in G^{\,}_{f}$.
A nonzero second entry in the equivalence class 
$[(\nu^{\,}_{\mathrm{B}},\rho^{\,}_{\mathrm{B}})]$ implies that there exists
at least one group element $g\in G^{\,}_{f}$ with $\rho^{\,}_{\mathrm{B}}(g)=1$,
i.e., the operator $\widehat{U}^{\,}_{\mathrm{B}}(g)$ is of odd fermion parity. 
If this is so, the boundary Hamiltonian $\widehat{H}^{\,}_{\mathrm{B}}$
must possess an emergent quantum mechanical supersymmetry. 
The supercharges associated with the boundary supersymmetry 
are constructed following Ref.\ ~\onlinecite{Behrends2020}.  
Assume without loss of
generality that all energy eigenvalues $\varepsilon^{\,}_{\alpha}$ of a
boundary Hamiltonian $\widehat{H}^{\,}_{\mathrm{B}}$ are shifted to
the positive energies, i.e., $\varepsilon^{\,}_{\alpha}>0$.  
Also assume that there exists a group element $g\in G^{\,}_{f}$
with $\rho^{\,}_{\mathrm{B}}(g)=1$.
For any orthonormal eigenstate $\ket{\psi^{\,}_{\alpha}}$
of $\widehat{H}^{\,}_{\mathrm{B}}$ with energy
$\varepsilon^{\,}_{\alpha}$, the state
\begin{subequations}
\label{eq:def supercharges mu0}
\begin{align}
\ket{\psi^{'}_{\alpha}}
\df
\widehat{U}^{\,}_{\mathrm{B}}(g)\,
\ket{\psi^{\,}_{\alpha}},
\label{eq:def supercharges mu0 a}
\end{align}
is also an orthonormal eigenstate of $\widehat{H}^{\,}_{\mathrm{B}}$
with the same energy but opposite fermion parity. Since the fermion parities of 
$\ket{\psi^{'}_{\alpha}}$ and $\ket{\psi^{\,}_{\alpha}}$ are different, they 
are orthogonal. Two supercharges can then be defined as
\begin{align}
&
\widehat{Q}^{\,}_{1}
\df
\sum_{\alpha}
\sqrt{\varepsilon^{\,}_{\alpha}}
\left[
\left(
\widehat{U}^{\,}_{\mathrm{B}}(g)\,
\ket{\psi^{\,}_{\alpha}}
\right)
\bra{\psi^{\,}_{\alpha}}
+
\ket{\psi^{\,}_{\alpha}}
\left(
\bra{\psi^{\,}_{\alpha}}\,
\widehat{U}^{\dag}_{\mathrm{B}}(g)
\right)
\right],
\label{eq:def supercharges mu0 b}
\\
&
\widehat{Q}^{\,}_{2}
\df
\sum_{\alpha}
\mathrm{i}
\sqrt{\varepsilon^{\,}_{\alpha}}
\left[
\left(
\widehat{U}^{\,}_{\mathrm{B}}(g)\,
\ket{\psi^{\,}_{\alpha}}
\right)
\bra{\psi^{\,}_{\alpha}}
-
\ket{\psi^{\,}_{\alpha}}
\left(
\bra{\psi^{\,}_{\alpha}}\,
\widehat{U}^{\dag}_{\mathrm{B}}(g)
\right)
\right].
\label{eq:def supercharges mu0 c}
\end{align}
Operators $\widehat{Q}^{\,}_{1}$ and $\widehat{Q}^{\,}_{2}$ are Hermitian, 
carry odd fermion parity, and satisfy the defining properties
\begin{align}
\left\{\widehat{Q}^{\,}_{i},\widehat{Q}^{\,}_{j}\right\}
=
2
\widehat{H}^{\,}_{\mathrm{B}}\,
\delta^{\,}_{i,j},
\qquad
\left[
\widehat{Q}^{\,}_{i},
\widehat{H}^{\,}_{\mathrm{B}}
\right]
=
0,
\qquad
i,j=1,2,
\label{eq:def supercharges mu0 d}
\end{align}
\end{subequations}
of fermionic supercharges. The precise 
number of supercharges on the boundary $\Lambda^{\,}_{\mathrm{B}}$
depends on the pair $[(\nu^{\,}_{\mathrm{B}},\rho^{\,}_{\mathrm{B}})]$
that characterizes
the number of symmetry operators $\widehat{U}^{\,}_{\mathrm{B}}(g)$ 
that carry odd fermion parity and their mutual algebra.

\subsection{The case of $[\mu]=1$}
\label{subsec: The case [mu]=1} 

When $[\mu]=1$, there are odd number of Majorana degrees of freedom 
localized on each disconnected component $\Lambda^{\,}_{\mathrm{L}}$ and
$\Lambda^{\,}_{\mathrm{R}}$
of the boundary $\Lambda^{\,}_{\mathrm{bd}}$
[recall definition \eqref{eq:def Lambda bd}].
In this case, the boundary Fock space $\mathfrak{F}^{\,}_{\Lambda^{\,}_{\mathrm{bd}}}$ 
spanned by Majorana degrees of freedom supported on $\Lambda^{\,}_{\mathrm{bd}}$
decomposes as
\begin{align}
\mathfrak{F}^{\,}_{\Lambda^{\,}_{\mathrm{bd}}}
=
\mathfrak{F}^{\,}_{\Lambda^{\,}_{\mathrm{L}}}
\otimes^{\,}_{\mathfrak{g}}
\mathfrak{F}^{\,}_{\Lambda^{\,}_{\mathrm{LR}}}
\otimes^{\,}_{\mathfrak{g}}
\mathfrak{F}^{\,}_{\Lambda^{\,}_{\mathrm{R}}},
\label{eq:Fock space decomp mu1}
\end{align}
where $\otimes^{\,}_{\mathfrak{g}}$ denotes a $\mathbb{Z}^{\,}_{2}$
graded tensor product. The Fock spaces
$\mathfrak{F}^{\,}_{\Lambda^{\,}_{\mathrm{B}}}$ with
B = L, R is spanned by all the Majorana operators localized 
at the disconnected components $\Lambda^{\,}_{\mathrm{B}}$ except one.
The two-dimensional Fock space $\mathfrak{F}^{\,}_{\Lambda^{\,}_{\mathrm{LR}}}$
is spanned by the two remaining Majorana operators with one localized
on the left boundary $\Lambda^{\,}_{\mathrm{L}}$
and the other localized on the right boundary $\Lambda^{\,}_{\mathrm{R}}$
of the open chain. Correspondingly, 
the pair of fermionic creation and annihilation operators that span
$\mathfrak{F}^{\,}_{\Lambda^{\,}_{\mathrm{LR}}}$
are nonlocal in the sense that they are formed by Majorana operators supported
on  opposite boundaries.
One can define Hamiltonians $\widehat{H}^{\,}_{\mathrm{L}}$ and
$\widehat{H}^{\,}_{\mathrm{R}}$
that are constructed out of Majorana operators localized at the boundaries 
$\Lambda^{\,}_{\mathrm{L}}$ and $\Lambda^{\,}_{\mathrm{R}}$. If so, the Hamiltonians 
$\widehat{H}^{\,}_{\mathrm{L}}$ and $\widehat{H}^{\,}_{\mathrm{R}}$
act on Fock spaces 
\begin{subequations}
\begin{align}
\mathfrak{F}^{\,}_{\Lambda^{\,}_{\mathrm{L}}}
\otimes^{\,}_{\mathfrak{g}}
\mathfrak{F}^{\,}_{\Lambda^{\,}_{\mathrm{LR}}},
\end{align}
and 
\begin{align}
\mathfrak{F}^{\,}_{\Lambda^{\,}_{\mathrm{R}}}
\otimes^{\,}_{\mathfrak{g}}
\mathfrak{F}^{\,}_{\Lambda^{\,}_{\mathrm{LR}}},
\end{align}
\end{subequations}
respectively.
By assumption, the Hamiltonians $\widehat{H}^{\,}_{\mathrm{L}}$ and
$\widehat{H}^{\,}_{\mathrm{R}}$
are invariant under the representations $\widehat{U}^{\,}_{\mathrm{L}}$ and 
$\widehat{U}^{\,}_{\mathrm{R}}$ of a given symmetry group $G^{\,}_{f}$, respectively.

On each boundary $\Lambda^{\,}_{\mathrm{B}}$,
there exists a local Hermitean and unitary operator
$\widehat{Y}^{\,}_{\mathrm{B}}$
that commutes with any other local operator supported on
$\Lambda^{\,}_{\mathrm{B}}$.
The operator $\widehat{Y}^{\,}_{\mathrm{B}}$ is defined by Eq.\
\eqref{eq:def re YB for mu1}
and is the representation of the
nontrivial central element of a Clifford algebra
$\mathrm{C}\ell^{\,}_{n}$ with $n$ an
odd number of generators.
It therefore carries an odd fermion parity and anticommutes
with the representation $\widehat{U}^{\,}_{\mathrm{B}}(p)$ 
of fermion parity.
It follows that $\widehat{Y}^{\,}_{\mathrm{B}}$ must commute with
$\widehat{H}^{\,}_{\mathrm{B}}$.
We label the simultaneous eigenstates of $\widehat{H}^{\,}_{\mathrm{B}}$
and $\widehat{Y}^{\,}_{\mathrm{B}}$ by $\ket{\psi^{\,}_{\mathrm{B},\alpha,\pm}}$, i.e.,
\begin{align}
\widehat{Y}^{\,}_{\mathrm{B}}\,
\ket{\psi^{\,}_{\mathrm{B},\alpha,\pm}}
=
\pm 
\ket{\psi^{\,}_{\mathrm{B},\alpha,\pm}},
\quad
\widehat{H}^{\,}_{\mathrm{B}}\,
\ket{\psi^{\,}_{\mathrm{B},\alpha,\pm}}
=
\varepsilon^{\,}_{\alpha}\,
\ket{\psi^{\,}_{\mathrm{B},\alpha,\pm}},
\end{align}
where $\varepsilon^{\,}_{\alpha}$ is the corresponding energy eigenvalue
which we assume without loss of generality to be strictly positive.
Hence, all eigenstates of $\widehat{H}^{\,}_{\mathrm{B}}$ 
are at least twofold degenerate.
Since $\widehat{Y}^{\,}_{\mathrm{B}}$ carries odd fermion parity,
the eigenstates 
$\ket{\psi^{\,}_{\mathrm{B},\alpha,\pm}}$ do not have definite 
fermion parities. The simultaneous eigenstates of $\widehat{H}^{\,}_{\mathrm{B}}$
and $\widehat{U}^{\,}_{\mathrm{B}}(p)$ must be the bonding and anti-bonding
linear combinations of
$\ket{\psi^{\,}_{\mathrm{B},\alpha,+}}$
and
$\ket{\psi^{\,}_{\mathrm{B},\alpha,-}}$
that are exchanged
under the action of $\widehat{Y}^{\,}_{\mathrm{B}}$.
The twofold degeneracy of $\widehat{H}^{\,}_{\mathrm{B}}$ when $[\mu]=1$
is due to the presence of the two-dimensional Fock space
$\mathfrak{F}^{\,}_{\mathrm{LR}}$.
This twofold degeneracy is of supersymmetric nature
and the associated supercharges are
\begin{subequations}
\label{eq:def supercharges mu1}
\begin{align}
&
\widehat{Q}^{\,}_{1}
\df
\sum_{\alpha}
\sqrt{\varepsilon^{\,}_{\alpha}}\,
\left(
\,
\ket{\psi^{\,}_{\alpha,+}}
\bra{\psi^{\,}_{\alpha,+}}
-
\ket{\psi^{\,}_{\alpha,-}}
\bra{\psi^{\,}_{\alpha,-}}\,
\right),
\label{eq:def supercharges mu1 a}
\\
&
\widehat{Q}^{\,}_{2}
\df
\sum_{\alpha}
\mathrm{i}
\sqrt{\varepsilon^{\,}_{\alpha}}\,
\left(
\,
\ket{\psi^{\,}_{\alpha,+}}
\bra{\psi^{\,}_{\alpha,-}}
-
\ket{\psi^{\,}_{\alpha,-}}
\bra{\psi^{\,}_{\alpha,+}}\,
\right).
\label{eq:def supercharges mu1 b}
\end{align}
Operators $\widehat{Q}^{\,}_{1}$ and $\widehat{Q}^{\,}_{2}$ are Hermitian.
They carry odd fermion parity since the operator
$\widehat{U}^{\,}_{\mathrm{B}}(p)$ 
exchanges the states $\ket{\psi^{\,}_{\alpha,\pm}}$ with
$\ket{\psi^{\,}_{\alpha,\mp}}$.
They satisfy the defining properties
\begin{align}
\left\{\widehat{Q}^{\,}_{i},\widehat{Q}^{\,}_{j}\right\}
=
2
\widehat{H}^{\,}_{\mathrm{B}}\,
\delta^{\,}_{i,j},
\qquad
\left[
\widehat{Q}^{\,}_{i},
\widehat{H}^{\,}_{\mathrm{B}}
\right]
=
0,
\qquad
i,j=1,2,
\label{eq:def supercharges mu1 d}
\end{align}
\end{subequations}
of fermionic supercharges.

There may be other supercharges in addition 
to the ones defined in Eq.\ \eqref{eq:def supercharges mu1}
due to the representation $\widehat{U}^{\,}_{\mathrm{B}}$ of the group $G^{\,}_{f}$.
The precise number of these additional supercharges on the boundary
$\Lambda^{\,}_{\mathrm{B}}$
depends on the pair $[(\nu^{\,}_{\mathrm{B}},\rho^{\,}_{\mathrm{B}})]$
that characterizes
the number of symmetry operators $\widehat{U}^{\,}_{\mathrm{B}}(g)$ 
that carry odd fermion parity and their mutual algebra.
They can be constructed in the same fashion as in Eq.\
\eqref{eq:def supercharges mu0}.

By definition,
the indices $([(\nu^{\,}_{\mathrm{L}},\rho^{\,}_{\mathrm{L}})],1)$
and $([(\nu^{\,}_{\mathrm{R}},\rho^{\,}_{\mathrm{R}})],1)$
associated to the representations 
$\widehat{U}^{\,}_{\mathrm{L}}$ and $\widehat{U}^{\,}_{\mathrm{R}}$,
respectively, satisfy
\begin{align}
([(\nu^{\,}_{\mathrm{L}},\rho^{\,}_{\mathrm{L}})],1)
\wedge
([(\nu^{\,}_{\mathrm{R}},\rho^{\,}_{\mathrm{R}})],1)
=
([(0,0)],0)
\end{align}
under the stacking rule \eqref{eq:stacking rules gen summ d}.
If we focus on a single boundary (denoted by B), 
the equivalence class $[(\nu^{\,}_{\mathrm{B}},\rho^{\,}_{\mathrm{B}})]$
characterizes the 
nontrivial projective nature of the boundary representation
$\widehat{U}^{\,}_{\mathrm{B}}$.
Whenever $[(\nu^{\,}_{\mathrm{B}},\rho^{\,}_{\mathrm{B}})]\neq [(0,0)]$,
it is guaranteed that there is
no state that is invariant under the action of $\widehat{U}^{\,}_{\mathrm{B}}(g)$
for all $g\in G^{\,}_{f}$. In other words, there is no state in the Fock space 
$\mathfrak{F}^{\,}_{\Lambda^{\,}_{\mathrm{B}}}$ that transforms as a singlet under 
the representation $\widehat{U}^{\,}_{\mathrm{B}}$.
Each eigenstate of a symmetric boundary Hamiltonian $\widehat{H}^{\,}_{\mathrm{B}}$
must carry degeneracies in addition to the twofold degeneracy
due to $[\mu]=1$. 
The degeneracy is protected by the particular 
representation $\widehat{U}^{\,}_{\mathrm{B}}$ of the symmetry group $G^{\,}_{f}$ 
and cannot be lifted without breaking the $G^{\,}_{f}$ symmetry.
The exact degeneracy protected by the representation depends 
on the explicit form of the group $G^{\,}_{f}$, 
and the boundary representation $\widehat{U}^{\,}_{\mathrm{B}}$ 
with the equivalence class $[(\nu^{\,}_{\mathrm{B}},\rho^{\,}_{\mathrm{B}})]$.

Since for $[\mu]=1$, the boundary representations $\widehat{U}^{\,}_{\mathrm{L}}$ 
and $\widehat{U}^{\,}_{\mathrm{R}}$ do not act on two decoupled Fock spaces.
The total protected ground-state degeneracy $\mathrm{GSD}^{[\mu]=1}_{\mathrm{bd}} $ 
when open boundary conditions are
imposed cannot be computed by taking the products of 
degeneracies associated with the Hamiltonians $\widehat{H}^{\,}_{\mathrm{L}}$ 
and $\widehat{H}^{\,}_{\mathrm{R}}$ separately. 
However, $\mathrm{GSD}^{[\mu]=1}_{\mathrm{bd}} $
can be computed by multiplying the ``naive''
protected ground state degeneracies of the Hamiltonians at the two 
boundaries and modding out the twofold degeneracy due to 
$\mathfrak{F}^{\,}_{\mathrm{LR}}$ shared by the two Hamiltonians, i.e.,
\begin{align}
\mathrm{GSD}^{[\mu]=1}_{\mathrm{bd}}
=
\frac{1}{2}
\times
\mathrm{GSD}^{[\mu]=1}_{\mathrm{L}}
\times
\mathrm{GSD}^{[\mu]=1}_{\mathrm{R}},
\label{eq:total degeneracy gen mu1}
\end{align}
where $\mathrm{GSD}^{[\mu]=1}_{\mathrm{L}}$ and $\mathrm{GSD}^{[\mu]=1}_{\mathrm{R}}$
are the protected ground state degeneracies of $\widehat{H}^{\,}_{\mathrm{L}}$ 
and $\widehat{H}^{\,}_{\mathrm{R}}$, respectively.

\section{Conclusion}
\label{sec:conclusion}

In this paper, we have studied
one-dimensional invertible fermionic topological (IFT) phases. 
By extending ideas presented in
Ref.\ \onlinecite{Fidkowski2011}, we have explicitly
constructed the boundary representations of any internal
fermionic symmetry group $G^{\,}_{f}$. To this end, we
have defined a triplet $([(\nu,\rho)],[\mu])$ that 
characterizes all inequivalent boundary representations
of $G^{\,}_{f}$. This index classifies all distinct
invertible fermionic topological phases with the internal
fermionic symmetry group $G^{\,}_{f}$.
We have also given an elementary derivation of the fermionic stacking rules.
These stacking rules dictate the group structure of one-dimensional
invertible fermionic topological phases 
given a symmetry group $G^{\,}_{f}$.
They agree with the stacking rules derived 
in Refs.\ \onlinecite{Turzillo2019,Bourne2021}, but disagree with
the ones derived in Ref.\ \onlinecite{Bultinck2017}.

Given an IFT phase in one-dimensional space
characterized by the triplet 
$([\nu,\rho]),[\mu])$, we have deduced the 
protected ground-state degeneracies on general grounds. 
In doing so, we have identified that an emergent supersymmetry 
at the boundaries is implied whenever the IFT phase is intrinsically
fermionic, i.e., either $\rho(g)=1$ for some $g\in G^{\,}_{f}$
or $[\mu]=1$ for the boundary projective representation.

Finally, we have given a concrete application of these results
by working out the IFT phases in symmetry class BDI from the tenfold way
in the supplemental material. By applying the Jordan-Wigner transformation
we can map the Majorana $c$ chains that we chose from the symmetry class BDI
to spin-1/2 cluster $c$ chains. We can then explain how IFT phases are turned
into bosonic symmetry protected topological phases of matter
by the non-local Jordan-Wigner transformation.

\section*{Acknowledgments}

We thank Alex Turzillo for many valuable comments.
\"O.M.A. was supported by the 
Swiss National Science Foundation (SNSF) 
under Grant No.\ 200021 184637.

\vfill

\appendix
\addappheadtotoc
\onecolumngrid

\section{Group cohomology}
\label{appsec:Group Cohomology}

\noindent Given two groups $G$ and $M$, an \textit{$n$-cochain} is the map
\begin{equation}
\begin{split}
\phi \colon G^{n} \to&\, M,
\\
(g^{\,}_{1},g^{\,}_{2},\cdots, g^{\,}_{n})\mapsto&\,
\phi(g^{\,}_{1},g^{\,}_{2},\cdots, g^{\,}_{n}),
\end{split}
\end{equation}
that maps an $n$-tuple $(g^{\,}_{1},g^{\,}_{2},\cdots, g^{\,}_{n})$
to an element $\phi(g^{\,}_{1},g^{\,}_{2},\dots,g^{\,}_{n})\in M$.
The set of all $n$-cochains from $G^{n}$ to $M$ 
is denoted by $C^{n}(G,M)$.
We define an $M$-valued 0-cochain to be an element of the group $M$
itself, i.e., $C^{0}(G,M)= M$.
Henceforth, we will denote 
the group composition rule in $G$ by $\cdot$
and the group composition rule in $M$ additively by $+$
($-$ denoting the inverse element).

Given the group homomorphism $\mathfrak{c}\colon G \to\left\{0,1\right\}$,
for any $g\in G$, we define the group action
\begin{equation}
\begin{split} 
\mathfrak{C}^{\,}_{g}
\colon 
M \to&M,
\\
m\ \mapsto&\
(-1)^{\mathfrak{c}(g)}\,m.
\end{split}
\end{equation}
The homomorphism $\mathfrak{c}$ indicates whether and element $g\in G$
is represented unitarily [$\mathfrak{c}(g)=0$] or antiunitarily
[$\mathfrak{c}(g)=1$]. 
We define the map $\delta^{n}_{\mathfrak{c}}$ 
\begin{subequations}\label{appeq:definition delta n}
\begin{equation}
\begin{split}
\delta^{n}_{\mathfrak{c}}\colon 
C^{n}(G,M)\ \to&\ C^{n+1}(G,M),
\\
\phi\ \mapsto&\ \left(\delta^{n}_{\mathfrak{c}}\phi\right),
\end{split}
\end{equation}
from $n$-cochains to $(n+1)$-cochains such that
\begin{align}
\label{eq:def coboundary operator}
\left(\delta^{n}_{\mathfrak{c}}\phi\right)
(g^{\,}_{1},\cdots,g^{\,}_{n+1})\df
\mathfrak{C}^{\,}_{g^{\,}_{1}}\!
\left(\phi(g^{\,}_{2},\cdots, g^{\,}_{n},g^{\,}_{n+1})\right)
+
\sum_{i=1}^{n}
(-1)^{i}
\phi(g^{\,}_{1},\cdots,g^{\,}_{i}\cdot g^{\,}_{i+1},
\cdots,g^{\,}_{n+1})
-
(-1)^{n}\,
\phi(g^{\,}_{1},\cdots,g^{\,}_{n}).
\end{align}
\end{subequations}
The map $\delta^{n}_{\mathfrak{c}}$
is called a \textit{coboundary operator}.  

\textbf{Example $n=2$:}
The coboundary operator $\delta^{2}_{\mathfrak{c}}$ is defined by
\begin{align}
\left(\delta^{2}_{\mathfrak{c}}\phi\right)
(g^{\,}_{1},g^{\,}_{2},g^{\,}_{3})=&\,
\mathfrak{C}^{\,}_{g^{\,}_{1}}\!
\left(\phi(g^{\,}_{2},g^{\,}_{3})\right)
+
(-1)^{1}
\phi(g^{\,}_{1}\cdot g^{\,}_{2},g^{\,}_{3})
+
(-1)^{2}
\phi(g^{\,}_{1}, g^{\,}_{2}\cdot g^{\,}_{3})
-
(-1)^{2}\,
\phi(g^{\,}_{1},g^{\,}_{2})
\nonumber\\
=&\,
(-1)^{
\mathfrak{c}(g^{\,}_{1})
     }\,
\phi(g^{\,}_{2},g^{\,}_{3})
-
\phi(g^{\,}_{1}\cdot g^{\,}_{2},g^{\,}_{3})
+
\phi(g^{\,}_{1}, g^{\,}_{2}\cdot g^{\,}_{3})
-
\phi(g^{\,}_{1},g^{\,}_{2}).
\label{appeq:example delta2}
\end{align}
We observe that
\begin{align}
\left(\delta^{2}_{\mathfrak{c}}\phi\right)
(g^{\,}_{1},g^{\,}_{2},g^{\,}_{3})=0
\ \Longleftrightarrow\
\phi(g^{\,}_{1},g^{\,}_{2})
+
\phi(g^{\,}_{1}\cdot g^{\,}_{2},g^{\,}_{3})=
\phi(g^{\,}_{1}, g^{\,}_{2}\cdot g^{\,}_{3})
+
(-1)^{
\mathfrak{c}(g^{\,}_{1})
     }\,
\phi(g^{\,}_{2},g^{\,}_{3})
\end{align}
is nothing but the 2-cocycle condition
\eqref{eq:def proj rep on one component bd d}
obeyed by
$\phi$.

\textbf{Example $n=1$:}
The coboundary operator $\delta^{1}_{\mathfrak{c}}$ is defined by
\begin{align}
\left(\delta^{1}_{\mathfrak{c}}\phi\right)
(g^{\,}_{1},g^{\,}_{2})=&\,
\mathfrak{C}^{\,}_{g^{\,}_{1}}\!
\left(\phi(g^{\,}_{2})\right)
+
(-1)^{1}
\phi(g^{\,}_{1}\cdot g^{\,}_{2})
-
(-1)^{1}
\phi(g^{\,}_{1})
\nonumber\\
=&\,
(-1)^{
\mathfrak{c}(g^{\,}_{1})
     }\,
\phi(g^{\,}_{2})
-
\phi(g^{\,}_{1}\cdot g^{\,}_{2})
+
\phi(g^{\,}_{1}).
\label{appeq:example delta1}
\end{align}
One verifies the important identity
\begin{equation}
\Phi(g^{\,}_{1},g^{\,}_{2})\df
\left(\delta^{1}_{\mathfrak{c}}\phi\right)
(g^{\,}_{1},g^{\,}_{2})
\ \implies \
(\delta^{2}_{\mathfrak{c}}\Phi)(g^{\,}_{1},g^{\,}_{2},g^{\,}_{3})=0.
\label{appeq:one-cochain always obeys the cocycle conditions}
\end{equation} 
Using the coboundary operator, we define two sets 
\begin{subequations}
\begin{align}
Z^{n}(G,M^{\,}_{\mathfrak{c}})\df
\mathrm{ker}(\delta^{n}_{\mathfrak{c}})=
\left\{
\phi \in C^{n}(G,M) \ |\
\delta^{n}_{\mathfrak{c}}\phi=0
\right\},
\end{align}
and
\begin{align}
B^{n}(G,M^{\,}_{\mathfrak{c}})\df
\mathrm{im}(\delta^{n-1}_{\mathfrak{c}})=
\left\{
\phi \in C^{n}(G,M) \ |\ 
\phi=\delta^{n-1}_{\mathfrak{c}}\phi',\,\,
\phi'\in C^{n-1}(G,M)
\right\}.
\end{align}
\end{subequations}
The cochains in $Z^{n}(G,M^{\,}_{\mathfrak{c}})$
are called \textit{$n$-cocycles}.
The cochains in $B^{n}(G,M^{\,}_{\mathfrak{c}})$
are called \textit{$n$-coboundaries}.
The action of the boundary operator on the elements of the group $M$
is sensitive to the homomorphism $\mathfrak{c}$.
For this reason, we label $M$ by
$\mathfrak{c}$ in $Z^{n}(G,M^{\,}_{\mathfrak{c}})$
and $B^{n}(G,M^{\,}_{\mathfrak{c}})$.
The importance of the coboundaries is that the identity
(\ref{appeq:one-cochain always obeys the cocycle conditions})
generalizes to
\begin{equation}
\phi=\delta^{n-1}_{\mathfrak{c}}\phi'
\ \implies\
\delta^{n}_{\mathfrak{c}}\phi=0.
\label{appeq:n-cochain always obeys the cocycle conditions}
\end{equation} 
The $n$th cohomology group is defined as the quotient of the
$n$-cocycles by the $n$-coboundaries, i.e.,
\begin{align}
H^{n}(G,M^{\,}_{\mathfrak{c}})\df
Z^{n}(G,M^{\,}_{\mathfrak{c}})/B^{n}(G,M^{\,}_{\mathfrak{c}}).
\end{align}
The $n$th cohomology group $H^{n}(G,M^{\,}_{\mathfrak{c}})$
is an additive Abelian group.
We denote its elements by $[\phi]\in H^{n}(G,M^{\,}_{\mathfrak{c}})$,
i.e., the equivalence class
of the $n$-cocycle $\phi$.

Finally, we define the following operation on the cochains.
Given two cochains $\phi\in C^{n}(G,N)$ and $\theta\in C^{m}(G,M)$,
we produce the cochain
$(\phi \cup \theta)\in C^{n+m}(G,N\times M)$
through
\begin{subequations}\label{eq:def cup product}
\begin{align}
(\phi \cup \theta)
(g^{\,}_{1},\cdots,g^{\,}_{n},g^{\,}_{n+1},\cdots,g^{\,}_{m})\df
\Big(
\phi(g^{\,}_{1},\cdots,g^{\,}_{n}),
\mathfrak{C}^{\,}_{g^{\,}_{1}\cdot g^{\,}_{2}\cdots g^{\,}_{n}}\!
\left(
\theta(g^{\,}_{n+1},\cdots,g^{\,}_{n+m})
\right)
\Big).
\label{eq:def cup product a}
\end{align}
If we compose operation (\ref{eq:def cup product a})
with the pairing map $f:N\times M \to M'$ where $M'$ is an Abelian group,
we obtain the cup product
\begin{align}
(\phi \smile \theta) (g^{\,}_{1},\cdots,g^{\,}_{n},g^{\,}_{n+1},\cdots,g^{\,}_{m})
\df
f
\Bigg(
\Big(
\phi(g^{\,}_{1},\cdots,g^{\,}_{n}),
\mathfrak{C}^{\,}_{g^{\,}_{1}\cdot g^{\,}_{2}\cdots g^{\,}_{n}}\!
\left(
\theta(g^{\,}_{n+1},\cdots,g^{\,}_{n+m})
\right)
\Big)
\Bigg).
\label{eq:def cup product b}
\end{align}
\end{subequations}
Hence,
$(\phi \smile \theta)\in C^{n+m}(G,M')$.
For our purposes, both $N$ and $M$ are subsets of the integer numbers,
$M'=\mathbb{Z}^{\,}_{2}$, while the pairing map $f$ is
\begin{align}
f
\Bigg(
\Big(
\phi(g^{\,}_{1},\cdots,g^{\,}_{n}),
\mathfrak{C}^{\,}_{g^{\,}_{1}\cdot g^{\,}_{2}\cdots g^{\,}_{n}}\!
\left(
\theta(g^{\,}_{n+1},\cdots,g^{\,}_{n+m})
\right)
\Big)
\Bigg)\df
\phi(g^{\,}_{1},\cdots,g^{\,}_{n})\,
\mathfrak{C}^{\,}_{g^{\,}_{1}\cdot g^{\,}_{2}\cdots g^{\,}_{n}}\!
\left(
\theta(g^{\,}_{n+1},\cdots,g^{\,}_{n+m})
\right)
\hbox{ mod 2}
\end{align}
where multiplication of cochains $\phi$ and $\theta$ is treated 
as multiplication of integers numbers modulo $2$.
For instance, for the cup product of a 1-cochain
$\alpha\in C^{1}\big(G,\mathbb{Z}^{\,}_{2}\big)$ 
and a 2-cochain $\beta \in C^{2}\big(G,\mathbb{Z}^{\,}_{2}\big)$, we write
\begin{align}
(\alpha \smile \beta ) (g^{\,}_{1},g^{\,}_{2},g^{\,}_{3})=
\alpha(g^{\,}_{1})\,
\mathfrak{C}^{\,}_{g^{\,}_{1}}\!
\left(
\beta(g^{\,}_{2},g^{\,}_{3})
\right)
=
\alpha(g^{\,}_{1})\,
\beta(g^{\,}_{2},g^{\,}_{3}),
\end{align}
where the cup product takes values in $\mathbb{Z}^{\,}_{2}=\left\{0,1\right\}$
and multiplication of $\alpha$ and $\beta$ is the multiplication of integers.
In reaching the last equality,
we have used the fact that the 2-cochain $\beta(g^{\,}_{2},g^{\,}_{3})$ 
takes values in $\mathbb{Z}^{\,}_{2}$ for which 
$\mathfrak{C}^{\,}_{g^{\,}_{1}}(\beta(g^{\,}_{2},g^{\,}_{3}))=
\beta(g^{\,}_{2},g^{\,}_{3})$ for any $g^{\,}_{1}$.
The cup product defined in Eq.\ \eqref{eq:def cup product b} satisfies
\begin{align}
\delta^{n+m}_{\mathfrak{c}} (\phi \smile \theta)
=
\left( \delta^{n}_{\mathfrak{c}} \phi \smile \theta \right)
+
(-1)^{n}\,
\left( \phi \smile \delta^{m}_{\mathfrak{c}}\theta \right),
\label{eq:boundary of a cup product}
\end{align}
given two cochains $\phi\in C^{n}(G,N)$ and $\theta\in C^{m}(G,M)$.
Hence, the cup product of two cocycles is again a cocycle as the
right-hand side of Eq.\ \eqref{eq:boundary of a cup product} vanishes.

\section{Construction of the fermionic symmetry Group $G^{\,}_{f}$}
\label{appsec:central extension review}

For quantum systems built out of Majorana degrees of freedom
the parity {(evenness or oddness)}
of the total fermion number is always
a constant of the motion. If $\widehat{F}$ denotes the
operator whose eigenvalues counts the total number of
local fermions in the Fock space, then the parity operator
$(-1)^{\widehat{F}}$ necessarily commutes with the
Hamiltonian that dictates the quantum dynamics, even though
$\widehat{F}$ might not, as is the case in any mean-field
treatment of superconductivity.

We denote the group of two elements $e$ and $p$
\begin{equation}
\mathbb{Z}^{\mathrm{F}}_{2}\df
\left\{e,p\,|\, e\, p=p\,e=p,\qquad e=e\,e=p\,p\right\},
\end{equation}
whereby $e$ is the identity element and we shall interpret
the quantum representation of $p$ as the fermion parity operator.
It is because of this interpretation of the group element $p$ that
we attach the upper index $F$ to the cyclic group $\mathbb{Z}^{\,}_{2}$.
In addition to the symmetry group $\mathbb{Z}^{\mathrm{F}}_{2}$,
we assume the existence of a second symmetry group
$G$ with the composition law $\cdot$ and the identity element
$\mathrm{id}$.
We would like to construct a new
symmetry group $G^{\,}_{f}$
out of the two groups $G$ and $\mathbb{Z}^{\mathrm{F}}_{2}$. Here,
the symmetry group $G^{\,}_{f}$ inherits the ``fermionic'' label
$f$ from its center $\mathbb{Z}^{\mathrm{F}}_{2}$. One possibility is to consider the Cartesian product
\begin{subequations}\label{eq:Cartesian product}
\begin{equation}
G\times\mathbb{Z}^{\mathrm{F}}_{2}\df
\left\{(g,h)\ |\ g\in G,\qquad h\in\mathbb{Z}^{\mathrm{F}}_{2}\right\}  
\label{eq:Cartesian product a}
\end{equation}
with the composition rule
\begin{equation}
(g^{\,}_{1},h^{\,}_{1})\circ
(g^{\,}_{2},h^{\,}_{2})\df
(g^{\,}_{1}\cdot g^{\,}_{2},h^{\,}_{1}\ h^{\,}_{2}).
\label{eq:Cartesian product b}
\end{equation}
\end{subequations}
The resulting group $G^{\,}_{f}$ is the direct product of $G$
and $\mathbb{Z}^{\mathrm{F}}_{2}$.
However, the composition rule (\ref{eq:Cartesian product b})
is not the only one compatible with the existence
of a neutral element, inverse, and associativity.
To see this, we assume first the existence of the map
\begin{subequations}
\label{eq:def ZF2}
\begin{equation}
\begin{split}
\gamma\colon
G\times G\ \to&\ \mathbb{Z}^{\mathrm{F}}_{2},
\\
(g^{\,}_{1},g^{\,}_{2})\ \mapsto&\ \gamma(g^{\,}_{1},g^{\,}_{2}),
\end{split}
\label{eq:def ZF2 a}
\end{equation}
whereby we impose the conditions
\begin{equation}	
\gamma(\mathrm{id},g)=
\gamma(g,\mathrm{id})=
e,
\qquad
\gamma(g^{-1},g)=
\gamma(g,g^{-1}),
\label{eq:def ZF2 b}
\end{equation}
for all $g\in G$ and
\begin{equation}
\gamma(g^{\,}_{1},g^{\,}_{2})\,
\gamma(g^{\,}_{1}\cdot g^{\,}_{2},g^{\,}_{3})=
\gamma(g^{\,}_{1}, g^{\,}_{2}\cdot g^{\,}_{3})\,
\gamma(g^{\,}_{2},g^{\,}_{3}),
\label{eq:def ZF2 c}
\end{equation}
for all $g^{\,}_{1},g^{\,}_{2},g^{\,}_{3}\in G$.
Second, we define $G^{\,}_{f}$ to be the set of all pairs
$(g,h)$ with $g\in G$ and $h\in\mathbb{Z}^{\mathrm{F}}_{2}$
obeying the composition rule
\begin{equation}
\begin{split}
\underset{\gamma}{\circ}\colon&
\left(G\times\mathbb{Z}^{\mathrm{F}}_{2}\right)
\times
\left(G\times\mathbb{Z}^{\mathrm{F}}_{2}\right)
\ \to\
G\times\mathbb{Z}^{\mathrm{F}}_{2},
\\
&
\Big(
(g^{\,}_{1},\,h^{\,}_{1}),
(g^{\,}_{2},\,h^{\,}_{2})
\Big)
\ \mapsto\,
(g^{\,}_{1},\,h^{\,}_{1})
\underset{\gamma}{\circ}
(g^{\,}_{2},\,h^{\,}_{2}),
\end{split}
\label{eq:def ZF2 d}
\end{equation}
where
\begin{equation}
(g^{\,}_{1},\,h^{\,}_{1})
\underset{\gamma}{\circ}
(g^{\,}_{2},\,h^{\,}_{2})
\df
\Big(
g^{\,}_{1}\cdot g^{\,}_{2},
h^{\,}_{1}
\,
h^{\,}_{2}
\,
\gamma(g^{\,}_{1},g^{\,}_{2})
\Big).
\label{eq:def ZF2 e}
\end{equation}
\end{subequations}
One verifies the following properties.
First, the order within the composition
$h^{\,}_{1}\, h^{\,}_{2}\,\gamma(g^{\,}_{1},g^{\,}_{2})$
is arbitrary since $\mathbb{Z}^{\mathrm{F}}_{2}$ is Abelian.
Second, conditions (\ref{eq:def ZF2 b}) and (\ref{eq:def ZF2 c})
ensure that $G^{\,}_{f}$ is a group with the
neutral element
\begin{subequations}
\begin{equation}
(\mathrm{id},e),
\end{equation}
the inverse to $(g,h)$ is
\begin{equation}
(g^{-1},[\gamma(g,g^{-1})]^{-1}\,h^{-1}),
\end{equation}
and the center
(those elements of the group that commute with all group elements)
given by
\begin{equation}
(\mathrm{id},\mathbb{Z}^{\mathrm{F}}_{2}),
\end{equation}
\end{subequations} 
i.e.,
the group $G^{\,}_{f}$ is a central extension of $G$ by $\mathbb{Z}^{\mathrm{F}}_{2}$.
Third, the map $\gamma$ can be equivalent to a map $\gamma^{\prime}$
of the form (\ref{eq:def ZF2 a})
in that they generate two isomorphic groups.
This is true if there exists the one-to-one map
\begin{subequations}
\begin{equation}
\begin{split}
\widetilde{\kappa}\colon
G\times\mathbb{Z}^{\mathrm{F}}_{2}\ \to&\ G\times\mathbb{Z}^{\mathrm{F}}_{2},
\\
(g,h)\ \mapsto&\ (g,\kappa(g)\,h)
\end{split}
\label{eq:def map kappa a}
\end{equation}
induced by the map
\begin{equation}
\begin{split}
\kappa\colon
G\ \to&\ \mathbb{Z}^{\mathrm{F}}_{2},
\\
g\ \mapsto&\ \kappa(g),
\end{split}
\label{eq:def map kappa b}
\end{equation}
\end{subequations}
such that the condition
\begin{equation}
\widetilde{\kappa}
\Big(
(g^{\,}_{1},\,h^{\,}_{1})
\,\underset{\gamma}{\circ}\,
(g^{\,}_{2},\,h^{\,}_{2})
\Big)=
\widetilde{\kappa}
\Big(
(g^{\,}_{1},\,h^{\,}_{1})
\Big)
\,\underset{\gamma'}{\circ}\,
\widetilde{\kappa}
\Big(
(g^{\,}_{2},\,h^{\,}_{2})
\Big)
\label{eq:condition on widetilde kappa for group isomorphism}
\end{equation}
holds for all $(g^{\,}_{1},\,h^{\,}_{1}),(g^{\,}_{2},\,h^{\,}_{2})
\in G\times\mathbb{Z}^{\mathrm{F}}_{2}$.
In other words, 
$\gamma$ and $\gamma'$
generate two isomorphic groups
if the identity
\begin{equation}
\kappa(g^{\,}_{1}\cdot g^{\,}_{2})\cdot\gamma(g^{\,}_{1},g^{\,}_{2})=
\kappa(g^{\,}_{1})\cdot\kappa(g^{\,}_{2})\cdot\gamma^{\prime}(g^{\,}_{1},g^{\,}_{2})
\label{eq:condition on kappa for group isomorphism}
\end{equation}
holds for all $g^{\,}_{1},g^{\,}_{2}\in G$.
This group isomorphism defines an equivalence relation.
We say that the group $G^{\,}_{f}$
obtained by extending the group $G$ with the group $\mathbb{Z}^{\mathrm{F}}_{2}$
through the map $\gamma$ splits when 
a map (\ref{eq:def map kappa b}) exists such that
\begin{equation}
\kappa(g^{\,}_{1}\cdot g^{\,}_{2})\cdot\gamma(g^{\,}_{1},g^{\,}_{2})=
\kappa(g^{\,}_{1})\cdot\kappa(g^{\,}_{2})
\label{eq: condition for Gf a split group}
\end{equation}
holds for all $g^{\,}_{1},g^{\,}_{2}\in G$, i.e.,
$G^{\,}_{f}$ splits when it is isomorphic to the direct product
(\ref{eq:Cartesian product}).

The task of classifying all the non-equivalent central
extensions of $G$ by $\mathbb{Z}^{\mathrm{F}}_{2}$ through $\gamma$
is achieved by enumerating all the elements of the second cohomology group 
$H^{2}\left(G,\mathbb{Z}^{\mathrm{F}}_{2}\right)$.
We define an index $[\gamma]\in H^{2}\left(G,\mathbb{Z}^{\mathrm{F}}_{2}\right)$
to represent such an equivalence class, whereby the index $[\gamma]=0$ 
is assigned to the case when $G^{\,}_{f}$ splits.

\section{Classification of projective representations of $G^{\,}_{f}$}
\label{appsec:classification of Gf}

It was described in Appendix
\ref{appsec:central extension review},
how a global symmetry group $G^{\,}_{f}$ for
a fermionic quantum system naturally contains the fermion{-}number parity
{symmetry group} $\mathbb{Z}^{\mathrm{F}}_{2}$
in its center, i.e., it is a central
extension of a group $G$ by $\mathbb{Z}^{\mathrm{F}}_{2}$. Such group extension
are classified by prescribing an element
$[\gamma]\in H^{2}(G,\mathbb{Z}^{\mathrm{F}}_{2})$,
such that we may think of $G^{\,}_{f}$
as the set of tuples $(g,h)\in G\times\mathbb{Z}^{\mathrm{F}}_{2}$ with
composition rule as in Eq.\ \eqref{eq:def ZF2 e}.  From this
perspective there is an implicit choice of trivialization
$\tau:G^{\,}_{f}\to\mathbb{Z}^{\mathrm{F}}_{2}$ and projection
$b: G^{\,}_{f} \to G$ such that
\begin{align}
\tau\big[(g,h)\big]=h,
\qquad
b\big[(g,h)\big]=g.
\end{align}
Importantly, $\tau$ is related to the extension class $\gamma$ that
defines the group extension via the relation
\begin{equation}
b^{\star}\gamma=\delta^{1}_{\mathfrak{c}}\tau
\end{equation}
where
$b^{\star}\gamma \in C^{2}(G^{\,}_{f},\mathbb{Z}^{\mathrm{F}}_{2})$
is the pullback of $\gamma$ via $b$.  

As explained in Sec.\ \ref{sec: Definition of indices}, we shall trade the
2-cocycle $\phi(g,h)\in Z^{2}(G,\mathrm{\mathrm{U}(1)}^{\,}_{\mathfrak{c}})$
with the tuple $(\nu,\rho) \in C^{2}\big[G,\mathrm{\mathrm{U}(1)}\big]\times
C^{1}\big(G,\mathbb{Z}^{\,}_{2}\big)$ that satisfy certain
cocycle and coboundary conditions.  To this end, it is convenient to
define the modified 2-coboundary operator
\begin{align}
\mathcal{D}^{2}_{\gamma}
\left(
\nu,\,
\rho
\right)
\df
\left(
\delta^{2}_{\mathfrak{c}} \nu -\pi\,\rho \smile \gamma,
\delta^{1}_{\mathfrak{c}} \rho
\right),
\label{eq:modified 2-coboundary}
\end{align}
acting on a tuple of cochains $(\nu,\rho) \in
C^{2}\big[G,\mathrm{\mathrm{U}(1)}\big]\times
C^{1}\big(G,\mathbb{Z}^{\,}_{2}\big)$
together with the modified
1-coboundary operator
\begin{align}
\mathcal{D}^{1}_{\gamma}
\left(
\alpha,
\beta
\right)
\df
\left(
\delta^{1}_{\mathfrak{c}} \alpha
+
\pi
\beta\smile\gamma,
\delta^{0}_{\mathfrak{c}} \beta
\right)
\label{eq:modified 1- coboundary}
\end{align}	
acting on a tuple of cochains $(\alpha,\,\beta) \in
C^{1}\big(G,\mathrm{\mathrm{U}(1)}\big)\times C^{0}\big(G,\mathbb{Z}^{\,}_{2}\big)$.
Being a 0-cochain $\beta$ does not take any arguments and takes values
in $\mathbb{Z}^{\,}_{2}$, i.e., $\beta \in \mathbb{Z}^{\,}_{2}$. 
Note that for the 0-cochain $\beta$,
the coboundary operator \eqref{eq:def coboundary operator} acts as 
\begin{align}
(\delta^{0}_{\mathfrak{c}}\beta)(g)=
\mathfrak{C}^{\,}_{g}(\beta)
-
\beta,
\end{align}
which in fact vanishes for any $g\in G$ 
since $\beta$ takes values in $\mathbb{Z}^{\,}_{2}$ and
$\mathfrak{C}^{\,}_{g}(\beta)=\beta$.
Using
Eq.\ \eqref{eq:boundary of a cup product} and the fact that $\gamma$
is a cocycle, i.e., $\delta^{2}_{\mathfrak{c}} \gamma = 0$, one verifies that
\begin{align}
\mathcal{D}^{2}_{\gamma}\,
\mathcal{D}^{1}_{\gamma}
(\alpha,\beta)
=
\left(
0,
0
\right)
\end{align}
for any tuple
$(\alpha,\,\beta)\in
C^{1}\big(G,\mathrm{\mathrm{U}(1)}\big)\times C^{0}\big(G,\mathbb{Z}^{\,}_{2}\big)$.

It was proved in
Ref.\ \onlinecite{Turzillo2019} that
one may assign to any 2-cocycle 
$[\phi]\in H^{2}\big(G^{\,}_{f},\mathrm{\mathrm{U}(1)}^{\,}_{\mathfrak{c}}\big)$
an equivalence class $[(\nu,\rho)]$ of those
tuples $(\nu,\rho)\in
C^{2}\big[G,\mathrm{\mathrm{U}(1)}\big]\times C^{1}\big(G,\mathbb{Z}^{\,}_{2}\big)$
that satisfy the cocycle condition
under the modified 2-coboundary operator \eqref{eq:modified 2-coboundary}
given by
\begin{align}
\mathcal{D}^{2}_{\gamma}
\left(
\nu,\rho
\right)
=
\left(
\delta^{2}_{\mathfrak{c}}\nu
-
\pi\,\rho\smile\gamma,\,
\delta^{1}_{\mathfrak{c}}\rho
\right)
=
(0,0).
\label{eq:Gf_cocycle_cond}
\end{align}
Indeed, two tuples $(\nu,\rho)$ and $(\nu', \rho')$
that satisfy Eq.\ \eqref{eq:Gf_cocycle_cond} are said to be
equivalent if there exists a tuple $(\alpha,\,\beta) \in C^{1}\big(G,\mathrm{\mathrm{U}(1)}\big)\times C^{0}\big(G,\mathbb{Z}^{\,}_{2}\big)$
such that
\begin{align}
(\nu,\,\rho)
=
(\nu',\,\rho')
+
\mathcal{D}^{1}_{\gamma}
(\alpha,\,\beta)
=
(
\nu'
+
\delta^{1}_{\mathfrak{c}} \alpha
+
\pi\,
\beta\smile\gamma,\,
\delta^{0}_{\mathfrak{c}} \beta).
\label{eq:Gf_equivalence_rel}
\end{align}
In other words, using this equivalence relation we define an
equivalence class $[(\nu,\rho)]$ of the tuple $(\nu,\rho)$ as an
element of the set
\begin{align}
[(\nu,\rho)]\in
\frac{\mathrm{ker}(\mathcal{D}^{2}_{\gamma})}
{\mathrm{im}(\mathcal{D}^{1}_{\gamma})}.
\end{align}
The proof of the one-to-one correspondence between $[\phi]$ and $[(\nu,\rho)]$
then follows in three steps which we sketch out below.
We refer the reader to Ref.\ \onlinecite{Turzillo2019} for more details.
\begin{enumerate}
\item
First, given a cocycle $\phi\in Z^{2}(G^{\,}_{f},\mathrm{\mathrm{U}(1)}^{\,}_{\mathfrak{c}})$, one can
define $\rho\in Z^{1}\big(G,\mathbb Z_{2}\big)$ via Eqs.\ \eqref{eq:def rho gen mu0} or \eqref{eq:def rho gen mu1}.
The fact that $\rho$ is a cocycle follows from that fact
that $\phi$ is a cocycle.
\item
Next, one can always find a representative $\phi$ in every cohomology
class $[\phi]\in H^{2}\big(G^{\,}_{f},\mathrm{\mathrm{U}(1)}^{\,}_{\mathfrak{c}}\big)$ that satisfies the relation
$\phi=\nu+\pi\rho\smile\tau$.
\item
Finally, the fact that $\delta^{2}_{\mathfrak{c}} \phi=0$ implies that
$\delta^{2}_{\mathfrak{c}}\nu=\pi\rho\smile\gamma$.
\end{enumerate}
We note that when the $[\gamma] = 0$, i.e., the group $G^{\,}_{f}$
splits as $G^{\,}_{f} = G\times \mathbb{Z}^{\mathrm{F}}_{2}$, the modified
coboundary operators \eqref{eq:modified 2-coboundary} and
\eqref{eq:modified 1- coboundary} reduce to the coboundary operator
\eqref{eq:def coboundary operator} with $n=2$ and $n=1$,
respectively. If so the cochains $\nu$ and $\rho$ are both cocycles,
i.e., $(\nu,\,\rho)\in Z^{2}(G,\mathrm{\mathrm{U}(1)}^{\,}_{\mathfrak{c}})\times
Z^{1}\big(G,\mathbb{Z}^{\,}_{2}\big)$. The equivalence classes $[(\nu,\rho)]$
of the tuple $(\nu,\rho)$ is then equal to the equivalence cohomology
classes of each of its components, i.e.,
\begin{align}
[(\nu,\rho)]
=
([\nu],[\rho])
\in 
H^{2}(G,\mathrm{\mathrm{U}(1)}^{\,}_{\mathfrak{c}})
\times
H^{1}\big(G,\mathbb{Z}^{\,}_{2}\big).
\end{align}
We use the notation $([\nu],[\rho])$ for the two indices
whenever the group $G^{\,}_{f}$ splits ($[\gamma]=0$). The
notation $[(\nu,\rho)]$ applies whenever the group $G^{\,}_{f}$ does not
split ($[\gamma]\neq0$). 

\section{Change in indices $(\nu,\rho)$ under group isomorphisms}
\label{appsec:group isomorphism on indices}

As explained in the Appendix \ref{appsec:central extension review},
the fermionic symmetry group $G^{\,}_{f}$ can be constructed as the
set of pairs $(g,h)\in G\times \mathbb{Z}^{F}_{2}$ with the
composition rule \eqref{eq:def ZF2} specified by the 2-cochain
$\gamma\in C^{2}(G,\mathbb{Z}^{F}_{2})$. The distinct central
extensions $G^{\,}_{f}$ of $G$ are then classified by the equivalence
classes $[\gamma]\in H^{2}(G,\mathbb{Z}^{F}_{2})$. In other words, the
central extension $G^{\,}_{f}$ is determined up to the group
isomorphisms \eqref{eq:def map kappa a} under which the equivalence
class $[\gamma]$ is invariant.

In Sec.\ \ref{sec: Definition of indices}, we defined the pair of
indices $(\nu,\rho)\in C^{2}(G,U(1))\times
C^{1}(G,\mathbb{Z}^{\,}_{2})$ for a given index $[\mu]=0,1$. The
definitions \eqref{eq:def 2-cochain nu [mu]=0} and
\eqref{eq:def 2-cochain nu [mu]=1} of $\nu$ and the definitions
\eqref{eq:def rho gen mu0} and \eqref{eq:def rho gen mu1}
are not invariant under group isomorphisms.
In particular, when restricting the domain of
definition of the 2-cochain $\phi$ from $G^{\,}_{f}$ to $G$, we made
an implicit choice of $\gamma$.  This choice is inherited from a
given bulk representation $\widehat{U}^{\,}_{\mathrm{bulk}}$ through
the consistency condition \eqref{eq:consistency cond general} since
under nontrivial group isomorphisms the transformation rules
implemented by at least one group element $g\in G^{\,}_{f}$ would
change.  In this appendix, we discuss how the pair $(\nu,\rho)\in
C^{2}(G,U(1))\times C^{1}(G,\mathbb{Z}^{\,}_{2})$ is shifted under the
group isomorphism \eqref{eq:def map kappa a} for the cases
$[\mu]=0,1$. 

Let $G^{\,}_{f}$ be a fermionic symmetry group obtained by centrally
extending the symmetry group $G$ by $\mathbb{Z}^{F}_{2}$ through the
2-cochain $\gamma$. We denote the elements of $G^{\,}_{f}$ by the
pairs $(g,h)\in G\times\mathbb{Z}^{F}_{2}$.  Let $G^{'}_{f}$ be a
fermionic symmetry group isomorphic to $G^{\,}_{f}$ through the group
isomorphism
\begin{subequations}
\begin{align}
\begin{split}
\widetilde{\kappa}\colon
G^{\,}_{f}\ \to&\ G^{\prime}_{f},
\\
(g,h)\ \mapsto&\ (g',h')=(g,p^{\kappa(g)}\,h)
\end{split}
\label{eq:def kappatilde group isomorphism}
\end{align}
where $\kappa(g)=0,1$ for any $g\in G$ and we introduced
the shorthand notation $p^{0}=e$ and 
$p^{1}=p$ for the elements in $\mathbb{Z}^{F}_{2}$. 
In other words, $G^{\prime}_{f}$ is the central extension 
of $G$ by $\mathbb{Z}^{F}_{2}$ through the 2-cochain $\gamma'$ such that
\begin{align}
\gamma^{\prime}(g^{\,}_{1},g^{\,}_{2})
=
\gamma(g^{\,}_{1},g^{\,}_{2})\,
p^{\kappa(g^{\,}_{1})+\kappa(g^{\,}_{2})+\kappa(g^{\,}_{1}\,g^{\,}_{2})},
\label{eq:gamma under isomorphism}
\end{align}
\end{subequations}
for any $g^{\,}_{1},g^{\,}_{2}\in G$.

One verifies that the pairs $(g',h')\in
G^{\prime}_{f}$ are identified with the pairs
$(g=g',p^{\kappa(g)}h')\in G^{\,}_{f}$ under the group isomorphism
$\widetilde{\kappa}$.  The identity
\begin{subequations}
\begin{align}
(g,h)
\underset{\gamma}{\circ}
(\mathrm{id},p^{\kappa(g)})=
(g,h\,p^{\kappa(g)}),
\label{eq:group iso comp ident}
\end{align}
which holds for any $g\in G$ and $h\in \mathbb{Z}^{F}_{2}$,
then suggests that the boundary representation
$\widehat{U}^{\prime}_{\mathrm{B}}$ of element $(g',h')\in
G^{\prime}_{f}$ is related to the boundary representation
$\widehat{U}^{\,}_{\mathrm{B}}$ of element
$(g=g',p^{\kappa(g)}\,h')\in G^{\,}_{f}$ via the relation
\begin{align}
\widehat{U}^{\prime}_{\mathrm{B}}\big((g',h')\big)
\propto
\widehat{U}^{\,}_{\mathrm{B}}\big((g=g',h')\big)\,
\left[
\widehat{U}^{\,}_{\mathrm{B}}\big((\mathrm{id},p)\big)
\right]^{\kappa(g)},
\label{eq:group isomorphism relation}
\end{align}
\end{subequations}
i.e., the operator
$\widehat{U}^{\prime}_{\mathrm{B}}\big((g',h')\big)$ must act up
to a multiplicative phase factor as the operator
$\widehat{U}^{\,}_{\mathrm{B}}\big((g,h')\big)$ composed with the
fermion parity operator
$\widehat{U}^{\,}_{\mathrm{B}}\big((\mathrm{id},p)\big)$ if
$\kappa(g)=1$.  Hereby, the exponent $\kappa(g)$ ensures that the
operators $\widehat{U}^{\prime}_{\mathrm{B}}\big((g',h')\big)$
and $\widehat{U}^{\,}_{\mathrm{B}}\big((g,h')\big)$ act identically,
if $\kappa(g)=0$.  Without loss of generality, we take the
proportionality in \eqref{eq:group isomorphism relation} to be
equality. We shall treat the cases of $[\mu]=0$ and $[\mu]=1$
separately.

\subsection{The case of $[\mu]=0$}

On the one hand, invoking the definition \eqref{eq:def 2-cochain nu [mu]=0}
for the 2-cochain $\nu'$ associated with the group $G^{\prime}_{f}$ delivers
\begin{subequations}
\begin{align}
\widehat{U}^{\prime}_{\mathrm{B}}\big((g^{\prime}_{1},e)\big)\,
\widehat{U}^{\prime}_{\mathrm{B}}\big((g^{\prime}_{2},e)\big)
&=
e^{\mathrm{i}\nu'(g^{\prime}_{1},g^{\prime}_{2})}\,
\widehat{U}^{\prime}_{\mathrm{B}}\Big(\big(g^{\prime}_{1}\,g^{\prime}_{2},
\gamma'(g^{\prime}_{1},g^{\prime}_{2})\big)\Big)
\nonumber\\
&=
e^{\mathrm{i}\nu'(g^{\,}_{1},g^{\,}_{2})}\,
\widehat{U}^{\,}_{\mathrm{B}}
\Big(\big(g^{\,}_{1}\,g^{\,}_{2},
\gamma(g^{\,}_{1},g^{\,}_{2})\,
p^{\kappa(g^{\,}_{1})+\kappa(g^{\,}_{2})+\kappa(g^{\,}_{1}\,g^{\,}_{2})}\big)
\Big)
\left[
\widehat{U}^{\,}_{\mathrm{B}}\big((\mathrm{id},p)\big)
\right]^{\kappa(g^{\,}_{1}\,g^{\,}_{2})},
\end{align}
where in reaching the last line we have used Eqs.\ \eqref{eq:gamma under isomorphism}
and \eqref{eq:group isomorphism relation}.
Applying the identity \eqref{eq:group iso comp ident}, we find
\begin{align}
\widehat{U}^{\prime}_{\mathrm{B}}\big((g^{\prime}_{1},e)\big)\,
\widehat{U}^{\prime}_{\mathrm{B}}\big((g^{\prime}_{2},e)\big)
&=
e^{\mathrm{i}\nu'(g^{\,}_{1},g^{\,}_{2})}\,
\widehat{U}^{\,}_{\mathrm{B}}
\Big(\big(g^{\,}_{1}\,g^{\,}_{2},
\gamma(g^{\,}_{1},g^{\,}_{2})\,
\Big)
\left[
\widehat{U}^{\,}_{\mathrm{B}}\big((\mathrm{id},p)\big)
\right]^{\kappa(g^{\,}_{1})+\kappa(g^{\,}_{2})},
\label{eq:isomorphism step 1 mu0}
\end{align}
where the equality holds up to a multiplicative phase factor that can be 
gauged away, reason for which it is omitted for convenience.
On the other hand, inserting Eq.\ \eqref{eq:group isomorphism relation}
on the left-hand side delivers
\begin{align}
\widehat{U}^{\prime}_{\mathrm{B}}\big((g^{\prime}_{1},e)\big)\,
\widehat{U}^{\prime}_{\mathrm{B}}\big((g^{\prime}_{2},e)\big)
&=
\widehat{U}^{\,}_{\mathrm{B}}\big((g^{\,}_{1},e)\big)\,
\left[
\widehat{U}^{\,}_{\mathrm{B}}\big((\mathrm{id},p)\big)
\right]^{\kappa(g^{\,}_{1})}\,
\widehat{U}^{\,}_{\mathrm{B}}\big((g^{\,}_{2},e)\big)\,
\left[
\widehat{U}^{\,}_{\mathrm{B}}\big((\mathrm{id},p)\big)
\right]^{\kappa(g^{\,}_{2})}
\nonumber\\
&=
e^{\mathrm{i}\nu(g^{\,}_{1},g^{\,}_{2})
+\mathrm{i}\pi\,\kappa(g^{\,}_{1})\rho(g^{\,}_{2}) }\,
\widehat{U}^{\,}_{\mathrm{B}}\big((g^{\,}_{1}\,g^{\,}_{2},
\gamma(g^{\,}_{1},g^{\,}_{2}))\big)
\left[
\widehat{U}^{\,}_{\mathrm{B}}\big((\mathrm{id},p)\big)
\right]^{\kappa(g^{\,}_{1}) + \kappa(g^{\,}_{2})},
\label{eq:isomorphism step 2 mu0}
\end{align}
\end{subequations}
where the phase factor $e^{\mathrm{i}\nu(g^{\,}_{1},g^{\,}_{2})}$
arises from the definition \eqref{eq:def 2-cochain nu [mu]=0} of
2-cochain $\nu$ and the phase factor
$e^{\mathrm{i}\kappa(g^{\,}_{1})\rho(g^{\,}_{2})}$ arises when the
operators $\widehat{U}^{\,}_{\mathrm{B}}\big((g^{\,}_{2},e)\big)\,$
and
$\big[\widehat{U}^{\,}_{\mathrm{B}}\big((\mathrm{id},p)\big)\big]^{\kappa(g^{\,}_{1})}$
are interchanged.
Comparing Eqs.\ \eqref{eq:isomorphism step 1 mu0}
and \eqref{eq:isomorphism step 2 mu0}, we make the identification
\begin{align}
\nu'(g^{\,}_{1},g^{\,}_{2})
=
\nu(g^{\,}_{1},g^{\,}_{2})
+
\pi\,(\kappa\smile\rho)(g^{\,}_{1},g^{\,}_{2}).
\label{eq:iso nu change mu0}
\end{align}

The index $\rho$ by definition \eqref{eq:def rho gen mu0 on Gf}
measures the fermion parity of the representation of the element
$(g,h)\in G^{\,}_{f}$.  One notes that the relation
\eqref{eq:group isomorphism relation}
implies that the representations
$\widehat{U}^{\,}_{\mathrm{B}}$ and
$\widehat{U}^{\prime}_{\mathrm{B}}$ have the same fermion
fermion parity since $\widehat{U}^{\,}_{\mathrm{B}}((\mathrm{id},p))$
is fermion parity even.  Hence, the indices $\rho$ and $\rho'$
associated with $G^{\,}_{f}$\ and $G^{\prime}_{f}$,
respectively, coincide.

We conclude that under the isomorphism
\eqref{eq:def kappatilde group isomorphism}
the pair of indices $((\nu',\rho'),0)$ and $((\nu,\rho),0)$ are related as 
\begin{align}
\left((\nu',\rho'),\,0\right) = 
\left((\nu + \pi\,(\kappa\smile\rho),\, \rho),0\right).
\label{eq:isomorp ind mu0}
\end{align}

\subsection{The case of $[\mu]=1$}

When $[\mu]=1$, the definition \eqref{eq:def 2-cochain nu [mu]=1} of
the index $\nu$ is the same as it is when $[\mu]=0$.  Hence, the
argument in the previous section follows through, i.e.,
Eq.\ \eqref{eq:iso nu change mu0} holds.  However, the definition
\eqref{eq:def rho gen mu1 on Gf} of index $\rho$ differs from its
definition in Eq.\ \eqref{eq:def rho gen mu0 on Gf} when $[\mu]=0$.
From Eqs.\ \eqref{eq:def rho gen mu1 on Gf} and \eqref{eq:group
  isomorphism relation}, one observes that under the isomorphism
\eqref{eq:def kappatilde group isomorphism} the index $\rho$ gets
shifted by $\kappa$, i.e.,
\begin{align}
\rho'(g)= \rho(g)+\kappa(g).
\end{align}
This is because computation of the index $\rho'$ involves an
additional conjugation of $\widehat{Y}^{\,}_{\mathrm{B}}$ by fermion
parity operator
$\widehat{U}^{\,}_{\mathrm{B}}\big((\mathrm{id},p)\big)$, which brings
an additional factor of $(-1)^{\kappa(g)}$. We conclude that under the
isomorphism \eqref{eq:def kappatilde group isomorphism} the pair of
indices $((\nu',\rho'),1)$ and $((\nu,\rho),1)$ are related as
\begin{align}
\left((\nu',\rho'),\,1\right)= 
\left((\nu + \pi\,(\kappa\smile\rho),\, \rho+\kappa),1\right).
\label{eq:isomorp ind mu1}
\end{align}
Under the group isomorphism
\eqref{eq:def kappatilde group isomorphism}
the values of the indices $(\nu,\rho)$ and their
respective equivalence classes may change (according to
Eqs.\ \eqref{eq:isomorp ind mu0} and \eqref{eq:isomorp ind mu1}).
However, the number of equivalence classes $([(\nu,\rho)],[\mu])$ and
their stacking rules remain the same, i.e., Eqs.\
\eqref{eq:stacked gen summ}
commute with the relations \eqref{eq:isomorp ind mu0} and
\eqref{eq:isomorp ind mu1}.

\twocolumngrid
\bibliography{References}

\onecolumngrid

\newpage


\section*{Supplementary material}

# Elementary derivation of the stacking rules of invertible fermionic topological phases in one dimension

Ömer M. Aksoy[1] and Christopher Mudry[1,2]

[1]*Condensed Matter Theory Group, Paul Scherrer Institute, CH-5232 Villigen PSI, Switzerland*
[2]*Institut de Physique, EPF Lausanne, CH-1015 Lausanne, Switzerland*
(Dated: July 19, 2022)

## I. OVERVIEW

In this supplement, we apply the toolkit developed in the main text to two well-known and closely related examples: one-dimensional invertible fermionic topological (IFT) phases with symmetry group $G_f = \mathbb{Z}_2^T \times \mathbb{Z}_2^F$ (class BDI) and one-dimensional bosonic symmetry protected topological (BSPT) phases with symmetry group $G = \mathbb{Z}_2^T \times \mathbb{Z}_2$.

We start with a review in Sec. II of the second cohomology group of $\mathbb{Z}_2^T \times \mathbb{Z}_2^F$ and the associated fermionic and bosonic stacking rules.

In Sec. III, we consider the time-reversal symmetric Majorana $c$ chains introduced in Ref. [1], that realize the representatives of the IFT phases in the symmetry class BDI. The label $c \in \mathbb{Z}$ counts the zero modes at the boundaries of noninteracting Bogoliubov-de-Gennes one-dimensional superconductors in the symmetry class BDI. We compute the indices associated with the left and the right boundaries of each Majorana chain and demonstrate that these indices form the cyclic group $\mathbb{Z}_8$ under the stacking rules (2.10) of the main text. We compute the protected ground-state degeneracy of each representative when open boundary conditions are imposed.

In Sec. IV, we consider the quantum spin-1/2 cluster $c$ chains introduced in Ref. [2], that realize the representatives of the BSPT phases with symmetry group $G = \mathbb{Z}_2^T \times \mathbb{Z}_2$. The cluster chains are intimately related to the time-reversal symmetric Majorana chains as they are related by a Jordan-Wigner transformation. This relation has been investigated in Ref. [3]. For each quantum spin-1/2 cluster $c$ chains, we compute the indices associated with the left and the right boundaries and show that they form the group $\mathbb{Z}_2 \times \mathbb{Z}_2$ under the bosonic stacking rules. We relate the indices of the BSPT phases to that of $\mathbb{Z}_4$ subgroup of the IFT phases in the symmetry class BDI, which is the group of fermionic symmetry protected topological (FSPT) phases in class BDI.

## II. REVIEW OF SECOND GROUP COHOMOLOGY OF $\mathbb{Z}_2^T \times \mathbb{Z}_2^F$

The split group $\mathbb{Z}_2^T \times \mathbb{Z}_2^F$ is generated by the two generators $t$ and $p$ corresponding to the cyclic subgroups $\mathbb{Z}_2^T \equiv \{e, t\}$ and $\mathbb{Z}_2^F \equiv \{e, p\}$, respectively. The upper index $T$ refers to the interpretation of $t$ as the generator of reversal of time and must be represented by an antiunitary operator. The upper index $F$ refers to the interpretation of $p$ as the fermion parity. The second group cohomology of $\mathbb{Z}_2^T \times \mathbb{Z}_2^F$ is given by

$$H^2(\mathbb{Z}_2^T \times \mathbb{Z}_2^F, \mathrm{U}(1)_\mathfrak{c}) = \mathbb{Z}_2 \times \mathbb{Z}_2. \tag{2.1a}$$

We denote the equivalent classes of $H^2(\mathbb{Z}_2^T \times \mathbb{Z}_2^F, \mathrm{U}(1)_\mathfrak{c})$ by the index $[(\nu, \rho)]$. Since $\mathbb{Z}_2^T \times \mathbb{Z}_2^F$ is a split group, i.e., a trivial extension of $\mathbb{Z}_2^T$ by $\mathbb{Z}_2^F$, we have the identity

$$[(\nu, \rho)] \equiv ([\nu], [\rho]) \in H^2(\mathbb{Z}_2^T, \mathrm{U}(1)_\mathfrak{c}) \times H^1(\mathbb{Z}_2^T, \mathbb{Z}_2) = \mathbb{Z}_2 \times \mathbb{Z}_2. \tag{2.1b}$$

Therefore, the two indices $[\nu] = 0, 1$ and $[\rho] = 0, 1$ characterizes the projective representations of the group $\mathbb{Z}^T \times \mathbb{Z}_2^F$, and therefore, the one-dimensional FSPT phases in class BDI. Given a representation $\widehat{U}$, the indices $[\nu]$ and $[\rho]$ are specified through the identities [see Eqs. (5.5) and (5.8) in the main text]

$$(-1)^{[\nu]} = \widehat{U}(t)\,\widehat{U}(t), \tag{2.2a}$$

$$(-1)^{[\rho]} = \begin{cases} \widehat{U}(t)\,\widehat{U}(p)\,\widehat{U}^\dagger(t)\,\widehat{U}(p), & \text{if } [\mu] = 0, \\ \widehat{U}(t)\,\widehat{Y}\,\widehat{U}^\dagger(t)\,\widehat{Y}, & \text{if } [\mu] = 1, \end{cases} \tag{2.2b}$$

where $\widehat{Y}$ is the nontrivial center of a Clifford algebra with an odd number of generators as must be the case when $[\mu] = 1$. Together with the index $[\mu] = 0, 1$ that specifies the parity of the number of boundary Majorana degrees of freedom, the triplet $([\nu], [\rho], [\mu]) \in \mathbb{Z}_2 \times \mathbb{Z}_2 \times \mathbb{Z}_2$ characterizes the eight IFT phases with the symmetry group $\mathbb{Z}^T \times \mathbb{Z}_2^F$.

TABLE I. The indices $([\nu], [\rho], [\mu])$ generated by stacking the indices $(0, 0, 1)$ $(c = 1)$ with itself form a $\mathbb{Z}_8$ under the stacking rules (2.4).

| $c$ | $([\nu], [\rho], [\mu])$ |
|-----|--------------------------|
| 1 | (0,0,1) |
| 2 | (0,1,0) |
| 3 | (1,1,1) |
| 4 | (1,0,0) |
| 5 | (1,0,1) |
| 6 | (1,1,0) |
| 7 | (0,1,1) |
| 8 | (0,0,0) |

Given two projective representations $\widehat{U}_1$ and $\widehat{U}_2$ of the group $G_f = \mathbb{Z}_2^T \times \mathbb{Z}_2^F$ acting on the Fock spaces $\mathfrak{F}_1$ and $\mathfrak{F}_2$, respectively, we now derive the indices associated with the projective representation $\hat{U}_\wedge$ acting on the Fock space $\mathfrak{F}_\wedge$ constructed from the graded tensor product of the Fock spaces $\mathfrak{F}_1$ and $\mathfrak{F}_2$. Using the stacking rules derived in the main text [Eq. (2.10)], we find

$$\nu_\wedge(t,t) = \begin{cases} \nu_1(t,t) + \nu_2(t,t) + \pi\,\rho_1(t)\,\rho_2(t), & \text{if } [\mu_\wedge] \equiv [\mu_1] + [\mu_2] = 0, \\ \nu_1(t,t) + \nu_2(t,t) + \pi\,\rho_1(t)\,\rho_2(t) + \pi\,\rho_1(t)\,\mathfrak{d}(t), & \text{if } [\mu_1] = 0,\ [\mu_2] = 1, \\ \nu_1(t,t) + \nu_2(t,t) + \pi\,\rho_1(t)\,\rho_2(t) + \pi\,\rho_2(t)\,\mathfrak{d}(t), & \text{if } [\mu_1] = 1,\ [\mu_2] = 0, \end{cases} \tag{2.3a}$$

for the value of the 2-cochain $\nu_\wedge(t,t)$, and

$$\rho_\wedge(t) = \begin{cases} \rho_1(t) + \rho_2(t) + \mathfrak{d}(t), & \text{if } [\mu_1] = 1,\ [\mu_2] = 1, \\ \rho_1(t) + \rho_2(t), & \text{otherwise,} \end{cases} \tag{2.3b}$$

for the value of the 1-cochain $\rho_\wedge(t)$. Assignments of indices $[\nu_1]$ and $[\rho_1]$ to the projective representations of the group $\mathbb{Z}_2^T \times \mathbb{Z}_2^F$ (as shown above) and the identity (2.3) imply that the indices of the tensor product representation are related to the indices of the constituent representations via

$$[\nu_\wedge] = \begin{cases} [\nu_1] + [\nu_2] + [\rho_1]\,[\rho_2], & \text{if } [\mu_\wedge] \equiv [\mu_1] + [\mu_2] = 0, \\ [\nu_1] + [\nu_2] + [\rho_1]\,[\rho_2] + [\rho_1], & \text{if } [\mu_1] = 0,\ [\mu_2] = 1, \\ [\nu_1] + [\nu_2] + [\rho_1]\,[\rho_2] + [\rho_2], & \text{if } [\mu_1] = 1,\ [\mu_2] = 0, \end{cases} \tag{2.4a}$$

for the value of the 2-cochain $\nu_\wedge(t,t)$, and

$$[\rho_\wedge] = \begin{cases} [\rho_1] + [\rho_2] + 1, & \text{if } [\mu_1] = 1,\ [\mu_2] = 1, \\ [\rho_1] + [\rho_2], & \text{otherwise,} \end{cases} \tag{2.4b}$$

where we have used the fact that $\mathfrak{d}(t) = 1$ since $\mathfrak{c}(t) = -1$. One thus finds that the one-dimensional IFT phases with symmetry group $\mathbb{Z}_2^T \times \mathbb{Z}_2^F$ form the cyclic group $\mathbb{Z}_8$ under the stacking rule (2.4). Without loss of generality, the generator of the group $\mathbb{Z}_8$ can be chosen as the IFT phase with indices $([\nu], [\rho], [\mu]) = (0, 0, 1)$. In Table I, the triplet $([\nu], [\rho], [\mu])$ for all elements of $\mathbb{Z}_8$ are computed using the stacking rules (2.4).

The one-dimensional BSPT phases protected by the group $\mathbb{Z}_2^T \times \mathbb{Z}_2$ are also classified by the second cohomology group (2.1) [4]. The two indices $[\nu] = 0, 1$ and $[\rho] = 0, 1$ generating $H^2(\mathbb{Z}_2^T \times \mathbb{Z}_2, \mathrm{U}(1)_\mathfrak{c})$ can be measured using Eq. (2.2) if we set $[\mu] = 0$ and make the identification between the fermion parity group $\mathbb{Z}_2^F$ and the cyclic group $\mathbb{Z}_2$ protecting the BSPT phases. There are three main differences as opposed to the fermionic case.

1. First, all bosonic invertible phases in one-dimension are BSPT phases. In particular, there is no analogue of the fermionic index $[\mu]$ for bosonic invertible phases.

2. Second, the cyclic subgroup $\mathbb{Z}_2$ for the BSPT phases is an ordinary $\mathbb{Z}_2$ as opposed to the fermion parity group $\mathbb{Z}_2^F$ in the sense that the former can be explicitly or spontaneously broken whereas the latter cannot.

3. Third, the stacking rules of BSPT phases follow the group composition rule of the second cohomology group (2.1), i.e., one-dimensional BSPT phases with $\mathbb{Z}_2^T \times \mathbb{Z}_2$ symmetry form the group $\mathbb{Z}_2 \times \mathbb{Z}_2$ under the stacking operation. In terms of the indices $[\nu]$ and $[\rho]$, we write

$$\left([\nu_\otimes], [\rho_\otimes]\right) \equiv ([\nu_1], [\rho_1]) \otimes ([\nu_2], [\rho_2]) \coloneqq ([\nu_1] + [\nu_2], [\rho_1] + [\rho_2]), \tag{2.5}$$

where $\otimes$ denotes the bosonic stacking operation.

## III. TIME-REVERSAL INVARIANT MAJORANA CHAINS

Let $c \in \mathbb{Z}$. We consider the family of Hamiltonians introduced in Ref. 1

$$\begin{aligned}
\widehat{H}_n &\coloneqq -\sum_{j=1}^{N-c} \left( \hat{c}_j^\dagger \hat{c}_{j+c} + \hat{c}_j^\dagger \hat{c}_{j+c}^\dagger + \text{H.c.} \right) \\
&= \sum_{j=1}^{N-c} \mathrm{i}\, \hat{\xi}_j\, \hat{\eta}_{j+c},
\end{aligned} \tag{3.1a}$$

where $\hat{c}_j$ and $\hat{c}_j^\dagger$ denote the annihilation and creation operators of spinless fermions and $\hat{\xi}_j$ and $\hat{\eta}_j$ denote the Majorana operators defined through the equations

$$\hat{c}_j =: \frac{1}{2}\left( \hat{\eta}_j + \mathrm{i}\hat{\xi}_j \right), \qquad \hat{c}_j^\dagger =: \frac{1}{2}\left( \hat{\eta}_j - \mathrm{i}\hat{\xi}_j \right), \qquad \hat{\eta}_j = \hat{\eta}_j^\dagger, \qquad \hat{\xi}_j = \hat{\xi}_j^\dagger, \tag{3.1b}$$

together with the algebra

$$\{\hat{c}_i, \hat{c}_j^\dagger\} = \delta_{ij} \qquad \{\hat{\eta}_i, \hat{\eta}_j\} = \{\hat{\xi}_i, \hat{\xi}_j\} = 2\delta_{ij}, \tag{3.1c}$$

with all other anticommutators vanishing. The bulk representation of the symmetry group $\mathbb{Z}_2^T \times \mathbb{Z}_2^F = \{e, t\} \times \{e, p\}$ is given by

$$\widehat{U}_{\text{bulk}}(t) \coloneqq \hat{\mathbb{1}}\, \mathsf{K}_{\text{bulk}}, \qquad \widehat{U}_{\text{bulk}}(p) \coloneqq \prod_j^N \left( 1 - 2\hat{c}_j^\dagger \hat{c}_j \right) = \prod_j^N \mathrm{i}\hat{\xi}_j \hat{\eta}_j. \tag{3.2a}$$

We demand that complex conjugation $\mathsf{K}_{\text{bulk}}$ acts on complex fermion operators $\hat{c}_j$ as

$$\mathsf{K}_{\text{bulk}}\, \hat{c}_j\, \mathsf{K}_{\text{bulk}} = \hat{c}_j, \qquad \mathsf{K}_{\text{bulk}}\, \hat{c}_j^\dagger\, \mathsf{K}_{\text{bulk}} = \hat{c}_j^\dagger, \tag{3.2b}$$

which implies

$$\widehat{U}_{\text{bulk}}(t)\, \hat{\eta}_j\, \widehat{U}_{\text{bulk}}^\dagger(t) = \mathsf{K}_{\text{bulk}}\, \hat{\eta}_j\, \mathsf{K}_{\text{bulk}} = \hat{\eta}_j, \qquad \widehat{U}_{\text{bulk}}(t)\, \hat{\xi}_j\, \widehat{U}_{\text{bulk}}^\dagger(t) = \mathsf{K}_{\text{bulk}}\, \hat{\xi}_j\, \mathsf{K}_{\text{bulk}} = -\hat{\xi}_j. \tag{3.2c}$$

In the family of Hamiltonians (3.1a) open boundary conditions are imposed. Consequently, the Majorana operators $\left\{\hat{\xi}_{N-c+1}, \hat{\xi}_{N-c+2}, \cdots, \hat{\xi}_N\right\}$ on the right and the Majorana operators $\{\hat{\eta}_1, \hat{\eta}_2, \cdots, \hat{\eta}_c\}$ on the left do not enter a Majorana $c$ open chain. These operators define the zero-energy Majorana degrees of freedom at the right- and left-boundaries, respectively.

We shall construct the boundary representations $\widehat{U}_L(g)$ and $\widehat{U}_R(g)$ that satisfy the compatibility conditions

$$\widehat{U}_{\text{bulk}}(g)\, \hat{\eta}_\alpha\, \widehat{U}_{\text{bulk}}^\dagger(g) = \widehat{U}_L(g)\, \hat{\eta}_\alpha\, \widehat{U}_L^\dagger(g), \qquad \widehat{U}_{\text{bulk}}(g)\, \hat{\xi}_\alpha\, \widehat{U}_{\text{bulk}}^\dagger(g) = \widehat{U}_R(g)\, \hat{\xi}_{N-\beta}\, \widehat{U}_R^\dagger(g), \tag{3.3}$$

for any $g \in \mathbb{Z}_2^T \times \mathbb{Z}_2^F$. To this end, we will construct the boundary representation for the case of $c = 1$. For all the remaining cases, we will stack the $c = 1$ representation with itself using the explicit form of the stacked representation for any $g \in \mathbb{Z}_2^T \times \mathbb{Z}_2^F$

$$\widehat{U}_\wedge(g) \coloneqq \begin{cases} \widehat{V}_1(g)\, \widehat{V}_2(g) \left[\widehat{U}_1(p)\right]^{\rho_2(g)} \left[\widehat{U}_2(p)\right]^{\rho_1(g)} \mathsf{K}_\wedge^{\epsilon(g)}, & \text{if } [\mu_1] = [\mu_2] = 0, \\ \widehat{V}_1(g)\, \widehat{V}_2(g)\, \widehat{Q}_2(g) \left[\widehat{U}_1(p)\, \hat{\gamma}_\infty^{(2)}\right]^{\rho_1(g)} \mathsf{K}_\wedge^{\epsilon(g)}, & \text{if } [\mu_1] = 0,\, [\mu_2] = 1, \\ (-\mathrm{i})^{\delta_{g,p}} \widehat{V}_1(g)\, \widehat{V}_2(g) \left[\widehat{U}_1(p)\right]^{\epsilon(g)+\rho_1(g)+\rho_2(g)} \left[\hat{\gamma}_{n_1}^{(1)}\right]^{q_1(g)+\rho_1(g)} \left[\hat{\gamma}_{n_2}^{(2)}\right]^{\epsilon(g)+q_2(g)+\rho_2(g)} \mathsf{K}_\wedge^{\epsilon(g)}, & \text{if } [\mu_1] = [\mu_2] = 1, \end{cases} \tag{3.4}$$

which is defined in the main text [Eq. (6.48)]. We will then compute the triplets $([\nu_L], [\rho_L], [\mu_L])$ and $([\nu_R], [\rho_R], [\mu_R])$ that are associated with the left and the right boundaries respectively. We will confirm that these indices are inverses of each other for any $c$ and follow the group structure in Table I dictated by the stacking rules (2.4).

### A. The case of $c = 1$

When $c = 1$, the sets of Majorana degrees of freedom at the left and the right boundaries are

$$\mathfrak{O}_\mathrm{L} = \{\hat{\eta}_1\}, \qquad \mathfrak{O}_\mathrm{R} = \left\{\hat{\xi}_N\right\}, \tag{3.5}$$

respectively. We choose the complex conjugation on the boundaries to act as

$$\mathsf{K}_\mathrm{L}\, \hat{\eta}_1\, \mathsf{K}_\mathrm{L} = +\hat{\eta}_1, \qquad \mathsf{K}_\mathrm{R}\, \hat{\xi}_N\, \mathsf{K}_\mathrm{R} = -\hat{\xi}_N. \tag{3.6}$$

The boundary representations that satisfy the conditions (3.3) are given by

$$\widehat{U}_\mathrm{L}(t) \coloneqq \mathsf{K}_\mathrm{L}, \qquad \widehat{U}_\mathrm{R}(t) \coloneqq \mathsf{K}_\mathrm{R}, \tag{3.7a}$$

$$\widehat{Y}_\mathrm{L} \coloneqq \hat{\eta}_1, \qquad \widehat{Y}_\mathrm{R} \coloneqq \hat{\xi}_N. \tag{3.7b}$$

Using the definitions (2.2) delivers the indices

$$([\nu_\mathrm{L}], [\rho_\mathrm{L}], [\mu_\mathrm{L}]) = (0, 0, 1), \qquad ([\nu_\mathrm{R}], [\rho_\mathrm{R}], [\mu_\mathrm{R}]) = (0, 1, 1), \tag{3.7c}$$

on the left and the right boundaries, respectively.

### B. The case of $c = 2$

When $c = 2$, the sets of Majorana degrees of freedom at the left and the right boundaries are

$$\mathfrak{O}_\mathrm{L} = \{\hat{\eta}_1, \hat{\eta}_2\}, \qquad \mathfrak{O}_\mathrm{R} = \left\{\hat{\xi}_{N-1}, \hat{\xi}_N\right\}, \tag{3.8}$$

respectively. We choose the complex conjugation on the boundaries to act as

$$\mathsf{K}_\mathrm{L}\, \hat{\eta}_\alpha\, \mathsf{K}_\mathrm{L} = (-1)^{\alpha+1}\hat{\eta}_\alpha, \qquad \mathsf{K}_\mathrm{R}\, \hat{\xi}_{N-\beta}\, \mathsf{K}_\mathrm{R} = (-1)^\beta \hat{\xi}_{N-\beta}, \qquad \alpha = 1, 2, \quad \beta = 0, 1. \tag{3.9}$$

Choosing $\widehat{U}_1(t)$ and $\widehat{U}_2(t)$ in Eq. (3.4) to be the $c = 1$ representations (3.7a), we obtain the $c = 2$ representations

$$\widehat{U}_\mathrm{L}(t) \coloneqq \hat{\eta}_1\, \mathsf{K}_\mathrm{L}, \qquad \widehat{U}_\mathrm{R}(t) \coloneqq \hat{\xi}_{N-1}\, \mathsf{K}_\mathrm{R}, \tag{3.10a}$$

$$\widehat{U}_\mathrm{L}(p) \coloneqq \mathrm{i}\hat{\eta}_2\, \hat{\eta}_1, \qquad \widehat{U}_\mathrm{L}(p) \coloneqq \mathrm{i}\hat{\xi}_{N-1}\, \hat{\xi}_N. \tag{3.10b}$$

Note that in the definition of the stacked operator (3.4) for odd-odd stacking, the operator $\gamma_{n_2}^{(2)}$ denotes the Majorana degree of freedom that is odd under stacked complex conjugation. In this case, these are the operators $\hat{\eta}_2$ and $\hat{\xi}_{N-1}$ according to Eq. (3.9). Using the definitions (2.2) delivers the indices

$$([\nu_\mathrm{L}], [\rho_\mathrm{L}], [\mu_\mathrm{L}]) = (0, 1, 0), \qquad ([\nu_\mathrm{R}], [\rho_\mathrm{R}], [\mu_\mathrm{R}]) = (1, 1, 0), \tag{3.10c}$$

on the left and the right boundaries, respectively.

### C. The case of $c = 3$

When $c = 3$, the sets of Majorana degrees of freedom at the left and the right boundaries are

$$\mathfrak{O}_\mathrm{L} = \{\hat{\eta}_1, \hat{\eta}_2, \hat{\eta}_3\}, \qquad \mathfrak{O}_\mathrm{R} = \left\{\hat{\xi}_{N-2}, \hat{\xi}_{N-1}, \hat{\xi}_N\right\}, \tag{3.11}$$

respectively. We choose the complex conjugation on the boundaries to act as

$$\mathsf{K}_\mathrm{L}\, \hat{\eta}_\alpha\, \mathsf{K}_\mathrm{L} = (-1)^{\alpha+1}\hat{\eta}_\alpha, \qquad \mathsf{K}_\mathrm{R}\, \hat{\xi}_{N-\beta}\, \mathsf{K}_\mathrm{R} = (-1)^\beta \hat{\xi}_{N-\beta}, \qquad \alpha = 1, 2, \quad \beta = 0, 1, \tag{3.12a}$$

$$\mathsf{K}_\mathrm{L}\, \hat{\eta}_3\, \mathsf{K}_\mathrm{L} = +\hat{\eta}_3, \qquad \mathsf{K}_\mathrm{R}\, \hat{\xi}_{N-2}\, \mathsf{K}_\mathrm{R} = -\hat{\xi}_{N-2}. \tag{3.12b}$$

Choosing $\widehat{U}_1(t)$ and $\widehat{U}_2(t)$ in Eq. (3.4) to be the $c = 2$ representation (3.10a) and the $c = 1$ representation (3.7a), respectively, we obtain the $c = 3$ representations

$$\widehat{U}_{\mathrm{L}}(t) := \hat{\eta}_2\,\hat{\xi}_{N-2}\,\mathsf{K}_{\mathrm{L}}, \qquad \widehat{U}_{\mathrm{R}}(t) := \hat{\xi}_N\,\hat{\eta}_3\,\mathsf{K}_{\mathrm{R}}, \tag{3.13a}$$

$$\widehat{Y}_{\mathrm{L}} := \mathrm{i}\hat{\eta}_3\,\hat{\eta}_2\,\hat{\eta}_1, \qquad \widehat{Y}_{\mathrm{R}} := \mathrm{i}\hat{\xi}_{N-2}\,\hat{\xi}_{N-1}\,\hat{\xi}_N. \tag{3.13b}$$

Using the definitions (2.2) delivers the indices

$$([\nu_{\mathrm{L}}], [\rho_{\mathrm{L}}], [\mu_{\mathrm{L}}]) = (1, 1, 1), \qquad ([\nu_{\mathrm{R}}], [\rho_{\mathrm{R}}], [\mu_{\mathrm{R}}]) = (1, 0, 1), \tag{3.13c}$$

on the left and the right boundaries, respectively.

### D. The case of $c = 4$

When $c = 4$, the sets of Majorana degrees of freedom at the left and the right boundaries are

$$\mathfrak{O}_{\mathrm{L}} = \{\hat{\eta}_1,\,\hat{\eta}_2,\,\hat{\eta}_3,\,\hat{\eta}_4\}, \qquad \mathfrak{O}_{\mathrm{R}} = \left\{\hat{\xi}_{N-3},\,\hat{\xi}_{N-2},\,\hat{\xi}_{N-1},\,\hat{\xi}_N\right\}, \tag{3.14}$$

respectively. We choose the complex conjugation on the boundaries to act as

$$\mathsf{K}_{\mathrm{L}}\,\hat{\eta}_\alpha\,\mathsf{K}_{\mathrm{L}} = (-1)^{\alpha+1}\hat{\eta}_\alpha, \qquad \mathsf{K}_{\mathrm{R}}\,\hat{\xi}_{N-\beta}\,\mathsf{K}_{\mathrm{R}} = (-1)^\beta\hat{\xi}_{N-\beta}, \qquad \alpha = 1,\cdots,4, \quad \beta = 0,\cdots,3, \tag{3.15a}$$

Choosing $\widehat{U}_1(t)$ and $\widehat{U}_2(t)$ in Eq. (3.4) to be the $c = 3$ representation (3.13a) and the $c = 1$ representation (3.7a), respectively, we obtain the $c = 4$ representations

$$\widehat{U}_{\mathrm{L}}(t) := \hat{\eta}_2\,\hat{\eta}_4\,\mathsf{K}_{\mathrm{L}}, \qquad \widehat{U}_{\mathrm{R}}(t) := \hat{\xi}_{N-2}\,\hat{\xi}_N\,\mathsf{K}_{\mathrm{R}}, \tag{3.16a}$$

$$\widehat{U}_{\mathrm{L}}(p) := \hat{\eta}_4\,\hat{\eta}_3\,\hat{\eta}_2\,\hat{\eta}_1, \qquad \widehat{U}_{\mathrm{L}}(p) := \hat{\xi}_{N-3}\,\hat{\xi}_{N-2}\,\hat{\xi}_{N-1}\,\hat{\xi}_N. \tag{3.16b}$$

Using the definitions (2.2) delivers the indices

$$([\nu_{\mathrm{L}}], [\rho_{\mathrm{L}}], [\mu_{\mathrm{L}}]) = (1, 0, 0), \qquad ([\nu_{\mathrm{R}}], [\rho_{\mathrm{R}}], [\mu_{\mathrm{R}}]) = (1, 0, 0), \tag{3.16c}$$

on the left and the right boundaries, respectively.

### E. The case of $c = 5$

When $c = 5$, the sets of Majorana degrees of freedom at the left and the right boundaries are

$$\mathfrak{O}_{\mathrm{L}} = \{\hat{\eta}_1,\,\hat{\eta}_2,\,\hat{\eta}_3,\,\hat{\eta}_4,\,\hat{\eta}_5\}, \qquad \mathfrak{O}_{\mathrm{R}} = \left\{\hat{\xi}_{N-4},\,\hat{\xi}_{N-3},\,\hat{\xi}_{N-2},\,\hat{\xi}_{N-1},\,\hat{\xi}_N\right\}, \tag{3.17}$$

respectively. We choose the complex conjugation on the boundaries to act as

$$\mathsf{K}_{\mathrm{L}}\,\hat{\eta}_\alpha\,\mathsf{K}_{\mathrm{L}} = (-1)^{\alpha+1}\hat{\eta}_\alpha, \qquad \mathsf{K}_{\mathrm{R}}\,\hat{\xi}_{N-\beta}\,\mathsf{K}_{\mathrm{R}} = (-1)^\beta\hat{\xi}_{N-\beta}, \qquad \alpha = 1,\cdots,4, \quad \beta = 0,\cdots,3, \tag{3.18a}$$

$$\mathsf{K}_{\mathrm{L}}\,\hat{\eta}_5\,\mathsf{K}_{\mathrm{L}} = +\hat{\eta}_5, \qquad \mathsf{K}_{\mathrm{R}}\,\hat{\xi}_{N-4}\,\mathsf{K}_{\mathrm{R}} = -\hat{\xi}_{N-4}. \tag{3.18b}$$

Choosing $\widehat{U}_1(t)$ and $\widehat{U}_2(t)$ in Eq. (3.4) to be the $c = 4$ representation (3.16a) and the $c = 1$ representation (3.7a), respectively, we obtain the $c = 5$ representations

$$\widehat{U}_{\mathrm{L}}(t) := \hat{\eta}_2\,\hat{\eta}_4\,\mathsf{K}_{\mathrm{L}}, \qquad \widehat{U}_{\mathrm{R}}(t) := \hat{\xi}_N\,\hat{\xi}_{N-2}\,\mathsf{K}_{\mathrm{R}}, \tag{3.19a}$$

$$\widehat{Y}_{\mathrm{L}} := \hat{\eta}_5\,\hat{\eta}_4\,\hat{\eta}_3\,\hat{\eta}_2\,\hat{\eta}_1, \qquad \widehat{Y}_{\mathrm{R}} := \hat{\xi}_{N-4}\,\hat{\xi}_{N-3}\,\hat{\xi}_{N-2}\,\hat{\xi}_{N-1}\,\hat{\xi}_N. \tag{3.19b}$$

Using the definitions (2.2) delivers the indices

$$([\nu_{\mathrm{L}}], [\rho_{\mathrm{L}}], [\mu_{\mathrm{L}}]) = (1, 0, 1), \qquad ([\nu_{\mathrm{R}}], [\rho_{\mathrm{R}}], [\mu_{\mathrm{R}}]) = (1, 1, 1), \tag{3.19c}$$

on the left and the right boundaries, respectively.

### F.  The case of $c = 6$

When $c = 6$, the sets of Majorana degrees of freedom at the left and the right boundaries are

$$\mathfrak{O}_\mathrm{L} = \{\hat{\eta}_1, \hat{\eta}_2, \hat{\eta}_3, \hat{\eta}_4, \hat{\eta}_5, \hat{\eta}_6\}, \qquad \mathfrak{O}_\mathrm{R} = \left\{\hat{\xi}_{N-5}, \hat{\xi}_{N-4}, \hat{\xi}_{N-3}, \hat{\xi}_{N-2}, \hat{\xi}_{N-1}, \hat{\xi}_N\right\}, \tag{3.20}$$

respectively. We choose the complex conjugation on the boundaries to act as

$$\mathsf{K}_\mathrm{L}\,\hat{\eta}_\alpha\,\mathsf{K}_\mathrm{L} = (-1)^{\alpha+1}\hat{\eta}_\alpha, \qquad \mathsf{K}_\mathrm{R}\,\hat{\xi}_{N-\beta}\,\mathsf{K}_\mathrm{R} = (-1)^\beta \hat{\xi}_{N-\beta}, \qquad \alpha = 1, \cdots, 6, \quad \beta = 0, \cdots, 5, \tag{3.21a}$$

Choosing $\widehat{U}_1(t)$ and $\widehat{U}_2(t)$ in Eq. (3.4) to be the $c = 5$ representation (3.19a) and the $c = 1$ representation (3.7a), respectively, we obtain the $c = 6$ representations

$$\widehat{U}_\mathrm{L}(t) \coloneqq \hat{\eta}_1\,\hat{\eta}_3\,\hat{\eta}_5\,\mathsf{K}_\mathrm{L}, \qquad \widehat{U}_\mathrm{R}(t) \coloneqq \hat{\xi}_{N-5}\,\hat{\xi}_{N-3}\,\hat{\xi}_{N-1}\,\mathsf{K}_\mathrm{R}, \tag{3.22a}$$

$$\widehat{U}_\mathrm{L}(p) \coloneqq \mathrm{i}\prod_{j=1}^{6}\hat{\eta}_{6-j+1}, \qquad \widehat{U}_\mathrm{L}(p) \coloneqq \mathrm{i}\prod_{j=1}^{6}\hat{\xi}_{N-6+j}. \tag{3.22b}$$

Using the definitions (2.2) delivers the indices

$$([\nu_\mathrm{L}], [\rho_\mathrm{L}], [\mu_\mathrm{L}]) = (1, 1, 0), \qquad ([\nu_\mathrm{R}], [\rho_\mathrm{R}], [\mu_\mathrm{R}]) = (0, 1, 0), \tag{3.22c}$$

on the left and the right boundaries, respectively.

### G.  The case of $c = 7$

When $c = 7$, the sets of Majorana degrees of freedom at the left and the right boundaries are

$$\mathfrak{O}_\mathrm{L} = \{\hat{\eta}_1, \hat{\eta}_2, \hat{\eta}_3, \hat{\eta}_4, \hat{\eta}_5, \hat{\eta}_6, \hat{\eta}_7\}, \qquad \mathfrak{O}_\mathrm{R} = \left\{\hat{\xi}_{N-6}, \hat{\xi}_{N-5}, \hat{\xi}_{N-4}, \hat{\xi}_{N-3}, \hat{\xi}_{N-2}, \hat{\xi}_{N-1}, \hat{\xi}_N\right\}, \tag{3.23}$$

respectively. We choose the complex conjugation on the boundaries to act as

$$\mathsf{K}_\mathrm{L}\,\hat{\eta}_\alpha\,\mathsf{K}_\mathrm{L} = (-1)^{\alpha+1}\hat{\eta}_\alpha, \qquad \mathsf{K}_\mathrm{R}\,\hat{\xi}_{N-\beta}\,\mathsf{K}_\mathrm{R} = (-1)^\beta \hat{\xi}_{N-\beta}, \qquad \alpha = 1, \cdots, 6, \quad \beta = 0, \cdots, 5, \tag{3.24a}$$

$$\mathsf{K}_\mathrm{L}\,\hat{\eta}_7\,\mathsf{K}_\mathrm{L} = +\hat{\eta}_7, \qquad \mathsf{K}_\mathrm{R}\,\hat{\xi}_{N-6}\,\mathsf{K}_\mathrm{R} = -\hat{\xi}_{N-6}. \tag{3.24b}$$

Choosing $\widehat{U}_1(t)$ and $\widehat{U}_2(t)$ in Eq. (3.4) to be the $c = 6$ representation (3.28a) and the $c = 1$ representation (3.7a), respectively, we obtain the $c = 7$ representations

$$\widehat{U}_\mathrm{L}(t) \coloneqq \hat{\eta}_2\,\hat{\eta}_4\,\hat{\eta}_6\,\hat{\xi}_{N-6}\,\mathsf{K}_\mathrm{L}, \qquad \widehat{U}_\mathrm{R}(t) \coloneqq \hat{\xi}_{N-4}\,\hat{\xi}_{N-2}\,\hat{\xi}_N\,\hat{\eta}_7\,\mathsf{K}_\mathrm{R}, \tag{3.25a}$$

$$\widehat{Y}_\mathrm{L} \coloneqq \mathrm{i}\prod_{j=1}^{7}\hat{\eta}_{7-j+1}, \qquad \widehat{Y}_\mathrm{R} \coloneqq \mathrm{i}\prod_{j=1}^{7}\hat{\xi}_{N-7+j}. \tag{3.25b}$$

Using the definitions (2.2) delivers the indices

$$([\nu_\mathrm{L}], [\rho_\mathrm{L}], [\mu_\mathrm{L}]) = (0, 1, 1), \qquad ([\nu_\mathrm{R}], [\rho_\mathrm{R}], [\mu_\mathrm{R}]) = (0, 0, 1), \tag{3.25c}$$

on the left and the right boundaries, respectively.

### H.  The case of $c = 8$

When $c = 8$, the sets of Majorana degrees of freedom at the left and the right boundaries are

$$\mathfrak{O}_\mathrm{L} = \{\hat{\eta}_1, \hat{\eta}_2, \hat{\eta}_3, \hat{\eta}_4, \hat{\eta}_5, \hat{\eta}_6, \hat{\eta}_7, \hat{\eta}_8\}, \qquad \mathfrak{O}_\mathrm{R} = \left\{\hat{\xi}_{N-7}, \hat{\xi}_{N-6}, \hat{\xi}_{N-5}, \hat{\xi}_{N-4}, \hat{\xi}_{N-3}, \hat{\xi}_{N-2}, \hat{\xi}_{N-1}, \hat{\xi}_N\right\}, \tag{3.26}$$

respectively. We choose the complex conjugation on the boundaries to act as

$$\mathsf{K}_{\mathrm{L}}\,\hat{\eta}_\alpha\,\mathsf{K}_{\mathrm{L}} = (-1)^{\alpha+1}\hat{\eta}_\alpha, \qquad \mathsf{K}_{\mathrm{R}}\,\hat{\xi}_{N-\beta}\,\mathsf{K}_{\mathrm{R}} = (-1)^\beta \hat{\xi}_{N-\beta}, \qquad \alpha = 1,\cdots,8, \quad \beta = 0,\cdots,7, \tag{3.27a}$$

Choosing $\widehat{U}_1(t)$ and $\widehat{U}_2(t)$ in Eq. (3.4) to be the $c = 7$ representation (3.25a) and the $c = 1$ representation (3.7a), respectively, we obtain the $c = 8$ representations

$$\widehat{U}_{\mathrm{L}}(t) \coloneqq \hat{\eta}_2\,\hat{\eta}_4\,\hat{\eta}_6\,\hat{\eta}_8\,\mathsf{K}_{\mathrm{L}}, \qquad \widehat{U}_{\mathrm{R}}(t) \coloneqq \hat{\xi}_{N-6}\,\hat{\xi}_{N-4}\,\hat{\xi}_{N-2}\,\hat{\xi}_N\,\mathsf{K}_{\mathrm{R}}, \tag{3.28a}$$

$$\widehat{U}_{\mathrm{L}}(p) \coloneqq \prod_{j=1}^{8} \hat{\eta}_{8-j+1}, \qquad \widehat{U}_{\mathrm{L}}(p) \coloneqq \prod_{j=1}^{8} \hat{\xi}_{N-8+j}. \tag{3.28b}$$

Using the definitions (2.2) delivers the indices

$$([\nu_{\mathrm{L}}], [\rho_{\mathrm{L}}], [\mu_{\mathrm{L}}]) = (0,0,0), \qquad ([\nu_{\mathrm{R}}], [\rho_{\mathrm{R}}], [\mu_{\mathrm{R}}]) = (0,0,0), \tag{3.28c}$$

on the left and the right boundaries, respectively. Note that the indices on both left and right-boundaries trivialize.

### I. Ground-state degeneracies for the symmetry class BDI

In this subsection, we shall calculate the protected ground-state degeneracy for the symmetry class BDI, for each value of the triplet $([\nu], [\rho], [\mu])$, see Table II.

#### 1. The case of $[\mu] = 0$ for the symmetry class BDI

The case $[\mu] = 0$ corresponds to the $c = 2, 4, 6, 8$ elements in the Table I, which form a $\mathbb{Z}_4$-group under the stacking rule (2.4). The index $[\nu_{\mathrm{B}}]$ dictates if the boundary representation $\widehat{U}_{\mathrm{B}}(t)$ squares to plus or minus the identity, as defined in Eq. (2.2). The index $[\rho_{\mathrm{B}}]$ characterizes the fermion parity of the representation $\widehat{U}_{\mathrm{B}}(t)$ of reversal of time, as defined in Eq. (2.2). Whenever $[\nu_{\mathrm{B}}] = 1$, it was shown in Sec. VIII of the main text that all states on the boundary $\Lambda_{\mathrm{B}}$ are at least twofold degenerate because of a quantum mechanical supersymmetry. Whenever $[\nu_{\mathrm{B}}] = 1$, there exists a twofold Kramer's degeneracy on the boundary $\Lambda_{\mathrm{B}}$. When $[\nu_{\mathrm{B}}] = [\rho_{\mathrm{B}}] = 1$, $\widehat{U}_{\mathrm{B}}(t)$ anticommutes with $\widehat{U}_{\mathrm{B}}(p)$, i.e., it is either the pair $\widehat{H}_{\mathrm{B}}$ and $\widehat{U}_{\mathrm{B}}(t)$ that can be simultaneously diagonalized with $\widehat{U}_{\mathrm{B}}(p)$ acting as a ladder operator or the pair $\widehat{H}_{\mathrm{B}}$ and $\widehat{U}_{\mathrm{B}}(p)$ that can be simultaneously diagonalized with $\widehat{U}_{\mathrm{B}}(t)$ acting as a ladder operator. Therefore, when $[\mu] = 0$, whenever the pair $([\nu_{\mathrm{B}}], [\rho_{\mathrm{B}}])$ is nontrivial, there exist a *twofold* protected degeneracy at the boundary $\Lambda_{\mathrm{B}}$. The ground state degeneracies are

$$\mathrm{GSD}_{\mathrm{L}}^{[\mu]=0} = 2^{[\nu_{\mathrm{L}}] + [\rho_{\mathrm{L}}] - [\nu_{\mathrm{L}}]\,[\rho_{\mathrm{L}}]}, \tag{3.29a}$$

$$\mathrm{GSD}_{\mathrm{R}}^{[\mu]=0} = 2^{[\nu_{\mathrm{R}}] + [\rho_{\mathrm{R}}] - [\nu_{\mathrm{R}}]\,[\rho_{\mathrm{R}}]}, \tag{3.29b}$$

$$\mathrm{GSD}_{\mathrm{bd}}^{[\mu]=0} = 2^{[\nu_{\mathrm{L}}] + [\rho_{\mathrm{L}}] - [\nu_{\mathrm{L}}]\,[\rho_{\mathrm{L}}]} \times 2^{[\nu_{\mathrm{R}}] + [\rho_{\mathrm{R}}] - [\nu_{\mathrm{R}}]\,[\rho_{\mathrm{R}}]}. \tag{3.29c}$$

We conclude that a one-dimensional nontrivial FSPT phase in the symmetry class BDI has *fourfold* protected ground-state degeneracy when open boundary conditions are imposed. Additional degeneracies are accidental.

#### 2. The case of $[\mu] = 1$ for the symmetry class BDI

The case $[\mu] = 1$ corresponds to the $c = 1, 3, 5, 7$ elements in the Table I. The index $[\nu_{\mathrm{B}}]$ dictates when the boundary representation $\widehat{U}_{\mathrm{B}}(t)$ squares to plus or minus identity as defined in Eq. (2.2). The index $[\rho_{\mathrm{B}}]$ dictates when the representation $\widehat{U}_{\mathrm{B}}(t)$ of reversal of time commutes or anticommutes with the central element $\widehat{Y}_{\mathrm{B}}$, as defined in Eq. (2.2). Whenever $[\nu_{\mathrm{B}}] = 1$, there exist a twofold Kramer's degeneracy at the boundary $\Lambda_{\mathrm{B}}$. Since by definition (4.10) in the main text, $\widehat{U}_{\mathrm{B}}(t)$ is constructed to have even fermion parity all Kramer's degenerate states carry the same fermion parity. When $[\rho_{\mathrm{B}}] = 1$, $\widehat{U}_{\mathrm{B}}(t)$ maps the eigenstates of $\widehat{Y}_{\mathrm{B}}$ with eigenvalues $\pm 1$ to the eigenstates with eigenvalues $\mp 1$. However, a nonzero $[\rho_{\mathrm{B}}]$ does not imply any additional protected degeneracy. Hence, any protected

TABLE II. The triplets $([\nu_B], [\rho_B], [\mu_B])$ on the left (B = L) and right (B = R) boundaries of a Majorana $c$ chain with $c = 1, \cdots, 8$. The last column GSD is the protected ground-state degeneracies defined in Eqs. (3.29) and (3.30) for $[\mu_B] = 0$ and $[\mu_B] = 1$, respectively.

| $c$ | $([\nu_L], [\rho_L], [\mu_L])$ | $([\nu_R], [\rho_R], [\mu_R])$ | GSD |
|-----|-----|-----|-----|
| 1 | $(0, 0, 1)$ | $(0, 1, 1)$ | 2 |
| 2 | $(0, 1, 0)$ | $(1, 1, 0)$ | 4 |
| 3 | $(1, 1, 1)$ | $(1, 0, 1)$ | 8 |
| 4 | $(1, 0, 0)$ | $(1, 0, 0)$ | 4 |
| 5 | $(1, 0, 1)$ | $(1, 1, 1)$ | 8 |
| 6 | $(1, 1, 0)$ | $(0, 1, 0)$ | 4 |
| 7 | $(0, 1, 1)$ | $(0, 0, 1)$ | 2 |
| 8 | $(0, 0, 0)$ | $(0, 0, 0)$ | 1 |

degeneracy in addition to that due to $[\mu] = 1$ comes from the nontrivial index $[\nu_B]$. The ground state degeneracies are

$$\mathrm{GSD}_L^{[\mu]=1} = 2 \times 2^{[\nu_L]}, \tag{3.30a}$$

$$\mathrm{GSD}_R^{[\mu]=1} = 2 \times 2^{[\nu_R]}, \tag{3.30b}$$

$$\mathrm{GSD}^{[\mu]=1} = \frac{1}{2} \times 2 \times 2^{[\nu_L]} \times 2 \times 2^{[\nu_R]} = 2 \times 2^{[\nu_L]} \times 2^{[\nu_R]}. \tag{3.30c}$$

Hence, we find that a one-dimensional nontrivial IFT phases in the symmetry class BDI with $c = 1, 3, 5, 7$ have protected ground-state degeneracy of 2, 8, 8, and 2, respectively, when open boundary conditions are imposed.

In Table II, we summarize the protected ground state degeneracies at each boundary and the total protected ground-state degeneracy of each one-dimensional IFT phase in symmetry class BDI.

## IV. SPIN-1/2 CLUSTER CHAINS

In this section, we consider a family of spin Hamiltonians that realize BSPT phases and are closely related to the Hamiltonians (3.1a) of the time-reversal symmetric Majorana chains. As was done in Sec. III, we will identify the set of gapless boundary degrees of freedom when open-boundary conditions are imposed and compute the associated indices that characterize the boundary projective representations.

We define the spin-1/2 cluster $c$ chains that are described by the family of Hamiltonians

$$\widehat{H}_c^{(b)} := -\sum_{j=1}^{2N-b|c|} \widehat{C}_j, \tag{4.1a}$$

where $b = 0, 1$ specifies the periodic ($b = 0$) or open ($b = 1$) boundary conditions and we have defined

$$\widehat{C}_j \coloneqq \begin{cases} \widehat{Z}_j \, \widehat{X}_{j+1} \cdots \widehat{X}_{j+c-1} \widehat{Z}_{j+c}, & \text{if } c > 0, \\ -\widehat{X}_j, & \text{if } c = 0, \\ \widehat{Y}_j \, \widehat{X}_{j+1} \cdots \widehat{X}_{j+|c|-1} \widehat{Y}_{j+|c|}, & \text{if } c < 0, \end{cases} \tag{4.1b}$$

with $\widehat{X}_j$, $\widehat{Y}_j$, and $\widehat{Z}_j$ the Pauli spin operators at site $j$ realizing the spin-1/2 representation of the $\mathfrak{su}(2)$ Lie algebra. The local terms $\widehat{C}_j$ are pairwise commuting and each has eigenvalues $\pm 1$. Therefore, the Hamiltonian (4.1) is a sum of pairwise commuting terms. We set the total number of sites to be even, i.e., $2N$.

The Hamiltonian (4.1) of the spin-1/2 cluster $c$ chains is related to the Hamiltonian (3.1a) of the time-reversal symmetric Majorana $c$ chains by the Jordan-Wigner (JW) transformation:

$$\hat{\eta}_j = \left( \prod_{i=1}^{j-1} \widehat{X}_i \right) \widehat{Z}_j, \qquad \hat{\xi}_i = \left( \prod_{i=1}^{j-1} \widehat{X}_i \right) \widehat{Y}_j, \qquad \mathrm{i}\hat{\xi}_j \, \hat{\eta}_j = -\widehat{X}_j, \tag{4.2a}$$

$$\widehat{Z}_j = \left( \prod_{i=1}^{j-1} \mathrm{i}\hat{\eta}_i \, \hat{\xi}_i \right) \hat{\eta}_j, \qquad \widehat{Y}_j = \left( \prod_{i=1}^{j-1} \mathrm{i}\hat{\eta}_i \, \hat{\xi}_i \right) \hat{\xi}_j, \qquad \widehat{X}_j = -\mathrm{i}\hat{\xi}_j \, \hat{\eta}_j. \tag{4.2b}$$

In what follows, we will first review the spectra and the internal symmetries of the Hamiltonian (4.1). We will then consider cluster chains with open boundary conditions and construct the boundary projective representations of the protecting symmetries. In doing so, we will also relate the boundary representations of the cluster chains to that of the Majorana chains that are obtained from JW transformation.

### A. Spectra and Internal Symmetries

We shall distinguish the cases of $c$ odd and $c$ even. For each cases, we shall single out an internal symmetry group of the spin-1/2 cluster $c$ chain.

#### 1. The case of $c$ even

When $c$ is an even integer, the Hamiltonian (4.1) has a nondegenerate and gapped ground state when periodic boundary conditions are imposed. This ground state is specified by being the eigenstate of each projector $\widehat{C}_j$ with the eigenvalue $+1$.

The Hamiltonian (4.1) is invariant under the global symmetry group $G_e = \mathbb{Z}_2^{T'} \times \mathbb{Z}_2 = \{e, t'\} \times \{e, g\}$ with the global representation

$$\widehat{U}_e(t') \coloneqq \hat{\mathbb{1}} \, \mathsf{K}, \qquad \widehat{U}_e(g) \coloneqq \prod_{j=1}^{2N} \widehat{X}_j, \tag{4.3a}$$

where $\hat{\mathbb{1}}$ is the identity operator and $\mathsf{K}$ is the complex conjugation. The representations (4.3a) act on the local spin-1/2 degrees of freedom as

$$\widehat{U}_e(t') \left( \widehat{X}_j \;\; \widehat{Y}_j \;\; \widehat{Z}_j \right)^{\mathsf{T}} \widehat{U}_e^{\dagger}(t') = \left( +\widehat{X}_j \;\; -\widehat{Y}_j \;\; +\widehat{Z}_j \right)^{\mathsf{T}}, \tag{4.3b}$$

$$\widehat{U}_e(g) \left( \widehat{X}_j \;\; \widehat{Y}_j \;\; \widehat{Z}_j \right)^{\mathsf{T}} \widehat{U}_e^{\dagger}(g) = \left( +\widehat{X}_j \;\; -\widehat{Y}_j \;\; -\widehat{Z}_j \right)^{\mathsf{T}}. \tag{4.3c}$$

When open boundary conditions are imposed, the Hamiltonian (4.1) may support gapless edge degrees of freedom that are protected by a boundary projective representation of the group $G_e$. Therefore, cluster $c$ chains with even $c$ realize the BSPT phases with $G_e$-symmetry. When JW transformation is applied on the Hamiltonian (4.1) with open boundary conditions, one obtains the Hamiltonian (3.1a). These are nothing but the FSPT phases (IFT phases with $[\mu] = 0$) in class BDI which has a $\mathbb{Z}_4$ classification. Under the JW transformation the operators $\widehat{U}_e(t)$ and $\widehat{U}_e(g)$ defined in Eq. (4.3a) are mapped to the reversal of time and fermion parity symmetries of the Majorana chains defined in Eq. (3.2), respectively.

### 2. The case of $c$ odd

When $c$ is an odd integer, the Hamiltonian (4.1) has a twofold degenerate ground-state manifold when periodic boundary conditions are imposed. This twofold degeneracy arises owing to the fact that product of all projectors $\widehat{C}_j$ is equal to the identity.

The Hamiltonian (4.1) is invariant under the global symmetry group $\mathrm{G_o} = \mathbb{Z}_2^T \times \mathbb{Z}_2^{T'} \times \mathbb{Z}_2 = \{e, t\} \times \{e, t'\} \times \{e, g\}$ with the global representations

$$\widehat{U}_{\mathrm{o}}(t) \coloneqq \left(\prod_{j=1}^{2N} \widehat{Y}_j\right) \mathsf{K}, \qquad \widehat{U}_{\mathrm{o}}(t') \coloneqq \widehat{\mathbb{1}}\, \mathsf{K}, \qquad \widehat{U}_{\mathrm{o}}(g) \coloneqq \prod_{j=1}^{2N} \widehat{X}_j. \tag{4.4a}$$

The actions of these generators on the local spin-1/2 degrees of freedom are

$$\widehat{U}_{\mathrm{o}}(t) \left(\widehat{X}_j \ \ \widehat{Y}_j \ \ \widehat{Z}_j\right)^{\mathsf{T}} \widehat{U}_{\mathrm{o}}^{\dagger}(t) = \left(-\widehat{X}_j \ \ -\widehat{Y}_j \ \ -\widehat{Z}_j\right)^{\mathsf{T}}, \tag{4.4b}$$

$$\widehat{U}_{\mathrm{o}}(t') \left(\widehat{X}_j \ \ \widehat{Y}_j \ \ \widehat{Z}_j\right)^{\mathsf{T}} \widehat{U}_{\mathrm{o}}^{\dagger}(t') = \left(+\widehat{X}_j \ \ -\widehat{Y}_j \ \ +\widehat{Z}_j\right)^{\mathsf{T}}, \tag{4.4c}$$

$$\widehat{U}_{\mathrm{o}}(g) \left(\widehat{X}_j \ \ \widehat{Y}_j \ \ \widehat{Z}_j\right)^{\mathsf{T}} \widehat{U}_{\mathrm{o}}^{\dagger}(g) = \left(+\widehat{X}_j \ \ -\widehat{Y}_j \ \ -\widehat{Z}_j\right)^{\mathsf{T}}. \tag{4.4d}$$

In the thermodynamic limit $N \to \infty$ at zero temperature, the symmetry group $G_o$ is spontaneously broken. Without loss of generality, we project on one of the symmetry breaking ground states and define the mean-field Hamiltonian

$$\widehat{H}_{\mathrm{MF},\, c \in 2\mathbb{Z}+1}^{(b)} \coloneqq -\sum_{j=1}^{2N-b(|c|-1)} \widehat{M}_j, \tag{4.5a}$$

where the monomial $\widehat{M}_j$ of odd order $|c|$ defined by

$$\widehat{M}_j \coloneqq \begin{cases} \widehat{Z}_j\, \widehat{Y}_{j+1}\, \widehat{Z}_{j+2}\, \widehat{Y}_{j+3}\, \cdots\, \widehat{Z}_{j+c-3}\, \widehat{Y}_{j+c-2}\, \widehat{Z}_{j+c-1}, & \text{if } c > 0, \\ \widehat{Y}_j\, \widehat{Z}_{j+1}\, \widehat{Y}_{j+2}\, \widehat{Z}_{j+3}\, \cdots\, \widehat{Y}_{j+|c|-3}\, \widehat{Z}_{j+|c|-2}\, \widehat{Y}_{j+|c|-1}, & \text{if } c < 0. \end{cases} \tag{4.5b}$$

This Hamiltonian has a nondegenerate and gapped ground state with closed boundary conditions (b=0). Hamiltonian (4.5a) breaks the $\mathbb{Z}_2^T$ symmetry under reversal of time for $\widehat{M}_j$ is odd under simultaneous sign reversal of all spin operators. In fact, we have the transformations laws

$$\widehat{U}_{\mathrm{o}}(t)\, \widehat{M}_j\, \widehat{U}_{\mathrm{o}}^{\dagger}(t) = -\widehat{M}_j, \qquad \widehat{U}_{\mathrm{o}}(t')\, \widehat{M}_j\, \widehat{U}_{\mathrm{o}}^{\dagger}(t') = (-1)^{\frac{c-1}{2}} \widehat{M}_j, \qquad \widehat{U}_{\mathrm{o}}(g)\, \widehat{M}_j\, \widehat{U}_{\mathrm{o}}^{\dagger}(g) = -\widehat{M}_j. \tag{4.6}$$

Nevertheless, by composing pairs of these three operations, it is possible to construct the subgroup

$$\mathrm{G_{MF}} = \widetilde{\mathbb{Z}}_2^{\tilde{T}'} \times \widetilde{\mathbb{Z}}_2, \qquad \widetilde{\mathbb{Z}}_2^{\tilde{T}'} \coloneqq \{e, \tilde{t}'\}, \qquad \widetilde{\mathbb{Z}}_2 \coloneqq \{e, \tilde{g}\}, \tag{4.7}$$

with the representations

$$\widehat{U}_{\mathrm{MF}}(\tilde{t}') \coloneqq \begin{cases} \widehat{\mathbb{1}}\, \mathsf{K}, & \text{if } (c-1)/2 = 0 \bmod 2, \\ \left(\prod_{j=1}^{2N} \widehat{Z}_j\right) \mathsf{K}, & \text{if } (c-1)/2 = 1 \bmod 2, \end{cases} \tag{4.8a}$$

$$\widehat{U}_{\mathrm{MF}}(\tilde{g}) \coloneqq \begin{cases} \prod_{j=1}^{2N} \widehat{Z}_j, & \text{if } (c-1)/2 = 0 \bmod 2, \\ \prod_{j=1}^{2N} \widehat{Y}_j, & \text{if } (c-1)/2 = 1 \bmod 2, \end{cases} \tag{4.8b}$$

such that

$$\widehat{U}_{\mathrm{MF}}(\tilde{t}') \left(\widehat{X}_j \ \ \widehat{Y}_j \ \ \widehat{Z}_j\right)^{\mathsf{T}} \widehat{U}_{\mathrm{MF}}^{\dagger}(\tilde{t}') = \begin{cases} \left(+\widehat{X}_j \ \ -\widehat{Y}_j \ \ +\widehat{Z}_j\right)^{\mathsf{T}}, & \text{if } (c-1)/2 = 0 \bmod 2, \\ \left(-\widehat{X}_j \ \ +\widehat{Y}_j \ \ +\widehat{Z}_j\right)^{\mathsf{T}}, & \text{if } (c-1)/2 = 1 \bmod 2, \end{cases} \tag{4.8c}$$

$$\widehat{U}_{\mathrm{MF}}(\tilde{g}) \left(\widehat{X}_j \ \ \widehat{Y}_j \ \ \widehat{Z}_j\right)^{\mathsf{T}} \widehat{U}_{\mathrm{MF}}^{\dagger}(\tilde{g}) = \begin{cases} \left(-\widehat{X}_j \ \ -\widehat{Y}_j \ \ +\widehat{Z}_j\right)^{\mathsf{T}}, & \text{if } (c-1)/2 = 0 \bmod 2, \\ \left(-\widehat{X}_j \ \ +\widehat{Y}_j \ \ -\widehat{Z}_j\right)^{\mathsf{T}}, & \text{if } (c-1)/2 = 1 \bmod 2, \end{cases} \tag{4.8d}$$

and

$$\widehat{M}_j = \widehat{U}_{\mathrm{MF}}(\tilde{t}')\,\widehat{M}_j\,\widehat{U}_{\mathrm{MF}}^\dagger(\tilde{t}') = \widehat{U}_{\mathrm{MF}}(\tilde{g})\,\widehat{M}_j\,\widehat{U}_{\mathrm{MF}}^\dagger(\tilde{g}), \qquad j = 1, \cdots, 2N. \tag{4.8e}$$

It follows that the group (4.7) is a symmetry group of the mean-field Hamiltonian (4.5a). After projecting the spin-1/2 cluster odd-$c$ Hamiltonian onto the mean-field Hamiltonian (4.5a), the spectrum and symmetries of the latter are the same and isomorphic, respectively, to those of the spin-1/2 cluster $[c - \mathrm{sgn}(c)]$ Hamiltonian. The consequences of the group cohomology that we study next are the same for the spin-1/2 cluster odd-$c$ Hamiltonian and the mean-field projection of the spin-1/2 cluster $[c - (-1)^{\lfloor \frac{c}{4} \rfloor} \mathrm{sgn}(c)]$ Hamiltonian.

### B. Symmetry fractionalization in spin-1/2 cluster $c$ chains

We are going to compute the indices classifying bosonic symmetry protected topological phases phases for the spin-1/2 cluster $c$ models defined in Hamiltonian (4.1). Without loss of generality, this is done as follows when $c$ is even [we consider the mean-field Hamiltonian (4.5a) when $c$ is odd] and the cardinality of the chain $\Lambda \coloneqq \{j = 1, \cdots, |\Lambda|\}$ is $|\Lambda|$.

1. When open boundary conditions are imposed, the cluster operators $\widehat{C}_{|\Lambda|-|c|+1}, \cdots, \widehat{C}_{|\Lambda|}$ are not present in Hamiltonian (4.1). Consequently, either on the left boundary

$$\Lambda_{\mathrm{L}} \coloneqq \{j = 1, 2, \cdots, |c|\} \tag{4.9}$$

or on the right boundary

$$\Lambda_{\mathrm{R}} \coloneqq \{j = 2N - |c| + 1, \cdots, 2N - 1, 2N\}, \tag{4.10}$$

the set of all operators commuting with $\widehat{H}_c^{(|c|)}$ is a Clifford algebra $\mathrm{C}\ell_{|c|}$ with $|c|$ generators represented by $2^{|c|}$-dimensional matrices acting on the Hilbert space $\mathfrak{h}_{\Lambda_{\mathrm{B}}} \coloneqq \mathbb{C}^{2^{|c|}}$ on either the left (B = L) or the right (B = R) boundary, respectively.

2. The Clifford algebra $\mathrm{C}\ell_{|c|}$ contains the Lie subalgebra

$$\underbrace{\mathfrak{su}(2) \oplus \cdots \oplus \mathfrak{su}(2)}_{|c| \text{ times}} \tag{4.11}$$

which is represented reducibly with $2^{|c|}$-dimensional matrices.

3. It is possible to represent the action of the protecting symmetries on either the left or right boundaries using the generators of the Lie subalgebra (4.11).

4. The same projective representation of the protecting symmetries is realized on either the left or right boundary.

5. We construct the boundary representations $\widehat{U}_{\mathrm{e,L}}$ on the boundary degrees of freedom by demanding the consistency condition

$$\widehat{U}_{\mathrm{e,L}}(g)\,\widehat{\mathfrak{S}}\,\widehat{U}_{\mathrm{e,L}}^\dagger(g) = \widehat{U}_e(g)\,\widehat{\mathfrak{S}}\,\widehat{U}_e^\dagger(g) \tag{4.12}$$

for any $g \in G_c$ and $\widehat{\mathfrak{S}} \in \mathfrak{su}(2) \oplus \cdots \oplus \mathfrak{su}(2)$. For the case of odd $c$, we shall use the global representations $\widehat{U}_{\mathrm{MF}}(g)$ of the group $G_{\mathrm{MF}}$.

We summarize the results of this exercise in Table III. It is instructive to compare the Table III with the Table II. We observe that the indices for cluster even-$c$ chains match the indices $([\nu_L], [\rho_L], 0)$ of the Majorana chains with $c$ even. On the other hand, the indices associated with the right boundary of the Majorana chains are not picked up by the corresponding cluster even-$c$ chains. This asymmetry is due to the fact that the JW transformation (4.2) is asymmetrical with respect to the left and the right boundaries [3]. The odd-$c$ cluster chains have the same indices as the even-$[c - (-1)^{\lfloor \frac{c}{4} \rfloor} \mathrm{sgn}(c)]$ chains. The protected ground state degeneracies of the cluster odd-$c$ chains match that of the Majorana $c = c$ chains if we also include the twofold degeneracy due to the spontaneous symmetry breaking.

TABLE III. The doublet $([\nu_{\mathrm{B}}], [\rho_{\mathrm{B}}])$ of a spin-1/2 cluster $c$ chain that defines the projective representation of the symmetry group (4.3) for even $c$ or (4.4) for odd $c$ that is realized on the left (B=L) or right (B=R) boundaries of an open chain. It is assumed that time-reversal symmetry is broken spontaneously when $c$ is odd. The column $\dim \mathfrak{h}_{\mathrm{B,min}}^{(c)}$ is the dimensionality of the smallest Hilbert space $\mathfrak{h}_{\mathrm{B,min}}^{(c)}$ for which it is possible to realize the projective algebra on either one of the left or right boundary. Squaring this number gives the topologically protected degeneracy $D(c)$ of the ground states of a spin-1/2 cluster $c$ open chain. The column $\dim \mathfrak{h}_{\mathrm{gs}}^{(c)}$ is the degeneracy of the ground state of the spin-1/2 cluster $c$ open chain that is protected by its integrability if spontaneous symmetry breaking is precluded.

| $c$ | $([\nu_{\mathrm{L}}], [\rho_{\mathrm{L}}])$ | $([\nu_{\mathrm{R}}], [\rho_{\mathrm{R}}])$ | $\dim \mathfrak{h}_{\mathrm{B,min}}^{(c)}$ | $D(c)$ | $\dim \mathfrak{h}_{\mathrm{gs}}^{(c)}$ |
|---|---|---|---|---|---|
| 0 | $(0,0)$ | $(0,0)$ | 1 | 1 | 1 |
| 1 | $(0,0)$ | $(0,0)$ | 1 | 1 | $2 = 2^1$ |
| 2 | $(0,1)$ | $(0,1)$ | 2 | 4 | $4 = 2^2$ |
| 3 | $(0,1)$ | $(0,1)$ | 2 | 4 | $8 = 2^3$ |
| 4 | $(1,0)$ | $(1,0)$ | 2 | 4 | $16 = 2^4$ |
| 5 | $(1,1)$ | $(1,1)$ | 2 | 4 | $32 = 2^5$ |
| 6 | $(1,1)$ | $(1,1)$ | 2 | 4 | $64 = 2^6$ |
| 7 | $(0,0)$ | $(0,0)$ | 1 | 1 | $128 = 2^7$ |
| 8 | $(0,0)$ | $(0,0)$ | 1 | 1 | $256 = 2^8$ |

### C. The case of $c$ even

We treat Hamiltonian (4.5a) with open boundary conditions $b = 1$ for $c = 0, \pm 2, +4$ and $c = -4$.

#### 1. The case of $c = 0$

When $c = 0$, Hamiltonian (4.1) with open boundary conditions becomes

$$\widehat{H}_0^{(1)} := -\sum_{j=1}^{2N} \widehat{X}_j = \widehat{H}_0^{(0)}, \tag{4.13}$$

which has a nondegenerate and gapped ground state. All $\widehat{X}_j$ with $j = 1, \cdots, 2N$ are present in $\widehat{H}_0^{(1)}$ and commute pairwise. In other words, the set of gapless boundary degrees of freedom is the empty set

$$\mathfrak{O}_{\mathrm{B},0} = \{\,\}, \qquad \mathrm{B = L, R}. \tag{4.14}$$

By convention, we associate the trivial indices

$$([\nu], [\rho]) = (0,0) \tag{4.15}$$

to the spin-1/2 cluster $c = 0$ chain.

#### 2. The case of $c = +2$

When $c = +2$, the choice of open boundary conditions implies that Hamiltonian

$$\widehat{H}_2^{(2)} = \sum_{j=1}^{2N-2} \widehat{Z}_j \, \widehat{X}_{j+1} \, \widehat{Z}_{j+2} \tag{4.16}$$

has a $2^2 = 4$-fold degenerate ground state separated from all excited states by a gap. Since open boundary conditions are selected, the pair of commuting operators $\widehat{Z}_{2N-1} \, \widehat{X}_{2N} \, \widehat{Z}_1$, $\widehat{Z}_{2N} \, \widehat{X}_1 \, \widehat{Z}_2$, are present in $\widehat{H}_2^{(0)}$ but are absent in $\widehat{H}_2^{(1)}$.

The set of all operators that commute with the Hamiltonian (4.16) and have support on the left boundary is the Clifford algebra $\mathrm{C}\ell_2$ spanned by the generators

$$\left\{ \widehat{X}_1\,\widehat{Z}_2, \quad \widehat{Z}_1 \right\}. \tag{4.17}$$

We define define the triplet

$$\widehat{S}_\mathrm{L}^x := \frac{1}{2}\,\widehat{X}_1\,\widehat{Z}_2, \quad \widehat{S}_\mathrm{L}^y := \frac{1}{2}\,\widehat{Y}_1\,\widehat{Z}_2, \quad \widehat{S}_\mathrm{L}^z := \frac{1}{2}\,\widehat{Z}_1, \tag{4.18}$$

of operators that obey the $\mathfrak{su}(2)$ Lie algebra. We deduce the transformation laws

$$\widehat{U}_\mathrm{e}(t')\,\left(\widehat{S}_\mathrm{L}^x\ \ \widehat{S}_\mathrm{L}^y\ \ \widehat{S}_\mathrm{L}^z\right)^\top\widehat{U}_\mathrm{e}^\dagger(t') = \left(+\widehat{S}_\mathrm{L}^x\ -\widehat{S}_\mathrm{L}^y\ +\widehat{S}_\mathrm{L}^z\right)^\top, \quad \widehat{U}_\mathrm{e}(g)\,\left(\widehat{S}_\mathrm{L}^x\ \ \widehat{S}_\mathrm{L}^y\ \ \widehat{S}_\mathrm{L}^z\right)^\top\widehat{U}_\mathrm{e}^\dagger(g) = \left(-\widehat{S}_\mathrm{L}^x\ +\widehat{S}_\mathrm{L}^y\ -\widehat{S}_\mathrm{L}^z\right)^\top. \tag{4.19}$$

We define the action $\mathsf{K}_\mathrm{L}$ of complex conjugation on the four-dimensional representation (4.18) of the $\mathfrak{su}(2)$ Lie algebra by demanding that

$$\mathsf{K}_\mathrm{L}\,\widehat{S}_\mathrm{L}^x\,\mathsf{K}_\mathrm{L} := +\widehat{S}_\mathrm{L}^x, \qquad \mathsf{K}_\mathrm{L}\,\widehat{S}_\mathrm{L}^y\,\mathsf{K}_\mathrm{L} := -\widehat{S}_\mathrm{L}^y, \qquad \mathsf{K}_\mathrm{L}\,\widehat{S}_\mathrm{L}^z\,\mathsf{K}_\mathrm{L} := +\widehat{S}_\mathrm{L}^z. \tag{4.20a}$$

Demanding the consistency condition (4.12) to hold, we find the boundary representation

$$\widehat{U}_\mathrm{e,L}(t') := \mathbb{1}_\mathrm{L}\,\mathsf{K}_\mathrm{L}, \qquad \widehat{U}_\mathrm{e,L}(g) := \widehat{S}_\mathrm{L}^y. \tag{4.20b}$$

Using the definition (2.2), we associate the pair of indices

$$([\nu_\mathrm{L}], [\rho_\mathrm{L}]) = (0, 1) \tag{4.20c}$$

to the spin-1/2 cluster $c = +2$ chain.

One verifies that the Clifford algebra spanned by the generators

$$\left\{ \widehat{Z}_{2N-1}\,\widehat{X}_{2N}, \quad \widehat{Z}_{2N} \right\} \tag{4.21a}$$

for the right boundary $\Lambda_\mathrm{R} := \{j = 2N-1, 2N\}$ delivers the pair of indices

$$([\nu_\mathrm{R}], [\rho_\mathrm{R}]) = (0, 1) \tag{4.21b}$$

for the projective representation of $t'$ and $g$ on the right boundary.

### 3. The case of $c = -2$

The case $c = -2$ is deduced from the case $c = +2$ by interchanging all the $\widehat{Z}_j$ and $\widehat{Y}_j$ operators for $j = 1, \cdots, 2N$. The set $\mathfrak{O}_\mathrm{L}$ of all operators that commute with the Hamiltonian $\widehat{H}_{-2}^{(2)}$ and have support on the boundary $\Lambda_\mathrm{L} := \{j = 1, 2\}$ is the Clifford algebra $\mathrm{C}\ell_2$ spanned by the generators

$$\left\{ \widehat{X}_1\,\widehat{Y}_2, \quad \widehat{Y}_1 \right\}. \tag{4.22}$$

The Clifford algebra with the generators (4.22) contains the $\mathfrak{su}(2)$ Lie algebra generated by the operators

$$\widehat{S}_\mathrm{L}^x := \frac{1}{2}\,\widehat{X}_1\,\widehat{Y}_2, \qquad \widehat{S}_\mathrm{L}^y := -\frac{1}{2}\,\widehat{Z}_1\,\widehat{Y}_2, \qquad \widehat{S}_\mathrm{L}^z := \frac{1}{2}\,\widehat{Y}_1. \tag{4.23}$$

We deduce the transformation laws

$$\widehat{U}_\mathrm{e}(t')\,\left(\widehat{S}_\mathrm{L}^x\ \ \widehat{S}_\mathrm{L}^y\ \ \widehat{S}_\mathrm{L}^z\right)^\top\widehat{U}_\mathrm{e}^\dagger(t') = \left(-\widehat{S}_\mathrm{L}^x\ -\widehat{S}_\mathrm{L}^y\ -\widehat{S}_\mathrm{L}^z\right)^\top, \quad \widehat{U}_\mathrm{e}(g)\,\left(\widehat{S}_\mathrm{L}^x\ \ \widehat{S}_\mathrm{L}^y\ \ \widehat{S}_\mathrm{L}^z\right)^\top\widehat{U}_\mathrm{e}^\dagger(g) = \left(-\widehat{S}_\mathrm{L}^x\ +\widehat{S}_\mathrm{L}^y\ -\widehat{S}_\mathrm{L}^z\right)^\top. \tag{4.24a}$$

We define the action $\mathsf{K}_\mathrm{L}$ of complex conjugation on the four-dimensional representation (4.23) of the $\mathfrak{su}(2)$ Lie algebra by demanding that

$$\mathsf{K}_\mathrm{L}\,\widehat{S}_\mathrm{L}^x\,\mathsf{K}_\mathrm{L} := +\widehat{S}_\mathrm{L}^x, \qquad \mathsf{K}_\mathrm{L}\,\widehat{S}_\mathrm{L}^y\,\mathsf{K}_\mathrm{L} := -\widehat{S}_\mathrm{L}^y, \qquad \mathsf{K}_\mathrm{L}\,\widehat{S}_\mathrm{L}^z\,\mathsf{K}_\mathrm{L} := +\widehat{S}_\mathrm{L}^z. \tag{4.25a}$$

Demanding the consistency condition (4.12) to hold, we find the boundary representation

$$\widehat{U}_{\mathrm{e,L}}(t') \coloneqq \widehat{S}_{\mathrm{L}}^{y}\, \mathsf{K}_{\mathrm{L}}, \qquad \widehat{U}_{\mathrm{e,L}}(g) \coloneqq \widehat{S}_{\mathrm{L}}^{y}. \tag{4.25b}$$

Using the definition (2.2), we associate the pair of indices

$$([\nu_{\mathrm{L}}], [\rho_{\mathrm{L}}]) = (1, 1) \tag{4.25c}$$

to the spin-1/2 cluster $c = -2$ chain. One verifies that the Clifford algebra spanned by the generators

$$\left\{ \widehat{Y}_{2N-1}\, \widehat{X}_{2N}, \quad \widehat{Y}_{2N} \right\} \tag{4.26a}$$

for the right boundary $\Lambda_{\mathrm{R}} \coloneqq \{ j = 2N - 1, 2N \}$ delivers the pair of indices

$$([\nu_{\mathrm{R}}], [\rho_{\mathrm{R}}]) = (1, 1) \tag{4.26b}$$

for the projective representation of $t'$ and $g$ on the right boundary.

### 4. The case of $c = +4$

When $c = +4$, the choice of open boundary conditions implies that Hamiltonian

$$\widehat{H}_{4}^{(1)} = \sum_{j=1}^{2N-4} \widehat{Z}_{j}\, \widehat{X}_{j+1}\, \widehat{X}_{j+2}\, \widehat{X}_{j+3}\, \widehat{Z}_{j+4} \tag{4.27}$$

has a $2^{4} = 16$-fold degenerate ground state separated from all excited states by a gap.

Since open boundary conditions are selected, the quadruplet of pairwise commuting operators

$$\begin{aligned} \widehat{Z}_{2N-3}\, \widehat{X}_{2N-2}\, \widehat{X}_{2N-1}\, \widehat{X}_{2N}\, \widehat{Z}_{1}, \qquad & \widehat{Z}_{2N-2}\, \widehat{X}_{2N-1}\, \widehat{X}_{2N}\, \widehat{X}_{1}\, \widehat{Z}_{2}, \\ \widehat{Z}_{2N-1}\, \widehat{X}_{2N}\, \widehat{X}_{1}\, \widehat{X}_{2}\, \widehat{Z}_{3}, \qquad & \widehat{Z}_{2N}\, \widehat{X}_{1}\, \widehat{X}_{2}\, \widehat{X}_{3}\, \widehat{Z}_{4}, \end{aligned} \tag{4.28}$$

are present in $\widehat{H}_{4}^{(0)}$ but are absent in $\widehat{H}_{4}^{(1)}$. The set $\mathfrak{O}_{\mathrm{L}}$ of all operators that commute with Hamiltonian (4.27) and have support on the boundary $\Lambda_{\mathrm{L}} \coloneqq \{ j = 1, 2, 3, 4 \}$ is the Clifford algebra $\mathrm{C}\ell_{4}$ spanned by the generators

$$\left\{ \widehat{X}_{1}\, \widehat{X}_{2}\, \widehat{X}_{3}\, \widehat{Z}_{4}, \quad \widehat{X}_{1}\, \widehat{X}_{2}\, \widehat{Z}_{3}, \quad \widehat{X}_{1}\, \widehat{Z}_{2}, \quad \widehat{Z}_{1} \right\}. \tag{4.29}$$

We define the pair of triplets

$$\widehat{S}_{\mathrm{L}}^{x} \coloneqq \frac{1}{2}\, \widehat{X}_{1}\, \widehat{X}_{2}\, \widehat{X}_{3}\, \widehat{Z}_{4}, \quad \widehat{S}_{\mathrm{L}}^{y} \coloneqq -\frac{1}{2}\, \widehat{X}_{1}\, \widehat{X}_{2}\, \widehat{Z}_{3}, \quad \widehat{S}_{\mathrm{L}}^{z} \coloneqq \frac{1}{2}\, \widehat{Y}_{3}\, \widehat{Z}_{4}, \tag{4.30a}$$

$$\widehat{J}_{\mathrm{L}}^{x} \coloneqq \frac{1}{2}\, \widehat{X}_{1}\, \widehat{Z}_{2}\, \widehat{Y}_{3}\, \widehat{Z}_{4}, \quad \widehat{J}_{\mathrm{L}}^{y} \coloneqq -\frac{1}{2}\, \widehat{Z}_{1}\, \widehat{Y}_{3}\, \widehat{Z}_{4}, \quad \widehat{J}_{\mathrm{L}}^{z} \coloneqq \frac{1}{2}\, \widehat{Y}_{1}\, \widehat{Z}_{2} \tag{4.30b}$$

of operators that obey the $\mathfrak{su}(2) \oplus \mathfrak{su}(2)$ Lie algebra. We deduce the transformation laws

$$\widehat{U}_{\mathrm{e}}(t') \left( \widehat{S}_{\mathrm{L}}^{x}\ \ \widehat{S}_{\mathrm{L}}^{y}\ \ \widehat{S}_{\mathrm{L}}^{z} \right)^{\mathsf{T}} \widehat{U}_{\mathrm{e}}^{\dagger}(t') = \left( +\widehat{S}_{\mathrm{L}}^{x}\ \ +\widehat{S}_{\mathrm{L}}^{y}\ \ -\widehat{S}_{\mathrm{L}}^{z} \right)^{\mathsf{T}}, \quad \widehat{U}_{\mathrm{e}}(g) \left( \widehat{S}_{\mathrm{L}}^{x}\ \ \widehat{S}_{\mathrm{L}}^{y}\ \ \widehat{S}_{\mathrm{L}}^{z} \right)^{\mathsf{T}} \widehat{U}_{\mathrm{e}}^{\dagger}(g) = \left( -\widehat{S}_{\mathrm{L}}^{x}\ \ -\widehat{S}_{\mathrm{L}}^{y}\ \ +\widehat{S}_{\mathrm{L}}^{z} \right)^{\mathsf{T}}, \tag{4.31a}$$

and

$$\widehat{U}_{\mathrm{e}}(t') \left( \widehat{J}_{\mathrm{L}}^{x}\ \ \widehat{J}_{\mathrm{L}}^{y}\ \ \widehat{J}_{\mathrm{L}}^{z} \right)^{\mathsf{T}} \widehat{U}_{\mathrm{e}}^{\dagger}(t') = \left( -\widehat{J}_{\mathrm{L}}^{x}\ \ -\widehat{J}_{\mathrm{L}}^{y}\ \ -\widehat{J}_{\mathrm{L}}^{z} \right)^{\mathsf{T}}, \quad \widehat{U}_{\mathrm{e}}(g) \left( \widehat{J}_{\mathrm{L}}^{x}\ \ \widehat{J}_{\mathrm{L}}^{y}\ \ \widehat{J}_{\mathrm{L}}^{z} \right)^{\mathsf{T}} \widehat{U}_{\mathrm{e}}^{\dagger}(g) = \left( -\widehat{J}_{\mathrm{L}}^{x}\ \ -\widehat{J}_{\mathrm{L}}^{y}\ \ +\widehat{J}_{\mathrm{L}}^{z} \right)^{\mathsf{T}}. \tag{4.31b}$$

We define the action $\mathsf{K}_{\mathrm{L}}$ of complex conjugation on the four-dimensional representation (4.30) of the $\mathfrak{su}(2) \oplus \mathfrak{su}(2)$ Lie algebra by demanding that

$$\mathsf{K}_{\mathrm{L}}\, \widehat{S}_{\mathrm{L}}^{x}\, \mathsf{K}_{\mathrm{L}} \coloneqq +\widehat{S}_{\mathrm{L}}^{x}, \qquad \mathsf{K}_{\mathrm{L}}\, \widehat{S}_{\mathrm{L}}^{y}\, \mathsf{K}_{\mathrm{L}} \coloneqq -\widehat{S}_{\mathrm{L}}^{y}, \qquad \mathsf{K}_{\mathrm{L}}\, \widehat{S}_{\mathrm{L}}^{z}\, \mathsf{K}_{\mathrm{L}} \coloneqq +\widehat{S}_{\mathrm{L}}^{z}, \tag{4.32a}$$

$$\mathsf{K}_{\mathrm{L}}\, \widehat{J}_{\mathrm{L}}^{x}\, \mathsf{K}_{\mathrm{L}} \coloneqq -\widehat{J}_{\mathrm{L}}^{x}, \qquad \mathsf{K}_{\mathrm{L}}\, \widehat{J}_{\mathrm{L}}^{y}\, \mathsf{K}_{\mathrm{L}} \coloneqq +\widehat{J}_{\mathrm{L}}^{y}, \qquad \mathsf{K}_{\mathrm{L}}\, \widehat{J}_{\mathrm{L}}^{z}\, \mathsf{K}_{\mathrm{L}} \coloneqq +\widehat{J}_{\mathrm{L}}^{z}. \tag{4.32b}$$

Demanding the consistency condition (4.12) to hold, we find the boundary representation

$$\widehat{U}_{\mathrm{e,L}}(t') \coloneqq \widehat{S}_{\mathrm{L}}^x \, \widehat{J}_{\mathrm{L}}^x \, \mathsf{K}_{\mathrm{L}}, \qquad \widehat{U}_{\mathrm{e,L}}(g) \coloneqq \widehat{S}_{\mathrm{L}}^z \, \widehat{J}_{\mathrm{L}}^z. \tag{4.32c}$$

Using the definition (2.2), we associate the pair of indices

$$([\nu_{\mathrm{L}}], [\rho_{\mathrm{L}}]) = (1,0) \tag{4.32d}$$

to the spin-1/2 cluster $c = +4$ chain. One verifies that the Clifford algebra spanned by the generators

$$\left\{ \widehat{Z}_{2N-3} \, \widehat{X}_{2N-2} \, \widehat{X}_{2N-1} \, \widehat{X}_{2N}, \quad \widehat{Z}_{2N-2} \, \widehat{X}_{2N-1} \, \widehat{X}_{2N}, \quad \widehat{Z}_{2N-1} \, \widehat{X}_{2N}, \quad \widehat{Z}_{2N} \right\} \tag{4.33a}$$

for the right boundary $\Lambda_{\mathrm{R}} \coloneqq \{j = 2N-3, 2N-2, 2N-1, 2N\}$ delivers the pair of indices

$$([\nu_{\mathrm{R}}], [\rho_{\mathrm{R}}]) = (1,0) \tag{4.33b}$$

for the projective representation of $t'$ and $g$ on the right boundary.

### 5. *The case of $c = -4$*

The case $c = -4$ is deduced from the case $c = +4$ by interchanging all the $\widehat{Z}_j$ and $\widehat{Y}_j$ operators for $j = 1, \cdots, 2N$. The set $\mathfrak{O}_{\mathrm{L}}$ of all operators that commute with the Hamiltonian $\widehat{H}_{-4}^{(4)}$ and have support on the boundary

$$\Lambda_{\mathrm{L}} \coloneqq \{j = 1, 2, 3, 4\} \tag{4.34a}$$

is the Clifford algebra $\mathrm{C}\ell_4$ spanned by the generators

$$\{\widehat{Y}_1, \quad \widehat{X}_1 \, \widehat{Y}_2, \quad \widehat{X}_1 \, \widehat{X}_2 \, \widehat{Y}_3, \quad \widehat{X}_1 \, \widehat{X}_2 \, \widehat{X}_3 \, \widehat{Y}_4 \}. \tag{4.34b}$$

The Clifford algebra with the generators (4.34b) contains the $\mathfrak{su}(2) \oplus \mathfrak{su}(2)$ Lie algebra generated by the operators

$$\widehat{S}_{\mathrm{L}}^x \coloneqq \tfrac{1}{2} \widehat{X}_1 \, \widehat{X}_2 \, \widehat{X}_3 \, \widehat{Y}_4, \quad \widehat{S}_{\mathrm{L}}^y \coloneqq \tfrac{1}{2} \widehat{X}_1 \, \widehat{X}_2 \, \widehat{Y}_3, \qquad \widehat{S}_{\mathrm{L}}^z \coloneqq \tfrac{1}{2} \widehat{Z}_3 \, \widehat{Y}_4, \tag{4.35}$$

$$\widehat{J}_{\mathrm{L}}^x \coloneqq \tfrac{1}{2} \widehat{X}_1 \, \widehat{Y}_2 \, \widehat{Z}_3 \, \widehat{Y}_4, \qquad \widehat{J}_{\mathrm{L}}^y \coloneqq \tfrac{1}{2} \widehat{Y}_1 \, \widehat{Z}_3 \, \widehat{Y}_4, \qquad \widehat{J}_{\mathrm{L}}^z \coloneqq \tfrac{1}{2} Z_1 \, Y_2. \tag{4.36}$$

We deduce the transformation laws

$$\widehat{U}_{\mathrm{e}}(t') \left( \widehat{S}_{\mathrm{L}}^x \ \widehat{S}_{\mathrm{L}}^y \ \widehat{S}_{\mathrm{L}}^z \right)^{\mathsf{T}} \widehat{U}_{\mathrm{e}}^\dagger(t') = \left( -\widehat{S}_{\mathrm{L}}^x \ -\widehat{S}_{\mathrm{L}}^y \ -\widehat{S}_{\mathrm{L}}^z \right)^{\mathsf{T}}, \quad \widehat{U}_{\mathrm{e}}(g) \left( \widehat{S}_{\mathrm{L}}^x \ \widehat{S}_{\mathrm{L}}^y \ \widehat{S}_{\mathrm{L}}^z \right)^{\mathsf{T}} \widehat{U}_{\mathrm{e}}^\dagger(g) = \left( -\widehat{S}_{\mathrm{L}}^x \ -\widehat{S}_{\mathrm{L}}^y \ +\widehat{S}_{\mathrm{L}}^z \right)^{\mathsf{T}}, \tag{4.37a}$$

and

$$\widehat{U}_{\mathrm{e}}(t') \left( \widehat{J}_{\mathrm{L}}^x \ \widehat{J}_{\mathrm{L}}^y \ \widehat{J}_{\mathrm{L}}^z \right)^{\mathsf{T}} \widehat{U}_{\mathrm{e}}^\dagger(t') = \left( +\widehat{J}_{\mathrm{L}}^x \ +\widehat{J}_{\mathrm{L}}^y \ -\widehat{J}_{\mathrm{L}}^z \right)^{\mathsf{T}}, \quad \widehat{U}_{\mathrm{e}}(g) \left( \widehat{J}_{\mathrm{L}}^x \ \widehat{J}_{\mathrm{L}}^y \ \widehat{J}_{\mathrm{L}}^z \right)^{\mathsf{T}} \widehat{U}_{\mathrm{e}}^\dagger(g) = \left( -\widehat{J}_{\mathrm{L}}^x \ -\widehat{J}_{\mathrm{L}}^y \ +\widehat{J}_{\mathrm{L}}^z \right)^{\mathsf{T}}. \tag{4.37b}$$

We define the action $\mathsf{K}_{\mathrm{L}}$ of complex conjugation on the sixteen-dimensional representation (4.23) of the $\mathfrak{su}(2) \oplus \mathfrak{su}(2)$ Lie algebra by demanding that

$$\mathsf{K}_{\mathrm{L}} \, \widehat{S}_{\mathrm{L}}^x \, \mathsf{K}_{\mathrm{L}} \coloneqq -\widehat{S}_{\mathrm{L}}^x, \qquad \mathsf{K}_{\mathrm{L}} \, \widehat{S}_{\mathrm{L}}^y \, \mathsf{K}_{\mathrm{L}} \coloneqq +\widehat{S}_{\mathrm{L}}^y, \qquad \mathsf{K}_{\mathrm{L}} \, \widehat{S}_{\mathrm{L}}^z \, \mathsf{K}_{\mathrm{L}} \coloneqq +\widehat{S}_{\mathrm{L}}^z, \tag{4.38a}$$

and

$$\mathsf{K}_{\mathrm{L}} \, \widehat{J}_{\mathrm{L}}^x \, \mathsf{K}_{\mathrm{L}} \coloneqq +\widehat{J}_{\mathrm{L}}^x, \qquad \mathsf{K}_{\mathrm{L}} \, \widehat{J}_{\mathrm{L}}^y \, \mathsf{K}_{\mathrm{L}} \coloneqq -\widehat{J}_{\mathrm{L}}^y, \qquad \mathsf{K}_{\mathrm{L}} \, \widehat{J}_{\mathrm{L}}^z \, \mathsf{K}_{\mathrm{L}} \coloneqq +\widehat{J}_{\mathrm{L}}^z. \tag{4.38b}$$

Demanding the consistency condition (4.12) to hold, we find the boundary representation

$$\widehat{U}_{\mathrm{e,L}}(t') \coloneqq \widehat{S}_{\mathrm{L}}^x \, \widehat{J}_{\mathrm{L}}^x \, \mathsf{K}_{\mathrm{L}}, \qquad \widehat{U}_{\mathrm{e,L}}(g) \coloneqq \widehat{S}_{\mathrm{L}}^z \, \widehat{J}_{\mathrm{L}}^z. \tag{4.38c}$$

Using the definition (2.2), we associate the pair of indices

$$([\nu_{\mathrm{L}}], [\rho_{\mathrm{L}}]) = (1,0) \tag{4.38d}$$

to the spin-1/2 cluster $c = -4$ chain. This is the same pair of indices as in Eq. (4.32d).

One verifies that the Clifford algebra spanned by the generators

$$\left\{ \widehat{Y}_{2N-3} \, \widehat{X}_{2N-2} \, \widehat{X}_{2N-1} \, \widehat{X}_{2N}, \quad \widehat{Y}_{2N-2} \, \widehat{X}_{2N-1} \, \widehat{X}_{2N}, \quad \widehat{Y}_{2N-1} \, \widehat{X}_{2N}, \quad \widehat{Y}_{2N} \right\} \tag{4.39a}$$

for the right boundary $\Lambda_{\mathrm{R}} \coloneqq \{j = 2N-3, 2N-2, 2N-1, 2N\}$ delivers the pair of indices

$$([\nu_{\mathrm{R}}], [\rho_{\mathrm{R}}]) = (1,0) \tag{4.39b}$$

for the projective representation of $t'$ and $g$ on the right boundary. This is the same pair of indices as in Eq. (4.33b).

### D. The case of $c$ odd

We treat Hamiltonian (4.5a) with open boundary conditions $b = 1$ for $c = \pm 1, \pm 3$ for which we verify explicitly that the case of spin-1/2 cluster odd-$(2c+1)$ chains is brought to the case of spin-1/2 even-$[(2c+1) - (-1)^{\lfloor \frac{2c+1}{4} \rfloor} \text{sgn}(2c+1)]$ chains.

#### 1. The cases of $c = \pm 1$

When $c = \pm 1$, the mean-field Hamiltonian (4.5a) becomes

$$\widehat{H}^{(1)}_{\text{MF}, \pm 1} = \begin{cases} -\sum\limits_{j=1}^{2N} \widehat{Z}_j, & \text{if } c = +1, \\[2ex] -\frac{\hbar\omega}{2} \sum\limits_{j=1}^{2N} \widehat{Y}_j, & \text{if } c = -1, \end{cases} \tag{4.40}$$

which is Hamiltonian (4.13) with $\widehat{X}_j$ replaced by $\widehat{Z}_j$ for $c = +1$ or $\widehat{Y}_j$ for $c = -1$.

Consequently, the set of gapless boundary degrees of freedom is empty, i.e.,

$$\mathfrak{O}_{\text{B}, \pm 1} = \{\,\}, \qquad \text{B} = \text{L}, \text{R}. \tag{4.41}$$

By convention, we associate the trivial indices

$$([\nu], [\rho]) = (0, 0) \tag{4.42}$$

to the spin-1/2 cluster $c = \pm 1$ chains.

#### 2. The case of $c = +3$

When $c = +3$, the mean-field Hamiltonian (4.5a) with open boundary conditions becomes

$$\widehat{H}^{(1)}_{\text{MF}, +3} = \sum_{j=1}^{2N-1} \widehat{Z}_j \widehat{Y}_{j+1} \widehat{Z}_{j+2}. \tag{4.43}$$

This mean-field Hamiltonian follows from Hamiltonian (4.16) upon replacing $\widehat{X}_j$ with $\widehat{Y}_j$.

On verifies the existence of a Clifford algebra $C\ell_2$ on the left boundary $\Lambda_{\text{L}} = \{j = 1, 2\}$ that (i) commutes with Hamiltonian (4.43) and (ii) contains an $\mathfrak{su}(2)$ Lie algebra whose generators are

$$\widehat{S}^x_{\text{L}} := \frac{1}{2} \widehat{Y}_1 \widehat{Z}_2, \qquad \widehat{S}^y_{\text{L}} := -\frac{1}{2} \widehat{X}_1 \widehat{Z}_2, \qquad \widehat{S}^z_{\text{L}} := \frac{1}{2} \widehat{Z}_1. \tag{4.44}$$

We deduce the transformation laws

$$\widehat{U}_{\text{MF}}(\tilde{t}') \left( \widehat{S}^x_{\text{L}} \ \ \widehat{S}^y_{\text{L}} \ \ \widehat{S}^z_{\text{L}} \right)^{\mathsf{T}} \widehat{U}^\dagger_{\text{MF}}(\tilde{t}') = \left( +\widehat{S}^x_{\text{L}} \ \ -\widehat{S}^y_{\text{L}} \ \ +\widehat{S}^z_{\text{L}} \right)^{\mathsf{T}}, \quad \widehat{U}_{\text{MF}}(\tilde{g}) \left( \widehat{S}^x_{\text{L}} \ \ \widehat{S}^y_{\text{L}} \ \ \widehat{S}^z_{\text{L}} \right)^{\mathsf{T}} \widehat{U}^\dagger_{\text{MF}}(\tilde{g}) = \left( -\widehat{S}^x_{\text{L}} \ \ +\widehat{S}^y_{\text{L}} \ \ -\widehat{S}^z_{\text{L}} \right)^{\mathsf{T}}. \tag{4.45a}$$

We define the action $\mathsf{K}_{\text{L}}$ of complex conjugation on the four-dimensional representation (4.44) of the $\mathfrak{su}(2)$ Lie algebra by demanding that

$$\mathsf{K}_{\text{L}} \, \widehat{S}^x_{\text{L}} \, \mathsf{K}_{\text{L}} := -\widehat{S}^x_{\text{L}}, \qquad \mathsf{K}_{\text{L}} \, \widehat{S}^y_{\text{L}} \, \mathsf{K}_{\text{L}} := +\widehat{S}^y_{\text{L}}, \qquad \mathsf{K}_{\text{L}} \, \widehat{S}^z_{\text{L}} \, \mathsf{K}_{\text{L}} := +\widehat{S}^z_{\text{L}}. \tag{4.46a}$$

Demanding the consistency condition (4.12) to hold, we find the boundary representation

$$\widehat{U}_{\text{MF}, \text{L}}(\tilde{t}') := \widehat{S}^z_{\text{L}} \, \mathsf{K}_{\text{L}}, \qquad \widehat{U}_{\text{MF}, \text{L}}(\tilde{g}) := \widehat{S}^y_{\text{L}}. \tag{4.46b}$$

Using the definition (2.2), we associate the pair of indices

$$([\nu_{\mathrm{L}}], [\rho_{\mathrm{L}}]) = (0, 1) \tag{4.46c}$$

to the spin-1/2 cluster $c = +3$ chain. This is the same pair of indices as in Eq. (4.20c).

One verifies that the pair of indices

$$([\nu_{\mathrm{R}}], [\rho_{\mathrm{R}}]) = (0, 1) \tag{4.47}$$

is found for the projective representation of $\tilde{t}'$ and $\tilde{g}$ on the right boundary $\Lambda_{\mathrm{R}} = \{2N - 1, 2N\}$.

### 3. The case of $c = -3$

The case of $c = -3$ is deduced from the case of $c = +3$ by interchanging all the $\widehat{Z}_j$ and $\widehat{Y}_j$ operators for $j = 1, \cdots, 2N$ in the mean-field Hamiltonian (4.43).

One verifies the existence of a Clifford algebra $\mathrm{C}\ell_2$ on the left boundary

$$\Lambda_{\mathrm{L}} = \{j = 1, 2\} \tag{4.48}$$

that (i) commutes with the mean-field Hamiltonian

$$\widehat{H}_{\mathrm{MF},-3}^{(1)} = \sum_{j=1}^{2N-1} \widehat{Y}_j \, \widehat{Z}_{j+1} \, \widehat{Y}_{j+2} \tag{4.49}$$

and (ii) contains an $\mathfrak{su}(2)$ Lie algebra whose generators are

$$\widehat{S}_{\mathrm{L}}^x \coloneqq \frac{1}{2} \widehat{Z}_1 \, \widehat{Y}_2, \qquad \widehat{S}_{\mathrm{L}}^y \coloneqq \frac{1}{2} \widehat{X}_1 \, \widehat{Y}_2, \qquad \widehat{S}_{\mathrm{L}}^z \coloneqq \frac{1}{2} \widehat{Y}_1. \tag{4.50}$$

We deduce the transformation laws

$$\widehat{U}_{\mathrm{MF}}(\tilde{t}') \left( \widehat{S}_{\mathrm{L}}^x \; \widehat{S}_{\mathrm{L}}^y \; \widehat{S}_{\mathrm{L}}^z \right)^{\mathsf{T}} \widehat{U}_{\mathrm{MF}}^\dagger(\tilde{t}') = \left( -\widehat{S}_{\mathrm{L}}^x \; -\widehat{S}_{\mathrm{L}}^y \; -\widehat{S}_{\mathrm{L}}^z \right)^{\mathsf{T}}, \quad \widehat{U}_{\mathrm{MF}}(\tilde{g}) \left( \widehat{S}_{\mathrm{L}}^x \; \widehat{S}_{\mathrm{L}}^y \; \widehat{S}_{\mathrm{L}}^z \right)^{\mathsf{T}} \widehat{U}_{\mathrm{MF}}^\dagger(\tilde{g}) = \left( -\widehat{S}_{\mathrm{L}}^x \; +\widehat{S}_{\mathrm{L}}^y \; -\widehat{S}_{\mathrm{L}}^z \right)^{\mathsf{T}}. \tag{4.51a}$$

Third, we denote by $\mathbb{1}_{\mathrm{L}}$ the unit $4 \times 4$ matrix and we define the action $\mathsf{K}_{\mathrm{L}}$ of complex conjugation on the four-dimensional representation (4.50) of the $\mathfrak{su}(2)$ Lie algebra by demanding that

$$\mathsf{K}_{\mathrm{L}} \, \widehat{S}_{\mathrm{L}}^x \, \mathsf{K}_{\mathrm{L}} \coloneqq -\widehat{S}_{\mathrm{L}}^x, \qquad \mathsf{K}_{\mathrm{L}} \, \widehat{S}_{\mathrm{L}}^y \, \mathsf{K}_{\mathrm{L}} \coloneqq +\widehat{S}_{\mathrm{L}}^y, \qquad \mathsf{K}_{\mathrm{L}} \, \widehat{S}_{\mathrm{L}}^z \, \mathsf{K}_{\mathrm{L}} \coloneqq +\widehat{S}_{\mathrm{L}}^z. \tag{4.52a}$$

Demanding the consistency condition (4.12) to hold, we find the boundary representation

$$\widehat{U}_{\mathrm{MF},\mathrm{L}}(\tilde{t}') \coloneqq \widehat{S}_{\mathrm{L}}^x \, \mathsf{K}_{\mathrm{L}}, \qquad \widehat{U}_{\mathrm{MF},\mathrm{L}}(\tilde{g}) \coloneqq \widehat{S}_{\mathrm{L}}^y. \tag{4.52b}$$

Using the definition (2.2), we associate the pair of indices

$$([\nu_{\mathrm{L}}], [\rho_{\mathrm{L}}]) = (1, 1) \tag{4.52c}$$

to the spin-1/2 cluster $c = -3$ chain. This is the same pair of indices as in Eq. (4.25c).

One verifies that the pair of indices

$$([\nu_{\mathrm{R}}], [\rho_{\mathrm{R}}]) = (1, 1) \tag{4.53}$$

is found for the projective representation of $\tilde{t}'$ and $\tilde{g}$ on the left boundary $\Lambda_{\mathrm{R}} = \{2N - 1, 2N\}$.

---



\end{document}